\documentclass[a4paper,usenames,dvipsnames,11pt]{article}
\pdfoutput=1

\usepackage{jheppub}
\usepackage{slashed}
\usepackage{mathrsfs,booktabs,multirow,tabularx}
\usepackage{stmaryrd}
\usepackage{xspace}
\usepackage{fancyvrb}
\usepackage[makeroom]{cancel}
\usepackage{amsmath}    
\usepackage{amssymb}    %
\usepackage{graphicx}   
\usepackage{verbatim}   
\usepackage{lscape}
\usepackage{subfig}
\usepackage{listings}
\usepackage{enumitem}
\usepackage{mathtools}
\usepackage{blkarray}

\def\beq{\begin{equation}}
\def\beqn{\begin{eqnarray}}
\def\eeq{\end{equation}}
\def\eeqn{\end{eqnarray}}
\def\abs#1{\left|#1\right|}
\def\ket#1{|#1\rangle}
\def\bra#1{\langle #1|}
\def\brat#1{\langle #1}
\def\eik#1#2{\Big[#1,#2\Big]}
\def\Feq#1{eq.~({\bf I}.#1)}
\def\binomial#1#2{\left(\begin{array}{cc}#1\\#2\end{array}\right)}
\def\stirlingS1#1#2{\left[\begin{array}{cc}#1\\#2\end{array}\right]}

\def\hult#1#2{{\cal S}_{\sss\rm H}\left(#1,#2\right)}
\def\Shult#1#2{{\cal T}_{\sss\rm H}\left(#1,#2\right)}
\def\Sset#1#2{S^{(#1)}_{#2}}
\def\bsigmaqRarg#1{\bar{\sigma}_{q{\rm R}#1}}
\def\bsigmaqLarg#1{\bar{\sigma}_{q{\rm L}#1}}
\def\bsigmapqLarg#1{\bar{\sigma}^\prime_{q{\rm L}#1}}
\def\nloops#1{\abs{#1}}

\DeclarePairedDelimiter\ceil{\lceil}{\rceil}
\DeclarePairedDelimiter\floor{\lfloor}{\rfloor}

\newcommand\sss{\scriptscriptstyle}
\newcommand\mydot{\!\cdot\!}

\newcommand\half{\frac{1}{2}}

\newcommand\as{\alpha_{\sss S}}
\newcommand\gs{g_{\sss S}}
\newcommand\aem{\alpha}

\newcommand\amp{{\cal A}}

\newcommand\ampn{\amp^{(n)}}

\newcommand\ampQGBgg{\amp^{(2q;n)}}

\newcommand\ampCS{\widehat{\cal A}}

\newcommand\ampCSn{\ampCS^{(n)}}

\newcommand\ampCSQGBgg{\ampCS^{(2q;n)}}
\newcommand\ampCSQGBggz{\ampCS^{(2q;0)}}

\newcommand\ampOS{\bar{\cal A}}
\newcommand\ampOSn{\ampOS^{(n)}}
\newcommand\ampOSQGBgg{\ampOS^{(2q;n)}}
\newcommand\ampsq{{\cal M}}

\newcommand\ampsqn{\ampsq^{(n)}}

\newcommand\ampsqnpoS{\ampsq^{(n+1)}_{\rm\sss SOFT}}

\newcommand\ampsqQGRS{\ampsq^{(2q;n+1)}_{\rm\sss SOFT}}
\newcommand\ampsqQGRSz{\ampsq^{(2q;1)}_{\rm\sss SOFT}}

\newcommand\ampsqQGBgg{\ampsq^{(2q;n)}}
\newcommand\ampsqQGBggz{\ampsq^{(2q;0)}}

\newcommand\Wampsq{\widetilde{\ampsq}}

\newcommand\seta{\{a_i\}}

\newcommand\setan{\{a_i\}_{i=1}^n}

\newcommand\setij{\{i_pj_p\}}
\newcommand\setijn{\{i_p,j_p\}_{p=1}^n}

\newcommand\setaQQ{\{a_i\}_{i=-2q}^{-1}}
\newcommand\setaQQf{\{a_i\}_{i=-4}^{-1}}
\newcommand\setaQG{\{a_i\}_{i=-2q}^n}

\newcommand\setaQGBgg{\{a_i\}_{i=-2q}^n}

\newcommand\amtq{a_{-2q}}
\newcommand\amoq{a_{-q}}

\newcommand\sigmap{\sigma^\prime}

\newcommand\gammap{\gamma^\prime}
\newcommand\igamma{\gamma^{-1}}
\newcommand\igammap{\gamma^{\prime^{-1}}}

\newcommand\CA{C_{\sss A}}
\newcommand\CF{C_{\sss F}}
\newcommand\TF{T_{\sss F}}

\newcommand\qb{\bar{q}}
\newcommand\ub{\bar{u}}

\newcommand\flowBgg{{\cal F}_{2q;n}}
\newcommand\flowBggz{{\cal F}_{2q;0}}
\newcommand\flowBggnk{{\cal F}_{2q;n-k}}
\newcommand\flowBggnp{{\cal F}_{2q;n-p}}

\newcommand\ident{{\cal I}}
\newcommand\Mp{-\!p}

\newcommand\spaceij{S_{\{ij\}}}
\newcommand\spacea{S_{\{a\}}}

\newcommand\loopset{{\cal L}}

\newcommand\bq{\bar{q}}
\newcommand\bsigma{\bar{\sigma}}
\newcommand\bsigmap{\bar{\sigma}^\prime}
\newcommand\sigmamo{\sigma^{-1}}
\newcommand\sigmapmo{\sigma^{\prime -1}}

\newcommand\bgamma{\bar{\gamma}}
\newcommand\bgammap{\bar{\gamma}^\prime}
\newcommand\hgamma{\hat{\gamma}}

\newcommand\LA{\Lambda^{(A)}}
\newcommand\LB{\Lambda^{(B)}}

\newcommand\bQ{\bar{Q}}
\newcommand\sigmaqR{\sigma_{q{\rm R}}}
\newcommand\sigmaqL{\sigma_{q{\rm L}}}
\newcommand\sigmapqL{\sigma^\prime_{q{\rm L}}}

\newcommand\setaQGtwo{\{a_i\}_{i=-4}^n}

\newcommand\proja{{\cal P}_{\parallel}}
\newcommand\projp{{\cal P}_{\perp}}
\newcommand\projap{\proja^{(p)}}
\newcommand\projpp{\projp^{(p)}}
\newcommand\projpo{\projp^{(1)}}
\newcommand\projpsk{\projp^{(s_k)}}

\newcommand\projapsq{\proja^{{(p)}^2}}
\newcommand\projppsq{\projp^{{(p)}^2}}
\newcommand\projpsksq{\projp^{{(s_k)}^2}}

\newcommand\setapQQ{\{a_i^\prime\}_{i=-2q}^{-1}}
\newcommand\setijp{\{i_p^\prime,j_p^\prime\}}
\newcommand\setijpn{\{i_p^\prime,j_p^\prime\}_{p=1}^n}
\newcommand\qRop{q{\rm R}}
\newcommand\qLop{q{\rm L}}
\newcommand\dipset{{\cal G}}
\newcommand\bdipset{\overline{{\cal G}}}
\newcommand\dipeik{{\cal D}}
\newcommand\flowexo{{\cal F}_{2;2}}
\newcommand\Sflowexot{{\cal F}_{2;2}}
\newcommand\Sflowexoo{{\cal F}_{2;1}}
\newcommand\Sflowexoz{{\cal F}_{2;0}}
\newcommand\Qa{-\!1}
\newcommand\Qb{-2}
\newcommand\bo{\bar{1}}
\newcommand\bt{\bar{2}}
\newcommand\eiksum{{\cal E}}
\newcommand\ampOSQGexo{\ampOS^{(2;2)}}
\newcommand\ampCSQGexo{\ampCS^{(2;2)}}
\newcommand\ampsqQGexo{\ampsq^{(2;2)}}
\newcommand\ampsqQGSexo{\ampsq^{(2;3)}_{\rm\sss SOFT}}

\newcommand\ampsqQGext{\ampsq^{(4;1)}}
\newcommand\ampsqQGSext{\ampsq^{(4;2)}_{\rm\sss SOFT}}

\title{The role of colour flows in matrix element computations 
and Monte Carlo simulations
}

\author[a,b]{Stefano Frixione,}
\affiliation[a]{INFN, Sezione di Genova, Via Dodecaneso 33, I-16146, 
Genoa, Italy}
\affiliation[b]{ PH Department, TH Division, CERN, CH-1211 Geneva 23, 
Switzerland}

\author[c]{Bryan R. Webber}
\affiliation[c]{Cavendish Laboratory, University of Cambridge,
  J.J.~Thomson Avenue, Cambridge CB3 0HE, U.K.}

\emailAdd{Stefano.Frixione@cern.ch}
\emailAdd{webber@hep.phy.cam.ac.uk}

\abstract{
We discuss how colour flows can be used to simplify the computation
of matrix elements, and in the context of parton shower Monte Carlos 
with accuracy beyond leading-colour. We show that, by systematically
employing them, the results for tree-level matrix elements and their soft
limits can be given in a closed form that does not require any colour
algebra. The colour flows that we define are a natural generalization
of those exploited by existing Monte Carlos; we construct their representations
in terms of different but conceptually equivalent quantities, namely colour 
loops and dipole graphs, and examine how these objects may help to extend 
the accuracy of Monte Carlos through the inclusion of subleading-colour
effects. We show how the results that we obtain can be used, with
trivial modifications, in the context of QCD+QED simulations, since
we are able to put the gluon and photon soft-radiation patterns on
the same footing. We also comment on some peculiar properties of
gluon-only colour flows, and their relationships with established
results in the mathematics of permutations.
}

\keywords{QCD, Monte Carlo}

\preprint{
\begin{flushright}
CERN-TH-2021-098\\
\end{flushright}
}

\begin{document}
\maketitle
\flushbottom

\section{Introduction\label{sec:intro}}
Barring an unlikely turn of events, searches for new physics at the LHC
will increasingly rely upon high-precision methods, on both the experimental
(with the collection of large-statistics samples and the reduction of
systematics) and the theoretical sides. An obvious benefit of high-precision 
results is that they also allow one to carry out progressively more
stringent checks on the (supposedly) well-known Standard Model physics,
which in turn help stress-test, and ultimately hone, theory predictions.

From the theoretical viewpoint, meeting the precision targets that
current and future experimental data demand almost always requires
performing computations that are perturbative in the series expansion
in terms of a coupling constant (the so-called fixed-order predictions)
at the highest possible accuracy. At the LHC, one is obviously dominated 
by QCD effects, but electroweak contributions must nowadays be considered 
as well, especially in the large transverse momentum regions where some 
characteristic new-physics phenomena are expected to show up. In the past 
few years, the progress in fixed-order technologies has been extremely
significant, for example with next-to-next-to-leading order (NNLO) results 
becoming routinely available. That being said, fixed-order predictions
cannot generally be directly compared to experimental data: even when they
are fully differential (a feature that is more difficult to achieve the
higher the perturbative accuracy of the computation), hadronization and
some acceptance corrections generally need be applied. 

It is therefore desirable that such corrections also be carried out
at the highest possible accuracy, so as not to spoil that of the 
underlying fixed-order calculations. This operation more often than
not involves the matching and/or merging of fixed-order results with
parton-shower Monte Carlo (PSMC) simulations. Unfortunately, the systematic
application of matching and merging techniques is currently limited to
next-to-leading order (NLO) at most. Furthermore, the PSMCs employed
in the context of these techniques are formally of leading-logarithmic
(LL) and leading-colour (LC) accuracy, although in practice they perform 
(in comparison to analytically-resummed computations) better than that.

Another issue that emerges when one considers large-statistics data samples
is that many-jet observables become accessible; loosely speaking, this may
also be viewed theoretically as a high-precision problem, although of
a different nature from those posed by higher-order computations. 
In particular, here one must be able to calculate the matrix 
elements\footnote{Note that throughout this paper we use the term 
``matrix element'' to denote the modulus-squared of the sum of 
amplitudes relevant to the computation of a cross section.} associated 
with a large number of partons in an efficient way. The complexity of this 
task grows rapidly, and sooner or later one must consider approximated, 
as opposed to exact, matrix-element results.

The primary aim of this paper is to illustrate how several of the problems
outlined thus far can be addressed by using colour flows.
This applies to both tree-level matrix element computations, and to the
inclusion of beyond-LC effects in PSMCs. We argue that by employing colour
flows at both the matrix-element and the PSMC level one has the 
advantage of a language common to the two cases, which should facilitate
the interplay between them in the attempt at increasing the overall
precision of the final predictions. Here, we focus mainly on the 
use of colour flows in tree-level matrix elements to provide a 
full-colour treatment of the initial conditions for parton showers. 
The same approach can in principle be applied in the subsequent 
evolution of the showers. We remark that the extension of
PSMCs beyond leading (logarithmic and/or colour) accuracy is the subject 
of much ongoing activity (see e.g.~refs.~\cite{Nagy:2012bt,Nagy:2015hwa,
Hamilton:2020rcu,DeAngelis:2020rvq,AngelesMartinez:2018cfz,Isaacson:2018zdi,
Giele:2011cb,Jadach:2011kc,Hartgring:2013jma,Gituliar:2014eba,Li:2016yez,
Hoche:2017iem,Hoche:2017hno,Dulat:2018vuy,Bewick:2019rbu,Dasgupta:2020fwr}) 
in directions that are possibly orthogonal to the methods discussed here; 
a comparison between all these developments is beyond the scope of the 
present work.

We point out that while the concept of colour flow is well known,
our work exploits it at the level of matrix elements (i.e.~of amplitudes
squared), in contrast to what is typically done in the literature (see 
e.g.~refs.~\cite{Mangano:1990by,DelDuca:1999rs,Maltoni:2002mq,
Johansson:2015oia,Melia:2015ika,Sjodahl:2018cca} and references therein), 
where the emphasis is at the amplitude 
level. By doing so, we are naturally led to introduce quantities (such
as closed flows and secondary flows) that help to express the matrix elements 
in a very compact way that does not require any further colour algebra, and 
is independent of whether the original amplitudes are written in terms 
of a fundamental or a colour-flow basis. An interesting question
that thus emerges is whether similar expressions could be obtained for
matrix elements beyond the tree-level (which, at one loop, would be
a natural complement of the real-emission subtraction procedure presented
in ref.~\cite{Frixione:2011kh}). While the general ideas on colour
flows we employ here are certainly valid beyond tree-level, there are 
specific aspects of the representations of loop amplitudes in terms
of colour bases (see e.g.~refs.~\cite{Bern:1990ux,Bern:1994zx,Bern:1994cg,
Kilian:2012pz,Reuschle:2013qna,Kalin:2017oqr}) that require an explicit 
analysis, and whose implications are therefore left to future work.

This paper is organized as follows. In view of the fact that we
often rely on the methods of ref.~\cite{Frixione:2011kh}, in 
sect.~\ref{sec:basic} we give a brief summary of the results of that
paper that are directly relevant here. In sect.~\ref{sec:loops},
we define the colour loops that stem from a given choice of colour flows. 
In sect.~\ref{sec:vspace} we discuss the so-called flow representation of 
matrix elements in geometrical terms. Sects.~\ref{sec:tree} and~\ref{sec:soft}
present the results for tree-level matrix elements and their soft limits,
respectively, expressed in terms of elementary quantities associated with 
colour flows. In sect.~\ref{sec:MCs} we show how the results obtained
for the matrix elements could be employed in the context of PSMCs
that include beyond-LC effects, and we propose two different but equivalent
ideas that generalize the concept of colour connections used in existing
PSMCs. In sect.~\ref{sec:mix} we illustrate how colour flows provide a
natural language to deal with the simultaneous perturbative expansion in 
two parameters, using the QCD+QED case as a definite example.
Finally, in sect.~\ref{sec:conc} we summarize our findings.
The appendices collect some technical material. In app.~\ref{sec:gflows}
miscellaneous results are presented that are relevant to gluon-only
colour flows, with the reinterpretation of some of their properties in the
context of the mathematics of permutations, whereas app.~\ref{sec:flvsdg}
gives some additional information on an idea introduced in sect.~\ref{sec:MCs}
that is particularly relevant to dipole-based showers.

\section{Summary of basic concepts\label{sec:basic}}
The present work makes extensive use of the results of ref.~\cite{Frixione:2011kh}.
For the reader who is not familiar with that paper, in this section we 
give a very brief summary of the ideas that underpin the derivations that 
will follow later. Henceforth, eq.~(x.y) of ref.~\cite{Frixione:2011kh} 
will be denoted by \Feq{x.y}.

All partons\footnote{Assumed to be massless; the generalization
to massive partons can be made along similar lines.} that participate in a 
hard scattering are regarded as outgoing. For a process with $q$ 
quark-antiquark pairs and $n$ gluons
\beq
0\;\longrightarrow\; 2q+n
\label{QGRproc}
\eeq
the labelling conventions for the $i^{th}$ parton are as follows:
\beqn
-2q\le i\le -q-1
\;\;\;\;&\longrightarrow&\;\;\;\;{\rm antiquarks}\,,
\label{qbars}
\\
-q\le i\le -1\phantom{-q}\;\;
\;\;\;\;&\longrightarrow&\;\;\;\;{\rm quarks}\,,
\label{qs}
\\
1\le i\le n\phantom{q-1\;}
\;\;\;\;&\longrightarrow&\;\;\;\;{\rm gluons}\,.
\label{glus}
\eeqn
Tree-level scattering amplitudes are written as sums of products of 
dual (or colour-ordered) amplitudes times colour factors. For any given 
choice of the representation of the $SU(N)$ colour algebra, colour
factors are in one-to-one correspondence with colour flows, which
are ordered sets of parton labels, and can be obtained from the
latter by means of simple rules. For the process of eq.~(\ref{QGRproc}),
the colour flows have the following form:
\beqn
\gamma&=&\bigcup_{p=1}^q \gamma_p\,,
\label{qgflow}
\\
\gamma_p&=&\Big(\Mp\,;\sigma(t_{p-1}+1),\ldots\sigma(t_p);\mu(-p-q)\Big)\,.
\label{qgflowp}
\eeqn
Here, $\sigma$ and $\mu$ denote permutations (including cyclic ones) of 
the first $n$ and $q$ integers\footnote{In eq.~(\ref{qgflowp}) and elsewhere,
$\mu(-p-q)$ is symbolic for $-\mu(p)-q$.}, respectively, while the set of $q+1$ 
integers $t_p$
\beqn
&&t=\left\{t_0,\ldots t_q\right\}\,,
\label{part1}
\\*&&
0=t_0\le t_1\le\ldots t_{q-1}\le t_q=n\,,
\label{part2}
\eeqn
is a partition of the ordered set of the first $n$ integers into
$q$ subsets of ordered integers, with the $p^{th}$ cell of the partition
defined to be\footnote{Equation~(\ref{pcell}) applies only to the non-trivial 
case $t_{p-1}<t_p$; when $t_{p-1}=t_p$, the $p^{th}$ cell of the partition 
is $(t_{p-1},t_p)$. See also footnote~\ref{ft:empty}.}
\beq
(t_{p-1}+1,t_{p-1}+2,\ldots t_p-1,t_p)\,,\;\;\;\;\;\;\;\;1\le p\le q\,.
\label{pcell}
\eeq
The set of $m$-integer permutations is denoted by $P_m$; the set of the
partitions of eqs.~(\ref{part1}) and~(\ref{part2}) is denoted by $T_{n|q}$.
Thus, the set of colour flows relevant to the process of eq.~(\ref{QGRproc})
is given by \Feq{3.27}
\beq
\flowBgg=\left(P_n,P_q,T_{n|q}\right)\,,
\label{Bflowdef}
\eeq
whose number of elements is equal to
\beq
\abs{\flowBgg}=\abs{P_n}\abs{P_q}\abs{T_{n|q}}=
n!\,q!\binomial{n+q-1}{q-1} = q\,(n+q-1)!\,.
\label{dimBflow}
\eeq
Each of the ordered sets in $\gamma_p$ in eq.~(\ref{qgflow}) is called an
open colour line, that connects quark $-p$ with antiquark $\mu(-p-q)$
and features the emission of gluon $\sigma(t_{p-1}+1)$ (which is
the closest to the quark) up to gluon $\sigma(t_p)$ (which is the closest
to the antiquark). Denoting by
\beqn
&&\left\{a_i\right\}_{i=1}^n\,,\;\;\;\;\;\;\;\;\;\;\,
a_i\,\in\,\left\{1,\ldots N^2-1\right\}\,,
\label{coladj}
\\
&&\left\{a_i\right\}_{i=-2q}^{-1}\,,\;\;\;\;\;\;\;
a_i\,\in\,\left\{1,\ldots N\right\}\,,
\label{colfund}
\eeqn
the colours of gluons, and of quarks and antiquarks, respectively,
the colour flow of eq.~(\ref{qgflow}) is associated with the following
colour factor in the fundamental representation\footnote{If an open
colour line does not feature any gluons, the corresponding colour factor 
is a Kronecker $\delta$ -- see~\Feq{3.23} and~\Feq{3.24}. Furthermore,
throughout this paper $\lambda$ matrices are normalized as the $SU(3)$ 
generators, per app.~A of ref.~\cite{Frixione:2011kh}.\label{ft:empty}.}
\beqn
\Lambda\left(\seta,\gamma\right)&=&
N^{-\rho(\gamma)}\,
\Big(\lambda^{a_{\sigma(t_0+1)}}\ldots
\lambda^{a_{\sigma(t_1)}}\Big)_{a_{-1}a_{\mu(-1-q)}}
\nonumber\\*&&\phantom{N^{aa}\;}\times
\Big(\lambda^{a_{\sigma(t_1+1)}}\ldots
\lambda^{a_{\sigma(t_2)}}\Big)_{a_{-2}a_{\mu(-2-q)}}
\nonumber\\*&&\phantom{N^{aa}\;}\times\ldots
\nonumber\\*&&\phantom{N^{aa}\;}\times
\Big(\lambda^{a_{\sigma(t_{q-1}+1)}}\ldots
\lambda^{a_{\sigma(t_q)}}\Big)_{\amoq a_{\mu(-2q)}}\,,
\label{Lambdagq}
\eeqn
where
\beq
\rho(\gamma)=\min\Big\{q-1,\,
\sum_{p=1}^q\delta\Big(\!\Mp-q,\mu(-p-q)\Big)\Big\}\,.
\label{rhodef}
\eeq
The integer $\rho(\gamma)$ is equal to the number of open colour lines
that coincide with flavour lines\footnote{In the unequal-flavour case;
see \Feq{3.30}--\Feq{3.32} when there are at least two equal flavour 
pairs.}, minus one in the case of maximal coincidence. Given a colour
configuration, i.e.~a specific choice for the colours of quarks, antiquarks,
and gluons, the scattering amplitude is written as was anticipated, i.e.
\beq
\ampQGBgg(\amtq,\ldots a_n)=
\sum_{\gamma\in\flowBgg}\,\Lambda\left(\seta,\gamma\right)\,
\ampCSQGBgg(\gamma)\,,
\label{mgCDampQG}
\eeq
where $\ampCSQGBgg(\gamma)$ is the dual amplitude associated with
the colour flow $\gamma$. In ref.~\cite{Frixione:2011kh} it has been
found convenient to represent amplitudes in a vector space, called
colour space, and denoted by $\spacea$. The set of all colour 
configurations corresponds to a complete basis of orthonormal
vectors in $\spacea$, thus\footnote{Throughout this paper, summation
and product indices may have ranges that include zero; we understand
the corresponding contributions to be null.}
\beqn
\ket{\amtq,\ldots a_n}&\in&\spacea\,,
\\
\brat{b_{-2q},\ldots b_n}\ket{\amtq,\ldots a_n}&=&
\prod_{i=-2q}^n\delta_{a_ib_i}\,,
\label{basisnorm}
\\
\sum_{\setaQG}\ket{\amtq,\ldots a_n}\bra{\amtq,\ldots a_n}&=&I\,.
\label{basisproj}
\eeqn
The colour space is separable into the subspaces $\spacea^{(p)}$ relevant
to each parton $p$ that enters the hard process
\beqn
&&\spacea=\bigotimes_{p=-2q}^n\spacea^{(p)}\,,
\label{vspaces}
\\
&&\ket{\amtq,\ldots a_n}=\bigotimes_{p=-2q}^n \ket{a_p}\,,
\;\;\;\;\;\;\;\;
\ket{a_p}\in\spacea^{(p)}\,.
\label{vecsep}
\eeqn
The dimensions of these vector spaces are
\beq
\abs{\spacea^{(p)}}=
\left\{\begin{array}{ll}
N\phantom{aaaaaa}-2q\le p\le -1\\
N^2-1\phantom{aaaaa}1\le p\le n
\end{array}\right.
\;\;\;\;\Longrightarrow\;\;\;\;
\abs{\spacea}=N^{2q}(N^2-1)^n\,.
\eeq
With this, one can introduce vectors associated with amplitudes
at given colour configurations or colour flows, thus
\beqn
&&\ket{\ampQGBgg(\amtq,\ldots a_n)}=
\ampQGBgg(\amtq,\ldots a_n)\,\ket{\amtq,\ldots a_n}\,,
\label{QGAcoldef}
\\
&&\ket{\ampQGBgg(\gamma)}=\sum_{\setaQGBgg}
\Lambda\left(\seta,\gamma\right)\,
\ampCSQGBgg(\gamma)\,\ket{\amtq,\ldots a_n}\,,
\label{QGAflowdef}
\eeqn
from which one can obtain the colour-summed amplitude
\beqn
\ket{\ampQGBgg}&=&\sum_{\setaQG}\ket{\ampQGBgg(\amtq,\ldots a_n)}
\label{ampCDQG}
\\
&=&\sum_{\gamma\in\flowBgg}\ket{\ampQGBgg(\gamma)}\,.
\label{ampflowQG}
\eeqn
By interfering a colour-summed amplitude with its complex conjugate
one obtains the matrix element that, multiplied by flux, average,
and phase-space factors gives the physical cross section
\beq
\ampsqQGBgg\equiv\brat{\ampQGBgg}\ket{\ampQGBgg}=
\sum_{\gammap,\gamma\in\flowBgg}\ampsqQGBgg(\gammap,\gamma)\,,
\label{ampsqQG}
\eeq
with
\beqn
\ampsqQGBgg(\gammap,\gamma)&=&
\brat{\ampQGBgg(\gammap)}\ket{\ampQGBgg(\gamma)}
\\*&=&
\ampCSQGBgg(\gammap)^\star \,C(\gammap,\gamma)\,\ampCSQGBgg(\gamma)\,.
\label{Mmflowsc}
\eeqn
The r.h.s.~of eq.~(\ref{Mmflowsc}) is expressed in terms of
scalar quantities, namely the dual amplitudes already introduced,
and the colour-flow matrix element
\beq
C(\gammap,\gamma)=\sum_{\setaQG}\Lambda\left(\seta,\gammap\right)^\star
\Lambda\left(\seta,\gamma\right)\,.
\label{CFQGmatdef}
\eeq
Equations~(\ref{ampsqQG})--(\ref{CFQGmatdef}) give us the opportunity to
stress a fact left implicit so far. Namely, a colour flow is a
quantity defined at the amplitude level. In the computation of cross
sections, what is relevant is always a pair of two flows, one on each
side of the cut. Such a pair, i.e.
\beq
(\gammap,\gamma)
\label{closeddefQG}
\eeq
in the expressions above, is called a {\em closed colour flow}. In 
eq.~(\ref{closeddefQG}), it is understood that $\gammap$ ($\gamma$) is 
the flow entering the amplitude on the left-hand side (right-hand side) 
of the cut. Because of this, when it is necessary to distinguish between
the colour flows on either side of the cut they will be called L-flows 
and R-flows.

The formulae written so far are relevant to the process of 
eq.~(\ref{QGRproc}). Although in principle that could be used to
describe purely gluonic processes as well (i.e.~when $q=0$), in practice
such a case entails significant simplifications that justify the
use of a slightly different notation. In particular, for a purely-gluonic
$n$-body process
\beq
0\;\longrightarrow\;n\,,
\label{ggBproc}
\eeq
the analogue of the colour flow of eq.~(\ref{qgflow}) features a single 
colour line, denoted as follows:
\beq
\sigma\equiv\left(\sigma(1),\ldots\sigma(n)\right)\,.
\eeq
Furthermore, in view of the fact that its associated colour factor,
i.e.~the analogue of eq.~(\ref{Lambdagq}), reads in the fundamental
representation
\beq
\Lambda\left(\seta,\sigma\right)=
{\rm Tr}\Big(\lambda^{a_{\sigma(1)}}\ldots\lambda^{a_{\sigma(n)}}\Big)\,,
\label{Lambdashort}
\eeq
the information embedded in the colour flow must be invariant under
cyclic permutations. Therefore, the set of the $n$-gluon colour
flows is the set of the non-cyclic $n$-integer permutations, which
we shall denote by
\beq
P_n^\prime\,,
\eeq
and which constitutes the analogue of the set introduced in
eq.~(\ref{Bflowdef}). These, and the formal replacements
\beqn
\ampQGBgg&\longrightarrow&\ampn\,,
\\
\ampsqQGBgg&\longrightarrow&\ampsqn\,,
\eeqn
are all that is needed in order to convert the expressions for the
quark-gluon matrix elements and related quantities into those
relevant to the gluon-only case.

\section{Colour loops\label{sec:loops}}
In the context of current parton shower Monte Carlos,
colour flows stemming from matrix element computations are responsible
for giving crucial information on the initial conditions for the showers
(and for hadronization); therefore, ultimately they have a major role in 
the determination of the radiation pattern.

In retrospect, this might appear strange for two reasons. Firstly, as
was discussed in sect.~\ref{sec:basic}, the quantities that are relevant
to matrix elements are not flows, but closed flows. Secondly, such flows
are Born-level (i.e.~$m$-body) objects, while even the first-emission 
radiation pattern is determined, in cross section computations, by
$(m+1)$-body objects. As far as the first item is concerned, the key
point is that existing PSMCs only consider closed flows for which the L-flow
coincides with the R-flow; in other words, only the diagonal of
the colour-flow matrix is taken into account, which is consistent with
the leading-$N$ nature of the MCs. Obviously, then, the information on
the closed flow is redundant, and that of the flow is used instead.
In turn, this immediately suggests that if one is interested in going
beyond the leading-$N$ approximation in MCs, retaining the information
on closed flows (e.g.~in determining colour connections) will be essential.
As far as the second observation above is concerned, this has again to
do with the leading-$N$ character of PSMCs, which is such that the 
radiation pattern at the $(m+1)$-body level can be described by
$m$-body quantities. Beyond the leading-$N$ this is not true any longer.
However, we shall show how one can circumvent this difficulty, and thus
to keep on working with $m$-body objects.

Before proceeding, we need to define precisely colour loops, which will
be used throughout this paper. Firstly, colour loops are properties
of a given closed colour flow. The set of all colour loops relevant
to the closed flow $(\gammap,\gamma)$ will be denoted by
\beq
\loopset(\gammap,\gamma)\,.
\label{loopsetdef}
\eeq
Each loop is defined iteratively, as follows:
\begin{enumerate}
\setcounter{enumi}{-1}
\item Write the L- and R-flow as two vertical sets of indices (the former
to the left of the latter). Each open colour line that belongs to these 
flows is a separate subset of such sets. The relative order of these subsets
is irrelevant.
\item
Start from either a quark or a gluon that belongs to the R-flow;
this is the first element of the loop.
\item
Go down one place in the R-flow; a gluon or an antiquark is found.
\item
Jump to the L-flow, landing on the element whose label is identical
to that of the gluon or the antiquark of the previous step.
\item
Go up one place in the L-flow; a gluon or a quark is found.
\item
Jump to the R-flow, landing on the element whose label is identical
to that of the gluon or the quark of the previous step.
\item When the landing element coincides with the first element of 
the loop then go to step 7, otherwise iterate steps 2--5.
\item A loop is the ordered set composed of the starting element and of all 
of the landing elements subsequently found, except for the last one (because 
it coincides with the starting element).
\end{enumerate}
Once a loop has been constructed, the next one (if it exists) can be
found by following again steps 1--7, with the condition that the starting
element in step 1 must not be already included as an R-element of
one of the loops previously constructed.
We call L- and R-elements of a given loop $\ell\in\loopset$ the landing
elements when jumping from right to left and from left to right, respectively; 
on top of those, the starting element of the loop is included in the set
of R-elements as well.

By construction, colour loops have the following
properties:
\begin{enumerate}[label=\roman{*}.]
\item
The sets of R-elements of any two loops are disjoint (i.e.~have an
empty intersection); the same is true for the L-elements.
\item
All quarks are R-elements, and all antiquarks are L-elements.
\item
A quark label will appear only once, i.e.~it belongs to a single loop;
the same is true for antiquarks.
\item
A gluon label will appear twice, either in the same loop or in two 
different loops. In the former case, one instance corresponds to an
R-element, and the other to an L-element (i.e.~the gluon label cannot
correspond to two L- or two R-elements).
\end{enumerate}
These properties will turn out to be important in deriving results
for the soft limit of matrix elements. We point out that the idea 
that underpins the construction of colour loops as given above is based 
on the usual procedure of following colour (as opposed to anticolour) 
indices\footnote{Needless to say, this is a purely conventional choice: 
one might choose the anticolour instead, and the procedure would be 
unchanged except for the formal replacements left vs right and up vs down.}. 
This is straightforward in the case of quarks and antiquarks; for gluons, 
one needs to consider the pair of $\lambda$ matrices associated with any 
gluon index, and to replace them with the first term of the Fierz identity. 
Roughly speaking, this corresponds to ``splitting'' colour lines by replacing 
each gluon there with a $q\bq$ pair (of a fictitious flavour,  different from 
all of the other flavours); this idea will play an important role starting
from sect.~\ref{sec:treeqg}.

We can also introduce the idea of distance within a colour loop.
More precisely, for any two parton labels $k$ and $l$ that belong
to a loop $\ell$
\beq
k\,,l\,\in\,\ell\,\in\,\loopset(\gammap,\gamma)\,,
\eeq
the distance between $k$ and $l$, denoted by
\beq
\delta_\ell(k,l)\,,
\label{loopdist}
\eeq
is defined as the number of partons in $\ell$ between $k$ and $l$;
thus, if $k$ and $l$ are contiguous, the distance is equal to zero.
Furthermore, according to item ii. above and taking into account
that colour loops alternate L- and R-elements, the distance between 
any quark and any antiquark is an even number, whereas the distance
between two quarks or two antiquarks is an odd number.
\begin{figure}[th!]
\begin{center}
\includegraphics[scale=0.15]{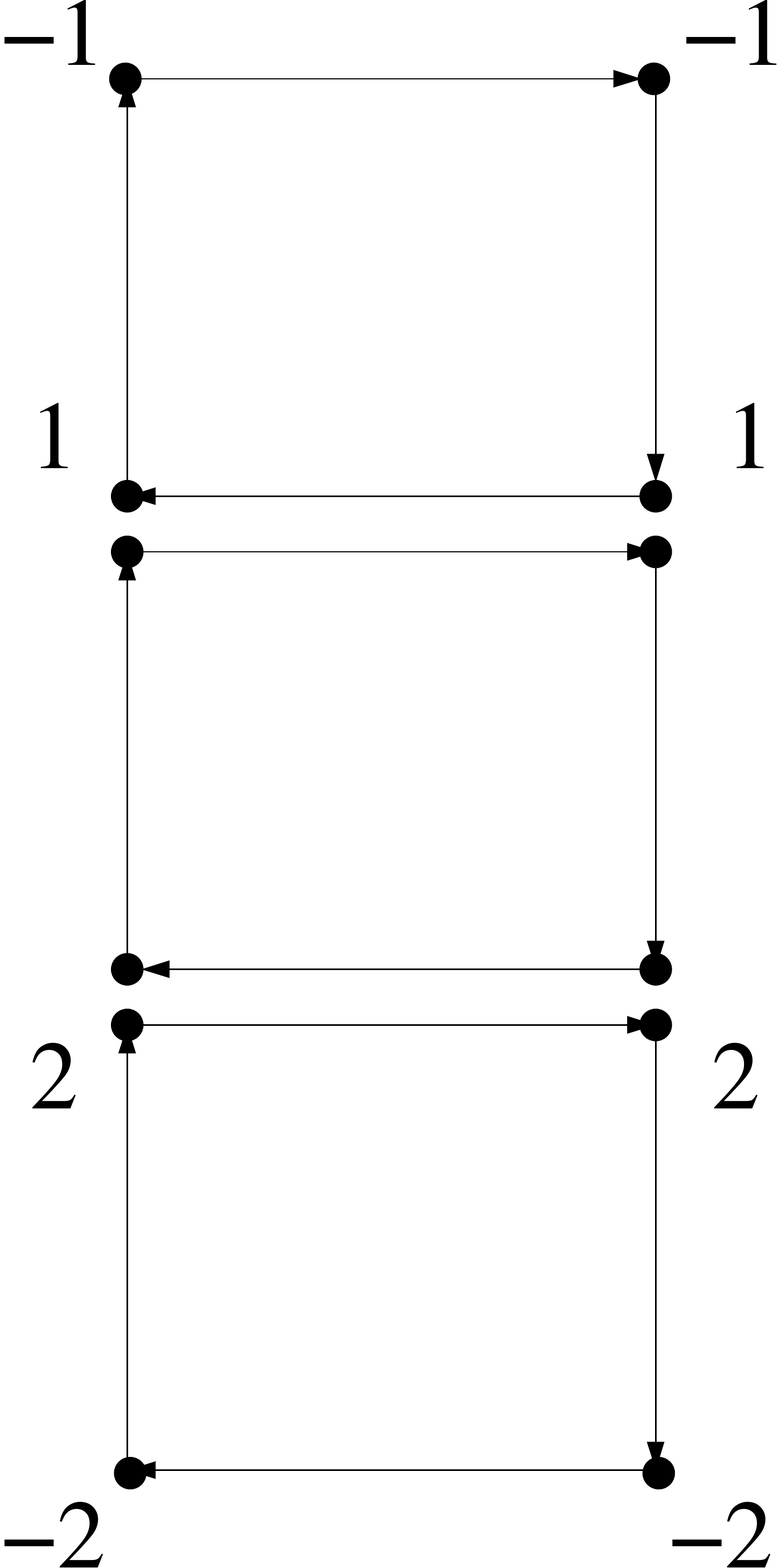}\hspace{5mm}
\includegraphics[scale=0.15]{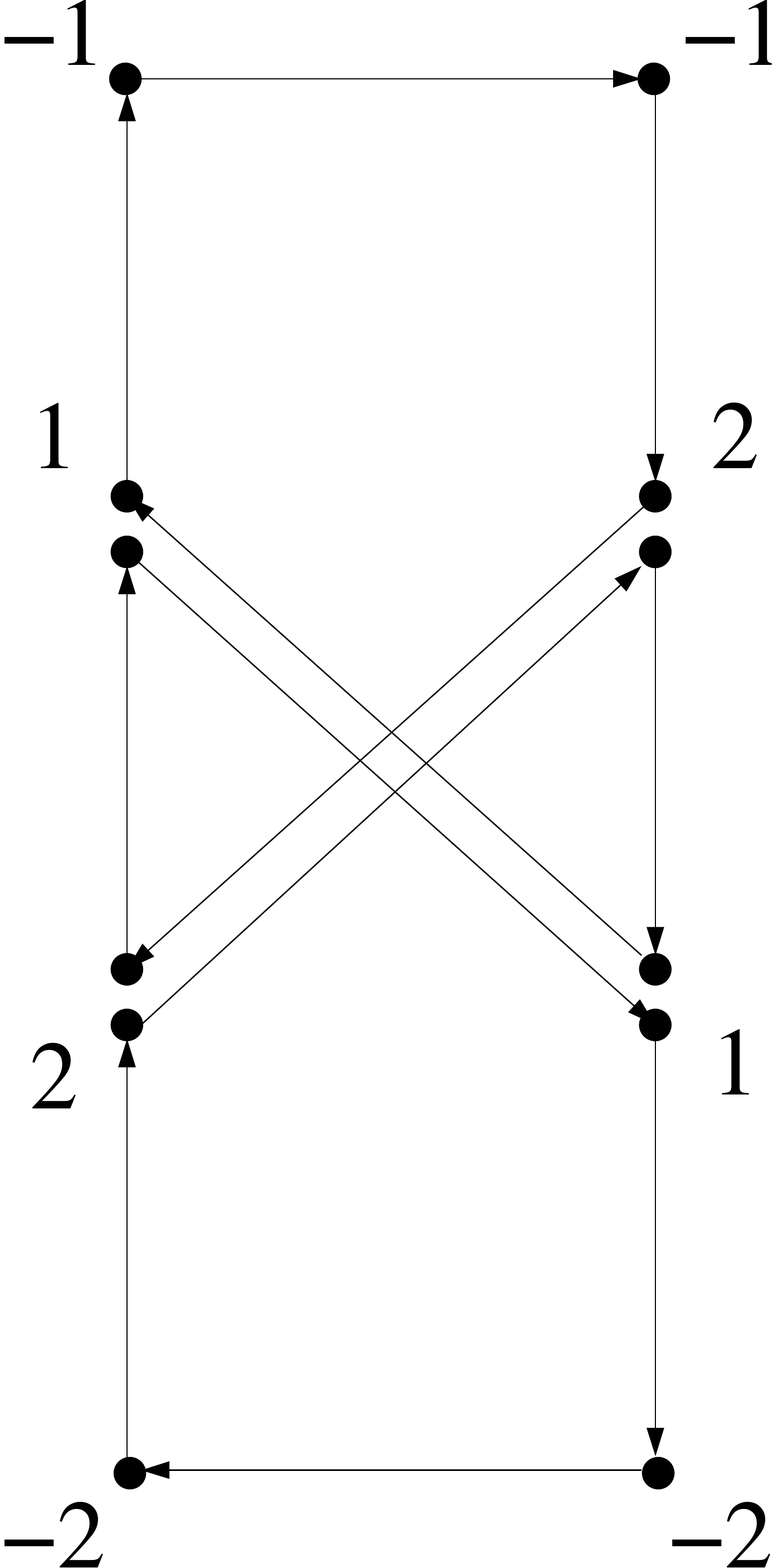}
\caption{\label{fig:CL1ex12}%
 Colour loop sets for $0\to q\qb gg$:
$\loopset (\gamma_1,\gamma_1)$ (left) and 
$\loopset (\gamma_1,\gamma_2)$ (right).}
\end{center}
\end{figure}

\vspace{2mm}
\noindent
$\bullet$ {\em Example.}  Consider the case of $0\to q \bq gg$.
According to the all-outgoing labeling of sect.~\ref{sec:basic}, this
process is $0\to q_{-1}\bq_{_-2} g_1g_2$ (i.e.~$\bq_{-1}q_{-2}\to g_1g_2$
in a physical configuration), with $q=1$ and $n=2$.  The set of flows, 
eq~(\ref{Bflowdef}), is
\beq
\flowexo=\Big\{\big(\Qa;1,2;\Qb\big),\big(\Qa;2,1;\Qb\big)\Big\}\equiv
\Big\{\gamma_1,\gamma_2\Big\}\,.
\label{Bflowex1}
\eeq
The colour loops corresponding to $\loopset
(\gamma_1,\gamma_1)$ and  $\loopset (\gamma_1,\gamma_2)$
are shown in Fig.~\ref{fig:CL1ex12}.  Clearly  the diagram for
$\loopset (\gamma_2,\gamma_2)$ is that for $\loopset
(\gamma_1,\gamma_1)$ with gluons 1 and 2 interchanged.
Thus the sets of colour loops, eq.~(\ref{loopsetdef}), associated with
these flows are
\beqn
\loopset(\gamma_{1},\gamma_{1})&=&
\Big\{\big(\Qa,1\big),\big(1,2\big),\big(2,\Qb\big)\Big\}
\;\;\;\;\Longrightarrow\;\;\;\;
\nloops{\loopset(\gamma_{1},\gamma_{1})}=3\,,
\label{CL01ex1}
\\
\loopset(\gamma_{1},\gamma_{2})&=&
\Big\{\big(\Qa,2,1,\Qb,2,1\big)\Big\}
\phantom{aaa}\,
\;\;\;\;\Longrightarrow\;\;\;\;
\nloops{\loopset(\gamma_{1},\gamma_{2})}=1\,,
\label{CL02ex1}
\\
\loopset(\gamma_{2},\gamma_{2})&=&
\Big\{\big(\Qa,2\big),\big(2,1\big),\big(1,\Qb\big)\Big\}
\;\;\;\;\Longrightarrow\;\;\;\;
\nloops{\loopset(\gamma_{2},\gamma_{2})}=3\,.
\label{CL3ex1}
\eeqn
For example, with
\beq
\ell=\Big\{\big(\Qa,2,1,\Qb,2,1\big)\Big\}
\eeq
we have
\beqn
&&\delta_\ell(\Qa,2)=0\,,\;\;\;
\delta_\ell(\Qa,1)=1\,,\;\;\;
\delta_\ell(\Qa,\Qb)=2\,,\;\;\;
\delta_\ell(\Qa,2)=3\,,\;\;\;
\delta_\ell(\Qa,1)=4\,,\phantom{aaaaa}
\\*
&&\delta_\ell(2,1)=0\,,\;\;\;
\delta_\ell(2,\Qb)=1\,,\;\;\;
\delta_\ell(2,2)=2\,,\;\;\;
\delta_\ell(2,1)=3\,,\;\;\;
\\*
&&\delta_\ell(1,\Qb)=0\,,\;\;\;
\delta_\ell(1,2)=1\,,\;\;\;
\delta_\ell(1,1)=2\,,\;\;\;
\\*
&&\delta_\ell(\Qb,2)=0\,,\;\;\;
\delta_\ell(\Qb,1)=1\,,\;\;\;
\\*
&&\delta_\ell(2,1)=0\,.
\eeqn
Consistently with the notation used throughout this paper for the number
of elements of a set (see e.g.~eq.~(\ref{dimBflow})), in 
eqs.~(\ref{CL01ex1})--(\ref{CL3ex1}) we have denoted by 
\beq
\nloops{\loopset(\gammap,\gamma)}
\label{numloops}
\eeq
the number of colour loops of the closed flow $(\gammap,\gamma)$;
this quantity will appear throughout this paper.

The construction of colour loops works also in the case of gluon-only
amplitudes. Given a closed flow $(\sigmap,\sigma)$, steps 0--7 previously
described are unchanged (with ``quark'' or ``antiquark'' there to be ignored),
except for the necessity of including a cyclicity condition, which is
simply achieved in the following way:
\label{thispage1}
\begin{itemize}
\item
Going down one place from the last element in the R-flow means
arriving at the first element of that flow.
\item
Going up one place from the first element in the L-flow means
arriving at the last element of that flow.
\end{itemize}

\section{Extended colour space, and the flow representation\label{sec:vspace}}
The colour-anticolour structure of gluons suggests the possibility
of working in a vector space larger than that introduced in
eq.~(\ref{vspaces}). Let $p$ be any gluon; we consider an 
$N^2$-dimensional vector space $\spaceij^{(p)}$ which embeds 
$\spacea^{(p)}$
\beq
\spacea^{(p)}\subset\spaceij^{(p)}\,,
\;\;\;\;\;\;\;\;
\abs{\spaceij^{(p)}}=N^2\,.
\eeq
The relationship between $\spacea$ and $\spaceij$ is established as
follows. Let
\beq
\big\{\ket{ij}\big\}_{i,j=1}^N\,\subset\,\spaceij^{(p)}
\eeq
be an orthonormal basis of $\spaceij^{(p)}$, so that
\beq
\brat{kl}\ket{ij}=\delta_{ik}\delta_{jl}\,,
\;\;\;\;\;\;\;\;
\sum_{ij}\ket{ij}\bra{ij}=I\,.
\label{baseij}
\eeq
The vectors $\ket{a}$ that constitute an orthonormal basis 
of $\spacea^{(p)}$ are then
\beq
\ket{a}=\TF^{-1/2}\sum_{ij}\lambda^{a}_{ij}\ket{ij}\,.
\label{baseadef}
\eeq
Therefore (see footnote~\ref{ft:empty} for the normalization)
\beq
\brat{b}\ket{a}=\TF^{-1}\sum_{ijkl}\lambda^{b}_{lk}\lambda^{a}_{ij}
\brat{kl}\ket{ij}=\TF^{-1}\,{\rm Tr}\left(\lambda^{b}\lambda^{a}\right)=
\delta^{ab}\,,
\eeq
which shows that eq.~(\ref{baseadef}) behaves as expected. Furthermore
\beqn
\projap\equiv\sum_a\ket{a}\bra{a}&=&
\TF^{-1}\sum_a\sum_{ijkl}\lambda^{a}_{lk}\lambda^{a}_{ij}\ket{kl}\bra{ij}
\nonumber
\\*&=&
\sum_{ijkl}
\left(\delta_{ik}\delta_{jl}-\frac{1}{N}\delta_{ij}\delta_{kl}\right)
\ket{kl}\bra{ij}\,.
\label{propa}
\eeqn
Equation~(\ref{propa}) suggests defining the vector
\beq
\ket{\!\perp}=\frac{1}{\sqrt{N}}\sum_i\ket{ii}
\;\;\;\;\Longrightarrow\;\;\;\;
\brat{\perp\!}\ket{\!\perp}=1\,,
\label{vperp}
\eeq
so that, given eq.~(\ref{baseij}), from eq.~(\ref{propa}) one obtains
\beqn
\projap&=&I-\projpp\,,
\\
\projpp&=&\ket{\!\perp}\bra{\perp\!}\,.
\eeqn
It also follows that
\beq
\projppsq=\projpp\,,
\;\;\;\;\;\;
\projapsq=\projap\,,
\;\;\;\;\;\;
\projap\projpp=\projpp\projap=0\,.
\label{projprop}
\eeq
Therefore, $\projap$ and $\projpp$ are projectors onto the (physical)
$\spacea^{(p)}$ and (unphysical) \mbox{$\spaceij^{(p)}\setminus\spacea^{(p)}$} 
subspaces of $\spaceij^{(p)}$, respectively. We stress that the definition of 
$\projap$ in eq.~(\ref{propa}) is usually an expression for the identity
(see eq.~(\ref{basisproj})). This is correct: by construction, $\projap$ 
{\em is} the identity in the subspace $\spacea^{(p)}$, but not in the 
larger space $\spaceij^{(p)}$.
For future use, we note that by direct computation one obtains
\beq
\projpp\ket{ij}=\frac{\delta_{ij}}{\sqrt{N}}\ket{\!\perp}
\;\;\;\;\Longrightarrow\;\;\;\;
\sum_{ij}\projpp\ket{ij}=\sqrt{N}\ket{\!\perp}=
\sum_{ij}\delta_{ij}\ket{ij}\,.
\label{tmp6}
\eeq
In sects.~\ref{sec:treeqg} and~\ref{sec:treeglu} we shall show
how the use of $\spaceij^{(p)}$ is in one-to-one correspondence
with what is usually called the flow representation for amplitudes,
which in the present context will acquire a natural geometrical meaning.

\section{Tree-level matrix elements\label{sec:tree}}
In this section, we present results for tree-level matrix elements,
regarded as quantities where all partons are hard and well separated
from each other.

\subsection{Quark-only matrix elements\label{sec:treeqrk}}
The processes of interest are those of eq.~(\ref{QGRproc}),
with $n=0$ there. Because of this, the colour factors are simply
products of Kronecker $\delta$'s (see footnote~\ref{ft:empty}) which,
when employed in the computation of the colour-flow matrix element
of eq.~(\ref{CFQGmatdef}), are contracted with each other, alternating
one Kronecker $\delta$ from the R-flow with one Kronecker $\delta$ from 
the L-flow. Eventually, one finds a $\delta$ whose right index is equal
to the left index of the $\delta$ from which the contraction had started,
thus resulting in the trace of the identity matrix, equal to $N$.
It is easy to convince oneself that the indices encountered when
contracting the $\delta$'s are nothing but the colour labels of the
partons that enter the colour loop to which any of these partons
belong (bear in mind that any quark or antiquark appears in a single loop).
Therefore, each loop is responsible for a contribution equal to $N$
to the colour-flow matrix element which implies, from eqs.~(\ref{Lambdagq})
and~(\ref{Mmflowsc})
\beq
\ampsqQGBggz(\gammap,\gamma)=
\ampCSQGBggz(\gammap)^\star \,\ampCSQGBggz(\gamma)\,
N^{-\rho(\gammap)-\rho(\gamma)}\,N^{\nloops{\loopset(\gammap,\gamma)}}\,,
\label{treeQres}
\eeq
where the quantity $\nloops{\loopset(\gammap,\gamma)}$ has been introduced
in eq.~(\ref{numloops}).
Summing eq.~(\ref{treeQres}) over flows is trivial: by using
eq.~(\ref{ampsqQG}) we obtain
\beq
\ampsqQGBggz=\sum_{\gammap,\gamma\in\flowBggz}
\ampCSQGBggz(\gammap)^\star \,\ampCSQGBggz(\gamma)\,
N^{-\rho(\gammap)-\rho(\gamma)}\,N^{\nloops{\loopset(\gammap,\gamma)}}\,.
\label{treeQressum}
\eeq

\subsection{Quark-gluon matrix elements\label{sec:treeqg}}
The remarkable simplicity of eq.~(\ref{treeQres}) stems from the 
trivial nature of the colour factor relevant to quark-only processes.
The presence of gluons complicates the picture; however, we shall show
in this section that results for quark-gluon matrix elements can
essentially be cast in the same form as eq.~(\ref{treeQres}), by
introducing quantities that we shall call secondary colour flows,
and that can be derived from the given colour flows.

The starting point is the vector $\ket{\ampQGBgg(\gamma)}\in\spacea$ that 
represents the amplitude for a given flow $\gamma$, eq.~(\ref{QGAflowdef}).
Consider one of the gluons that enter the process; it is not restrictive 
to assume its label to be $1$. There exists a quark with label $-p_1$ 
such that
\beq
\gamma=\Big(-p_1;\ldots,1,\ldots;\mu(-p_1-q)\Big)\,\hgamma\,.
\label{qgRfl}
\eeq
In other words, gluon $1$ belongs to the colour line that begins at
quark $-p_1$ and ends at antiquark $\mu(-p_1-q)$. Write the colour
factor associated with $\gamma$ as follows:
\beq
\Lambda(\gamma)=N^{-\rho(\gamma)}\,
\Big(\LA\lambda^{a_1}\LB\Big)_{a_{-p_1}a_{\mu(-p_1-q)}}
\,\hat{\Lambda}(\hgamma)\,,
\label{lamlam}
\eeq
with $\LA$ and $\LB$ suitable products of $\lambda$ matrices; the
definition of $\hat{\Lambda}(\hgamma)$ can be derived from 
eq.~(\ref{Lambdagq}), but will not be needed here.
Equation~(\ref{QGAflowdef}) can then be re-written more compactly
as follows:
\beqn
&&\ket{\ampQGBgg(\gamma)}=\sum_{\setaQGBgg}
\Big(\LA\lambda^{a_1}\LB\Big)_{a_{-p_1}a_{\mu(-p_1-q)}}
\ket{a_1}\otimes\ket{v}
\label{chrep0}
\\&&\phantom{aaaaa}
=\TF^{-1/2}\sum_{ij}\sum_{\setaQGBgg}
\Big(\LA\lambda^{a_1}\LB\Big)_{a_{-p_1}a_{\mu(-p_1-q)}}
\lambda^{a_1}_{ij}\ket{ij}\otimes\ket{v}\,,
\nonumber
\eeqn
where
\beq
\ket{v}=N^{-\rho(\gamma)}\,\hat{\Lambda}(\hgamma)\ampCSQGBgg(\gamma)
\bigotimes_{p\ne 1}\ket{a_p}\,,
\label{ketv}
\eeq
and we have used eq.~(\ref{baseadef}), understanding
$\ket{ij}\in\spaceij^{(1)}$. By means of the Fierz identity
one obtains
\beqn
&&\TF^{-1/2}\sum_{ij}\sum_{a_1}
\Big(\LA\lambda^{a_1}\LB\Big)_{a_{-p_1}a_{\mu(-p_1-q)}}
\lambda^{a_1}_{ij}\ket{ij}
\label{chrep1}
\\*&&\phantom{aaa}
=2^{-1/2}\sum_{ij}\left(
\LA_{a_{-p_1}j}\LB_{ia_{\mu(-p_1-q)}}
-\frac{1}{N}
\Big(\LA\LB\Big)_{a_{-p_1}a_{\mu(-p_1-q)}}\delta_{ij}\right)\ket{ij}
\nonumber
\\*&&\phantom{aaa}
=2^{-1/2}\sum_{ij}\left(
\LA_{a_{-p_1}j}\LB_{ia_{\mu(-p_1-q)}}
-\frac{1}{N}
\Big(\LA\LB\Big)_{a_{-p_1}a_{\mu(-p_1-q)}}\projpo\right)\ket{ij}\,,
\phantom{aa}
\nonumber
\eeqn
having used eq.~(\ref{tmp6}).

The vector on the l.h.s.~of eq.~(\ref{chrep1}) belongs to $\spacea^{(1)}$.
It is written in the r.h.s.~as the difference of two vectors, the second
of which belongs to \mbox{$\spaceij^{(1)}\setminus\spacea^{(1)}$}. In other
words, a physical vector is written as the difference of two unphysical
vectors, constructed so that their unphysical components cancel each other.
The colour factors associated with these two vectors are also interesting.
In that of the first term on the rightmost side of eq.~(\ref{chrep1}) the
gluon is replaced by a $Q\bQ$ pair with colour and anticolour indices
equal to $i$ and $j$, respectively, whereas in that of the second term 
the gluon disappears altogether. Thus, these colour factors are those
one would obtain by employing eq.~(\ref{Lambdagq}) with flows\footnote{For
the sake of clarity, with abuse of notation in eq.~(\ref{qgRfla}) we have 
used the same symbol $Q$ for a quark and an antiquark, and their labels. 
This is temporary, and a more  precise notation will be adopted soon.}
\beqn
\bgamma_a&=&\Big(-p_1;\ldots,\bQ\Big)\Big(Q,\ldots;\mu(-p_1-q)\Big)\,\hgamma\,,
\label{qgRfla}
\\
\bgamma_b&=&\Big(-p_1;\ldots,\cancel{1},\ldots;\mu(-p_1-q)\Big)\,\hgamma\,,
\label{qgRflb}
\eeqn
potentially up to an overall power of $N$, which can however be easily
fixed. Namely, the original colour flow $\gamma$ has a pre-factor equal
to $N^{-\rho(\gamma)}$, which here is part of the definition of
$\ket{v}$ in eq.~(\ref{ketv}). For $\rho(\gamma)$ to be equal to
$\rho(\bgamma_a)$, it is sufficient that the flavour of the fictitious
pair $Q\bQ$ be different from all of the other quark flavours relevant
to the current process, which is a condition that we can impose by 
construction. Conversely, since the flavour structure of $\bgamma_b$ is 
identical to that of $\gamma$, then \mbox{$\rho(\gamma)=\rho(\bgamma_b)$}.
This however does not take into account the factor $-1/N$ that
appears in eq.~(\ref{chrep1}); indeed, it will turn out to be convenient
(in particular, in view of its negative sign) to account for that
factor separately from $\rho(\gamma)$.

The procedure that leads one from eq.~(\ref{chrep0}) to eq.~(\ref{chrep1}) 
can be iterated, by considering all of the other gluons one after another.
At each step, the number of contributions will grow by a factor of two (two 
being the number of terms on the r.h.s.~of the Fierz identity), with each
gluon either replaced by a fictitious quark-antiquark pair, or eliminated;
with $n$ gluons, the overall number of contributions will be thus equal
to $2^n$. In order to enumerate them, and to construct the analogues
of the flows in eqs.~(\ref{qgRfla}) and~(\ref{qgRflb}), we introduce
the following notation. Let
\beq
\Sset{n}{k}
\label{Ssetdef}
\eeq
be the set of all unordered subsets of \mbox{$\{1,\ldots n\}$} of 
length $k$. In other words
\beq
s_k\in\Sset{n}{k}\;\;\;\;\Longrightarrow\;\;\;\;
s_k=\big\{j_1,\ldots j_k\big\}\,\subseteq\,\big\{1,\ldots n\big\}\,.
\label{skdef}
\eeq
Note that the definition of $\Sset{n}{k}$ includes the special cases
\beq
\Sset{n}{0}=\Big\{\{\}\Big\}\,,
\;\;\;\;\;\;\;\;
\Sset{n}{n}=\Big\{\big\{1,\ldots n\big\}\Big\}\,.
\label{specials}
\eeq
Note also that
\beq
\abs{\Sset{n}{k}}=\binomial{n}{k}\,,
\label{dimSnk}
\eeq
and thus that
\beq
\sum_{k=0}^n\abs{\Sset{n}{k}}=2^n\,.
\label{countsecfl}
\eeq
The set $\Sset{n}{k}$ encompasses all possible choices of $k$ gluon 
labels out of those of the given $n$ gluons; an element $s_k\in\Sset{n}{k}$ 
is one specific such choice. Then, for a given colour flow $\gamma$
consider the $2^n$-element set
\beq
\bigcup_{k=0}^n\bigcup_{s_k\in\Sset{n}{k}}\Big\{\bgamma_{\cancel{s_k}}\Big\}\,.
\label{secflnew}
\eeq
Each $\bgamma_{\cancel{s_k}}$ element is called a {\em secondary
colour flow}, and is constructed by taking $\gamma$ and removing 
from it the $k$ gluons associated with the labels in $s_k$. As far
as the other $n-k$ gluons are concerned, eq.~(\ref{qgRfla}) tells us
that they need be replaced by fictitious quark-antiquark pairs. Since
such pairs are essentially a bookkeeping device, it is easier to 
continue using the same gluon labels, with a bar on top. For future
reference, we point out that according to the procedure of 
sect.~\ref{sec:loops} exactly the same colour loops will emerge from 
closed secondary flows regardless of whether quark-antiquark labels,
or barred gluon labels, are employed.

It is a matter of simple algebra to see that the iteration of
eq.~(\ref{chrep1}) leads to the following result:
\beqn
\ket{\ampQGBgg(\gamma)}&=&\ampCSQGBgg(\gamma)
\sum_{k=0}^n\sum_{s_k\in\Sset{n}{k}}
\frac{(-1)^k}{2^{n/2}N^k}\!\!
\sum_{\setaQQ}\sum_{\setijn}
\Lambda\left(\{aij\},\bgamma_{\cancel{s_k}}\right)
\nonumber
\\*&&\phantom{aa}
\times
\projpsk\,\ket{i_1j_1,\ldots i_nj_n}\bigotimes_{p=-2q}^{-1}\ket{a_p}\,,
\label{QGAflowproj}
\eeqn
with the sets $\Sset{n}{k}$ introduced in eq.~(\ref{Ssetdef}),
and having defined
\beq
\projpsk=\bigotimes_{p\in s_k}\projpp\,.
\eeq
The special case (see eq.~(\ref{specials}))
\beq
\projp^{(s_0)}=I\,,
\;\;\;\;\;\;\;\;
s_0=\big\{\big\}
\label{sspecial}
\eeq
is understood. Note that, apart from a factor $N^{-\rho(\gamma)}$,
the colour factor $\Lambda$ in eq.~(\ref{QGAflowproj}) is a string of
\mbox{$n-k+q$} Kronecker $\delta$s. Concerning the $-1/N$ factors 
mentioned above, they collectively give rise to the \mbox{$(-1)^k/N^k$}
factor on the r.h.s.~of eq.~(\ref{QGAflowproj}). As one can see,
it is extremely easy to keep track of them, by means of the summation
index $k$, which essentially counts the number of gluons whose labels
are removed (thanks to the second term in the Fierz identity) from the 
secondary flows. As the colour factors render clear, such gluons
become colourless; therefore, they are what in the literature are
usually called $U(1)$ gluons. Equation~(\ref{QGAflowproj}) gives
a geometrical interpretation to such objects. Namely, $U(1)$ gluons 
are associated with vectors that live in the 
\beq
\bigotimes_{p\in s_k}\spaceij^{(p)}\setminus\spacea^{(p)}
\eeq
subspace, which one projects onto by means of $\projpsk$. Therefore,
$U(1)$-gluon vectors are orthogonal to physical-gluon vectors, which belong 
to the $\spacea^{(p)}$ subspaces; it is therefore natural to expect
that the interferences between such terms vanish.

The $k=0$ term in eq.~(\ref{QGAflowproj}) is what is usually called
the ``flow representation'' of the given amplitude. That equation clarifies
the extent to which this is an improper terminology (strictly, two 
representations are two different ways for writing the {\em same} object). 
As can be seen there, the fundamental and flow representations differ 
by non-null vectors, which have at least one component in the $U(1)$-gluon 
subspaces.

We now turn to computing the analogue of eq.~(\ref{treeQres}), i.e.~the 
interference of two amplitudes
\beq
\ampsqQGBgg(\gammap,\gamma)=
\brat{\ampQGBgg(\gammap)}\ket{\ampQGBgg(\gamma)}\,.
\label{ampSQint}
\eeq
Equation~(\ref{QGAflowproj}) implies an immediate simplification. 
By using it to write symbolically
\beq
\bra{\ampQGBgg(\gammap)}=\sum_{k=0}^n\sum_{s_k\in\Sset{n}{k}}
\bra{w(s_k)}\projpsk\,,
\eeq
one obtains (see eq.~(\ref{sspecial}))
\beq
\ampsqQGBgg(\gammap,\gamma)=\brat{w(s_0)}\ket{\ampQGBgg(\gamma)}\,,
\label{ampsqsimp}
\eeq
since
\beq
\projpsk\ket{\ampQGBgg(\gamma)}=0\;\;\;\;\;\;\;\;
\forall\,s_k\ne s_0
\label{projofA}
\eeq
owing to the fact that
\beq
\bigotimes_{p\in s_k}\projap\ket{\ampQGBgg(\gamma)}=\ket{\ampQGBgg(\gamma)}\,,
\eeq
and eq.~(\ref{projprop}).
From eq.~(\ref{QGAflowproj}) one can read the explicit form of $\bra{w(s_0)}$
\beqn
\bra{w(s_0)}&=&\ampCSQGBgg(\gammap)^\star
\,\frac{1}{2^{n/2}}\!\sum_{\setapQQ}\sum_{\setijpn}
\Lambda\left(\{a^\prime i^\prime j^\prime\},
\bgammap_{\cancel{s_0}}\right)^\star
\nonumber
\\*&&\phantom{aa}
\times
\bra{i_1^\prime j_1^\prime,\ldots i_n^\prime j_n^\prime}
\bigotimes_{p=-2q}^{-1}\bra{a_p^\prime}\,.
\label{omega0}
\eeqn
In the computation of eq.~(\ref{ampsqsimp}), one therefore finds
\beq
\prod_{p=-2q}^{-1}\brat{a_p^\prime}\ket{a_p}=
\prod_{p=-2q}^{-1}\delta_{a_p a_p^\prime}
\eeq
and
\beqn
&&\bra{i_1^\prime j_1^\prime,\ldots i_n^\prime j_n^\prime}\,
\projpsk\,\ket{i_1j_1,\ldots i_nj_n}
\nonumber
\\*&&\phantom{aaaaaaa}
=\bra{i_1^\prime j_1^\prime,\ldots i_n^\prime j_n^\prime}\,
\projpsksq\,\ket{i_1j_1,\ldots i_nj_n}
\nonumber
\\*&&\phantom{aaaaaaa}
=\frac{1}{N^k}\prod_{p^\prime\in s_k}
\delta_{i_{p^\prime}^\prime j_{p^\prime}^\prime}
\prod_{p\in s_k}\delta_{i_p j_p}
\prod_{r\in s_k}\brat{\perp_r\!}\ket{\!\perp_r}
\prod_{s\notin s_k}\brat{i_s^\prime j_s^\prime}\ket{i_s j_s}
\nonumber
\\*&&\phantom{aaaaaaa}
=\frac{1}{N^k}\prod_{p^\prime\in s_k}
\delta_{i_{p^\prime}^\prime j_{p^\prime}^\prime}
\prod_{p\in s_k}\delta_{i_p j_p}
\prod_{s\notin s_k}\delta_{i_s^\prime i_s}\delta_{j_s^\prime j_s}\,,
\label{indices}
\eeqn
having employed the leftmost part of eq.~(\ref{tmp6}). The colour factor
\mbox{$\Lambda(\{aij\},\bgamma_{\cancel{s_k}})$} in 
eq.~(\ref{QGAflowproj}) does not depend on any of the indices
$i_p$ and $j_p$ if \mbox{$p\in s_k$}. One can therefore carry out the
sums over these indices, exploiting the second Kronecker $\delta$
in eq.~(\ref{indices}): this results in a factor $N^k$, that cancels
an identical factor in the denominator of eq.~(\ref{indices}). Conversely,
the dependence on the indices that appear in the leftmost Kronecker
$\delta$ in eq.~(\ref{indices}) is non trivial, owing to the colour
factor in eq.~(\ref{omega0}). However, it can be simplified by observing
that
\beq
\sum_{\setijpn}\!\!
\Lambda\left(\{a^\prime i^\prime j^\prime\},
\bgammap_{\cancel{s_0}}\right)^\star
\prod_{p\in s_k}\delta_{i_p^\prime j_p^\prime}=
\sum_{\setijp_{p\notin s_k}}\!\!
\Lambda\left(\{a^\prime i^\prime j^\prime\},
\bgammap_{\cancel{s_k}}\right)^\star\,.
\label{redoflam}
\eeq
By putting everything back together, and by exploiting the rightmost 
Kronecker $\delta$ in eq.~(\ref{indices}) one finally obtains
\beqn
\ampsqQGBgg(\gammap,\gamma)&=&
\ampCSQGBgg(\gammap)^\star\ampCSQGBgg(\gamma)
\sum_{k=0}^n\sum_{s_k\in\Sset{n}{k}}
\frac{(-1)^k}{2^n}\,N^{-k}
\nonumber
\\*&\times&\!\!\!
\sum_{\setaQQ}\sum_{\setij_{p\notin s_k}}
\Lambda\left(\{aij\},\bgammap_{\cancel{s_k}}\right)^\star
\Lambda\left(\{aij\},\bgamma_{\cancel{s_k}}\right)\,.
\phantom{aaa}
\label{ampsqfin}
\eeqn
The colour factor in eq.~(\ref{ampsqfin}) has the same form as
any colour factor in a quark-only process. By using the result
of sect.~\ref{sec:treeqrk}, one obtains
\beqn
&&
\sum_{\setaQQ}\sum_{\setij_{p\notin s_k}}
\Lambda\left(\{aij\},\bgammap_{\cancel{s_k}}\right)^\star
\Lambda\left(\{aij\},\bgamma_{\cancel{s_k}}\right)
\nonumber
\\*&&\phantom{aaaa}
=N^{-\rho(\gamma)-\rho(\gammap)}
N^{\nloops{\loopset(\bgammap_{\cancel{s_k}},\bgamma_{\cancel{s_k}})}}\,,
\eeqn
and therefore the sought final result
\beqn
\ampsqQGBgg(\gammap,\gamma)&=&
\ampCSQGBgg(\gammap)^\star\ampCSQGBgg(\gamma)
N^{-\rho(\gamma)-\rho(\gammap)}
\nonumber\\*&&\phantom{aaa}\times
\sum_{k=0}^n\sum_{s_k\in\Sset{n}{k}}
\frac{(-1)^k}{2^n}\,N^{-k}
N^{\nloops{\loopset(\bgammap_{\cancel{s_k}},\bgamma_{\cancel{s_k}})}}\,.
\label{treeQGres}
\eeqn
The similarity of this result to that of eq.~(\ref{treeQres})
is striking. It implies that quark-gluon matrix elements have
essentially the same structure as quark-only ones, provided one
expresses them by means of secondary colour flows, thereby introducing
$U(1)$ gluons into the picture. The role of the latter is quite prominent
in eq.~(\ref{treeQGres}): their number is equal to the summation index $k$,
and they induce alternating signs ($(-1)^k$) in the summands. We stress
that this is true also on the diagonal of the colour-flow matrix
(i.e.~when $\gammap=\gamma$), which is a real number squared.
Furthermore, $U(1)$ gluons give eq.~(\ref{treeQGres}) a hierarchy
in powers of $1/N$. Unfortunately, this hierarchy is not absolute
for off-diagonal elements. This is due to the fact that while
\beq
\nloops{\loopset(\bgamma_{\cancel{s_k}},\bgamma_{\cancel{s_k}})}\,\ge\,
\nloops{\loopset(\bgamma_{\cancel{s_l}},\bgamma_{\cancel{s_l}})}\;\;\;\;\;\;
{\rm when}\;\;k<l\,,
\eeq
an analogous inequality does not hold true in general for a generic
off-diagonal $(\gammap,\gamma)$ closed flow.

\subsubsection{Summing over flows\label{sec:QGfsum}}
The matrix elements relevant to a given closed flow given in
eq.~(\ref{treeQGres}) need be summed, in most physical applications,
over flows. Owing to the summations over secondary flows, this operation 
is less trivial than its quark-only counterpart, which had led immediately
to the result of eq.~(\ref{treeQressum}). We perform it in this section,
showing in the process that it leads naturally to the definition of
linear combinations of dual amplitudes where $U(1)$ gluons play
a special role.

The relevant summations can be carried out directly on the scalar
quantities on the r.h.s.~of eq.~(\ref{treeQGres}); this implies two
sums, as both the L- and R-flows must be summed. It is therefore
more convenient to arrive at the sought result by summing amplitudes,
as opposed to interferences of amplitudes, since this entails carrying
out a single sum; in other words, we exploit eq.~(\ref{ampflowQG})
rather than the rightmost side of eq.~(\ref{ampsqQG}). Thus,
we consider
\beq
\ket{\ampQGBgg}=\sum_{\gamma\in\flowBgg}\ket{\ampQGBgg(\gamma)}\,,
\label{ampflowQG2}
\eeq
where the summands on the r.h.s.~are given in eq.~(\ref{QGAflowproj}).
The key idea is the following: so far, secondary colour flows are 
quantities derived from some given colour flows. However, they have
a structure which is perfectly well-defined in its own right; therefore,
when a sum over flows has to be carried out, it is possible to exchange
the order in which the sums over flows and secondary flows are performed,
thus inverting the logic, since the former are now seen as quantities 
derived from the latter. 

In order to do this, one starts by introducing a notation for the sets 
of $q$-quark, \mbox{$(n-k)$}-gluon colour flows as follows:
\beqn
&&s_k=\big\{j_1,\ldots j_k\big\}\in\Sset{n}{k}
\;\;\longrightarrow
\nonumber
\\*&&\phantom{aaaaaa}
\flowBggnk^{(s_k)}\equiv\flowBggnk\Big(\,{\rm gluon~labels~in}~
\big\{1,\ldots n\big\}\setminus\big\{j_1,\ldots j_k\big\}\Big)\,.
\label{Fnksets}
\eeqn
Thus, $\flowBggnk^{(s_k)}$ is identical to $\flowBggnk$ up to the
labels of the gluons that enter its flows, which must be different
from the indices $j_i$ that belong to the given set $s_k$. This
implies
\beq
\bgamma_{\cancel{s_k}}\,\in\,\flowBggnk^{(s_k)}\,.
\label{flinFnk}
\eeq
With this, one can generalize the operator $I_+$ of 
ref.~\cite{Frixione:2011kh} (see app.~B there). In particular
\beqn
&&s_k\in\Sset{n}{k}\,,\;\;\;\;
\bgamma\in\flowBggnk^{(s_k)}\,,\;\;\;\;
\big\{i_1,\ldots i_k\big\}\subseteq
\big\{-1,\ldots -q\big\}\cup\big\{1,\ldots n\big\}\phantom{aaaa}
\nonumber
\\*&&\phantom{aaaaaaaa}
\longrightarrow\;\;\;\;
I_+^{(s_k)}(i_1,\ldots i_k)\bgamma =
I_+^{(j_1)}(i_1)I_+^{(j_2)}(i_2)\ldots I_+^{(j_k)}(i_k)\bgamma\,,
\label{Ipdef}
\eeqn
for any $k\ge 1$; when $k=0$, $I_+$ is defined to be the identity.
The single-label $I_+$ operator employed $k$ times on the rightmost side 
of eq.~(\ref{Ipdef}) coincides with that ref.~\cite{Frixione:2011kh},
except for a minor notational detail: here, the upper index $j_\alpha$
indicates the label of the gluon to be inserted to the right of position 
$i_\alpha$; this index was not necessary in ref.~\cite{Frixione:2011kh}, 
because $j_\alpha=n+1$ there. Implicit in eq.~(\ref{Ipdef}) is the fact 
that the $I_+$ operators are to be used in a right-to-left order: first 
the gluon with label $j_k$ is inserted, then that with label $j_{k-1}$,
and so forth up to $j_1$. As far as the values that the position indices 
$i_\alpha$ can assume are concerned: for the first gluon to be inserted ($j_k$)
one has $q+n-k$ possibilities (i.e.~to the right of each quark and of each
gluon in the given flow $\bgamma$); for the second one, the possibilities
are $q+n-k+1$, since that gluon is now inserted in the flow 
\mbox{$I_+^{(j_k)}(i_k)\bgamma$} which has one additional gluon
w.r.t.~the original one $\bgamma$. The procedure is then iterated till
the last gluon to be inserted ($j_1$). Since the $I_+$ operator of
eq.~(\ref{Ipdef}) is single-valued given the position parameters $i_\alpha$,
the dimension of the set that contains all possible flows generated by
$I_+$ acting on a given $\bgamma$ is equal to the dimension of the codomain
of the acceptable values of the $i_\alpha$ parameters. We have therefore
\beq
\abs{\bigcup_{i_1\ldots i_k}\Big\{I_+^{(s_k)}(i_1,\ldots i_k)\bgamma\Big\}}=
(n+q-1)(n+q-2)\ldots (n+q-k)\,,
\label{dimcodi}
\eeq
for any given flow $\bgamma\in\flowBggnk^{(s_k)}$. Also
\beq
\bgamma\,,\bgammap\in\flowBggnk^{(s_k)}\,,\;\;\;\;
\bgamma\ne\bgammap\;\;\;\;\Longrightarrow\;\;\;\;
I_+^{(s_k)}(i_1,\ldots i_k)\bgamma\,\ne\,
I_+^{(s_k)}(i_1,\ldots i_k)\bgamma\,.
\eeq
Finally
\beq
I_+^{(s_k)}(i_1,\ldots i_k)\bgamma\,\in\,\flowBgg\,.
\label{Ipbgam}
\eeq
Taken together, the above facts imply that
\beq
\flowBgg=\bigcup_{\bgamma\in\flowBggnk^{(s_k)}}
\bigcup_{i_1\ldots i_k}
\Big\{I_+^{(s_k)}(i_1,\ldots i_k)\bgamma\Big\}\,,
\;\;\;\;
\forall\;0\le k\le n\;\;{\rm and}\;\;s_k\in\Sset{n}{k}\,,
\label{flBbvsflB}
\eeq
and
\beq
\left(I_+^{(s_k)}(i_1,\ldots i_k)\bgamma\right)_{\cancel{s_k}}=\bgamma\,,
\;\;\;\;\;\;
\forall\;\bgamma\in\flowBggnk^{(s_k)}
\;\;{\rm and}\;\;
\big\{i_1,\ldots i_k\big\}\,.
\label{Ipcancel}
\eeq
Therefore, for any function $F$ that is summed over all flows and
depends on a $q$-quark, $n$-gluon flow and on one of its secondary flows, 
the following identity holds:
\beq
\sum_{\gamma\in\flowBgg}F\left(\gamma,\bgamma_{\cancel{s_k}}\right)=
\sum_{\bgamma\in\flowBggnk^{(s_k)}}\sum_{i_1\ldots i_k}
F\left(I_+^{(s_k)}(i_1,\ldots i_k)\bgamma,\bgamma\right)\,,
\label{Fiden}
\eeq
for any $k$ and $s_k\in\Sset{n}{k}$; the number of summands on the
two sides of eq.~(\ref{Fiden}) is the same, owing to eqs.~(\ref{dimBflow}),
(\ref{dimcodi}), and~(\ref{flBbvsflB}).

We now consider the sum on the r.h.s.~of eq.~(\ref{ampflowQG2}),
by replacing the amplitude there with its expression given in 
eq.~(\ref{QGAflowproj}). There are no cross-constraints between
the sum over $\gamma$ and those over $k$ and $s_k$, and this implies
the latter two can be used as the outermost ones, which leads us to
the following result:
\beqn
\ket{\ampQGBgg}&=&
\sum_{k=0}^n\sum_{s_k\in\Sset{n}{k}}
\frac{(-1)^k}{2^{n/2}N^k}\!\!
\sum_{\gamma\in\flowBgg}\ampCSQGBgg(\gamma)
\sum_{\setaQQ}\sum_{\setijn}
\Lambda\left(\{aij\},\bgamma_{\cancel{s_k}}\right)
\nonumber
\\*&&\phantom{aa}
\times
\projpsk\,\ket{i_1j_1,\ldots i_nj_n}\bigotimes_{p=-2q}^{-1}\ket{a_p}\,.
\label{QGAflowproj2}
\eeqn
In this expression, the part to the right of the sum over $\flowBgg$
(including it) is in the same form as the l.h.s.~of eq.~(\ref{Fiden}).
Thus, by employing the r.h.s.~of that equation instead we arrive at
\beqn
\ket{\ampQGBgg}&=&
\sum_{k=0}^n\sum_{s_k\in\Sset{n}{k}}
\frac{(-1)^k}{2^{n/2}N^k}\!\!
\sum_{\bgamma\in\flowBggnk^{(s_k)}}
\!\!\ampOSQGBgg(\bgamma)
\sum_{\setaQQ}\sum_{\setijn}
\!\!\Lambda\left(\{aij\},\bgamma\right)
\nonumber
\\*&&\phantom{aa}
\times
\projpsk\,\ket{i_1j_1,\ldots i_nj_n}\bigotimes_{p=-2q}^{-1}\ket{a_p}\,,
\label{QGAflowsum}
\eeqn
where we have defined the following linear combinations of the
standard (because of eq.~(\ref{Ipbgam})) dual amplitudes:
\beq
\bgamma\in\flowBggnk^{(s_k)}
\;\;\;\;\Longrightarrow\;\;\;\;\
\ampOSQGBgg(\bgamma)=
\sum_{i_1\ldots i_k}
\ampCSQGBgg\left(I_+^{(s_k)}(i_1,\ldots i_k)\bgamma\right)\,.
\label{bAdef}
\eeq
In words: for a given $q$-quark-antiquark, $(n-k)$-gluon colour
flow $\bgamma$, an amplitude $\ampOSQGBgg(\bgamma)$ is associated with 
it that is obtained by summing the dual amplitudes relevant to the colour
flows constructed by inserting $k$ gluons in $\bgamma$ in all possible
manners. Those extra gluons are colourless as far as the total
amplitude is concerned, which is consistent with their (implicit)
role in $\bgamma$ as $U(1)$ gluons.

Starting from eq.~(\ref{QGAflowsum}), we can repeat the procedure
that has brought us from eq.~(\ref{ampSQint}) to eq.~(\ref{treeQGres}),
to obtain
\beqn
\ampsqQGBgg&=&\sum_{k=0}^n
\sum_{s_k\in\Sset{n}{k}}
\frac{(-1)^{k}}{2^n}\,N^{-k}\,
\nonumber
\\*&&\phantom{aaa}\times\!\!\!\!
\sum_{\bgamma,\bgammap\in\flowBggnk^{(s_k)}}
\ampOSQGBgg(\bgammap)^\star\,\ampOSQGBgg(\bgamma)\,
N^{-\rho(\bgamma)-\rho(\bgammap)}
N^{\nloops{\loopset(\bgammap,\bgamma)}}\,.
\phantom{aa}
\label{matelfin}
\eeqn
As was already the case for the matrix elements at given closed
colour flow (eq.~(\ref{treeQGres})), eq.~(\ref{matelfin}) is very
similar to its quark-only counterpart, eq.~(\ref{treeQressum}).
The same comments made about eq.~(\ref{treeQGres}) vs eq.~(\ref{treeQres})
in this regard apply here, with the additional observations on the
$\ampOSQGBgg(\bgamma)$ amplitudes of eq.~(\ref{bAdef}). Note that
eqs.~(\ref{treeQGres}) and~(\ref{matelfin}) share an important 
feature, namely: not only do quantities associated with different
numbers of $U(1)$ gluons not interfere with each other but also,
for the interference to have a non-zero result, the identities of
such gluons must be the same on both sides of the cut (technically,
the set $s_k$ is relevant to both the L- and R-flow). 

We remark that expressions whose content is analogous to that of
eq.~(\ref{QGAflowsum}) appear in ref.~\cite{Maltoni:2002mq}, where
use is made of the same linear combinations of dual amplitudes
as in eq.~(\ref{bAdef}). What we have shown here is that: {\em a)}
at the level of amplitudes, the structure of eq.~(\ref{QGAflowsum}) 
is already present at fixed colour flow (eq.~(\ref{QGAflowproj}));
{\em b)} the number of summands associated with different sets of 
$U(1)$ gluons (the differences being in their numbers and/or identities)
is equal to $2^n$ (see eq.~(\ref{countsecfl})) at the level of both the
amplitudes (eqs.~(\ref{QGAflowproj}) and~(\ref{QGAflowsum})) {\em and the
matrix elements} (eqs.~(\ref{treeQGres}) and~(\ref{matelfin})). This
fact is non-trivial: one can show that by naively (i.e.~without using
the properties of projectors) squaring eq.~(\ref{QGAflowsum}) to obtain
eq.~(\ref{matelfin}), an individual term with $k$ $U(1)$ gluons in the
latter expression results from the sum of $3^k$ terms, \mbox{$3^k-1$}
of which mutually cancel.

\subsection{Gluon-only matrix elements\label{sec:treeglu}}
The results of sect.~\ref{sec:treeqg} show that ultimately gluons can be 
made to behave as extra quark-antiquark pairs in quark-gluon amplitudes.
This is a strong hint that the same property will hold true also
in the case of gluon-only amplitudes, i.e.~when there are no quarks
or antiquarks to start with. In this section, we shall prove that this
is indeed the case. At the end of the day this will not matter when summing 
over colour flows, since in that case Ward identities will ensure that 
$U(1)$-gluon contributions vanish. However, at fixed flows
$U(1)$ gluons do play a non-trivial role, which puts gluon-only
amplitudes on the same footing as quark-only ones, as has already
happened with quark-gluon amplitudes.

Let
\beq
\sigma=\left(\sigma(1),\ldots\sigma(n)\right)
\label{Rflnc}
\eeq
be a generic $n$-gluon colour flow. We define two operators, $\qRop$ and 
$\qLop$, that by acting on $\sigma$ result in the following 
$2$-quark-antiquark, $n$-gluon colour flows:
\beqn
q{\rm R}\big(\sigma\big)\equiv
\sigmaqR&=&\Big(-1;\sigma(1),\ldots\sigma(n);-4\Big)
\Big(-2;-3\Big)\,,
\label{qRdef2}
\\
q{\rm L}\big(\sigma\big)\equiv
\sigmaqL&=&\Big(-1;-4\Big)
\Big(-2;\sigma(1),\ldots\sigma(n);-3\Big)\,.
\label{qLdef2}
\eeqn
Here, following the conventions of eqs.~(\ref{qbars})--(\ref{glus}), we have 
introduced two quark-antiquark pairs of different flavours, labelled
by $(-1,-3)$ and $(-2,-4)$. Thus, according to eq.~(\ref{Lambdagq}) 
(see also footnote~\ref{ft:empty}), the associated colour factors are
\beqn
\Lambda(\seta,\sigmaqR)&=&\Big(\lambda^{a_1}\lambda^{a_{\sigma(2)}}
\ldots\lambda^{a_{\sigma(n)}}\Big)_{a_{-1}a_{-4}}
\delta_{a_{-2}a_{-3}}\,,
\label{colfR}
\\
\Lambda(\seta,\sigmaqL)&=&\delta_{a_{-1}a_{-4}}
\Big(\lambda^{a_1}\lambda^{a_{\sigma(2)}}
\ldots\lambda^{a_{\sigma(n)}}\Big)_{a_{-2}a_{-3}}\,,
\label{colfL}
\eeqn
which among other things rely on the fact that
\beq
\rho\Big(q{\rm R}\big(\sigma\big)\Big)=
\rho\Big(q{\rm L}\big(\sigma\big)\Big)=0
\label{rhoeqz}
\eeq
for any $\sigma$. With these definitions, for any closed colour
flow $(\sigmap,\sigma)$ we have
\beqn
C(\sigmap,\sigma)&\equiv&
\sum_{\setan}\Lambda\left(\seta,\sigmap\right)^\star
\Lambda\left(\seta,\sigma\right)
\label{CFmatdef}
\\
&=&C(\sigmapqL,\sigmaqR)\equiv
\sum_{\setaQGtwo}\!\!
\Lambda(\seta,\sigmapqL)^\star\,\Lambda(\seta,\sigmaqR)\,.
\label{Cnglueq}
\eeqn
Furthermore, the definitions of eqs.~(\ref{qRdef2}) and~(\ref{qLdef2}) 
respect the colour-loop structure. In other words, by using the
results of sect.~\ref{sec:loops}, it is straightforward
to prove the following properties:
\begin{itemize}
\item The closed colour flow $(\sigmap,\sigma)$ has the same number
of colour loops as $(\sigmapqL,\sigmaqR)$
\beq
\nloops{\loopset(\sigmap,\sigma)}=\nloops{\loopset(\sigmapqL,\sigmaqR)}\,.
\label{lseq}
\eeq
\item The colour loops in $\loopset(\sigmap,\sigma)$ and 
$\loopset(\sigmapqL,\sigmaqR)$ are pairwise identical, except for
that which contains $\sigma(n)$ as an R-element, and that which 
contains $\sigmap(1)$ as an L-element\footnote{These loops may actually
coincide.}; these differ {\em only} because the loops in 
$\loopset(\sigmapqL,\sigmaqR)$ include the fictitious quarks and
antiquarks labelled \mbox{$-1,\ldots -4$}.
\item Given any loop $\ell_1\in\loopset(\sigmap,\sigma)$, let
$\ell_2\in\loopset(\sigmapqL,\sigmaqR)$ be the loop corresponding 
to it according to the previous item, so that for any two gluon
labels $k$ and $l$ we have
\beq
k,l\in\ell_1\;\;\;\Longleftrightarrow\;\;\;k,l\in\ell_2\,.
\eeq
Then the distance introduced in eq.~(\ref{loopdist}) fulfills the property
\beq
\delta_{\ell_1}(k,l)=\delta_{\ell_2}(k,l)+2i\,,
\label{distparity}
\eeq
with $i$ an integer. In other words, the distance between $k$ and $l$ has 
the same parity in the $\ell_1$ and $\ell_2$ loops.
\end{itemize}
Equations~(\ref{CFmatdef})--(\ref{distparity}) are sufficient to allow
one to repeat, with only notational changes, what has been done
in sect.~\ref{sec:treeqg}. The analogue of eq.~(\ref{QGAflowproj})
reads as follows:
\beqn
\ket{\ampn(\sigma)}&=&\ampCSn(\sigma)
\sum_{k=0}^n\sum_{s_k\in\Sset{n}{k}}
\frac{(-1)^k}{2^{n/2}N^k}\!\!
\sum_{\setaQQf}\sum_{\setijn}
\Lambda\left(\{aij\},\bsigmaqRarg{\cancel{s_k}}\right)
\nonumber
\\*&&\phantom{aa}
\times
\projpsk\,\ket{i_1j_1,\ldots i_nj_n}\bigotimes_{p=-4}^{-1}\ket{a_p}\,.
\label{GGAflowproj}
\eeqn
Note a technical peculiarity: because of the necessity to use the
$\qRop$ and $\qLop$ operators, $\ket{\ampn(\sigma)}$ is a vector
that lives in the
\beq
\bigotimes_{p=-4}^{-1}\spacea^{(p)}\,
\bigotimes_{p=1}^{n}\spaceij^{(p)}
\eeq
vector space. However, the fictitious-quark degrees of freedom only 
play the role of giving the correct normalization and index-contraction
conditions when an interference of amplitudes is considered.
More significantly, the contributions to eq.~(\ref{GGAflowproj})
with $k\ge 1$ (i.e.~with at least one $U(1)$ gluon) do {\em not}
vanish. This is contrary to what one normally expects, but in fact
that expectation is based on results fully summed over colour degrees
of freedom. In other words, not summing over those, and fixing
e.g.~the colour flow, the contributions due to $U(1)$ gluons are non-trivial.
The implication of this is that, at fixed flows, the flow and the fundamental
representations do not coincide for gluon-only amplitudes, thus making 
the latter behave exactly as their quark-gluon counterparts. As we shall
see later, this symmetry is lifted upon summing over flows, and this
thanks to the dual Ward identities, which have no analogue in the
quark-gluon case.

By interfering the amplitude of eq.~(\ref{GGAflowproj}) with an
analogous one on the left-hand side of the cut (relevant to an
L-flow equal to $\sigmap$), one obtains the analogue of 
eq.~(\ref{treeQGres}), namely
\beqn
\ampsqn(\sigmap,\sigma)&=&
\ampCSn(\sigmap)^\star\ampCSn(\sigma)
\nonumber\\*&&\phantom{aaa}\times
\sum_{k=0}^n\sum_{s_k\in\Sset{n}{k}}
\frac{(-1)^k}{2^n}\,N^{-k}
N^{\nloops{\loopset(\bsigmapqLarg{\cancel{s_k}},
\bsigmaqRarg{\cancel{s_k}})}}\,,
\label{treeGGres}
\eeqn
having used eq.~(\ref{rhoeqz}). As was the case at the amplitude level,
$U(1)$ gluons do contribute non-trivially to eq.~(\ref{treeGGres}).

\subsubsection{Summing over flows\label{sec:GGfsum}}
The sum over colour flows of the results of eqs.~(\ref{GGAflowproj})
and~(\ref{treeGGres}) follows the same steps as in sect.~\ref{sec:QGfsum}.
Therefore, one readily arrives at the analogue of eq.~(\ref{QGAflowsum}),
namely
\beqn
\ket{\ampn}&=&
\sum_{k=0}^n\sum_{s_k\in\Sset{n}{k}}
\frac{(-1)^k}{2^{n/2}N^k}\!\!
\sum_{\bsigma\in P_{n-k}^{\prime(s_k)}}
\!\!\ampOSn(\bsigma)
\sum_{\setaQQf}\sum_{\setijn}
\!\!\Lambda\left(\{aij\},\bsigma\right)
\nonumber
\\*&&\phantom{aa}
\times
\projpsk\,\ket{i_1j_1,\ldots i_nj_n}\bigotimes_{p=-4}^{-1}\ket{a_p}\,,
\label{GGAflowsum}
\eeqn
where, similarly to what was done in eq.~(\ref{bAdef}), we have been
led to define
\beq
\bsigma\in P_{n-k}^{\prime(s_k)}
\;\;\;\;\Longrightarrow\;\;\;\;\
\ampOSn(\bsigma)=
\sum_{i_1\ldots i_k}
\ampCSn\left(I_+^{(s_k)}(i_1,\ldots i_k)\bsigma\right)\,.
\label{bAdefgg}
\eeq
In eqs.~(\ref{GGAflowsum}) and~(\ref{bAdefgg}), by $P_{n-k}^{\prime(s_k)}$
we have denoted the analogue of $\flowBggnk^{(s_k)}$, i.e.~the set of 
$(n-k)$-gluon colour flows with indices not equal to any of the integers
that belong to the set $s_k$ (the presence of the fictitious quark labels
is understood, and ignored notation-wise). Furthermore, the $I_+$
operators that appear on the r.h.s.~of eq.~(\ref{bAdefgg}) are defined
in a manner formally identical to that of eq.~(\ref{Ipdef}), with the
condition that the indices $i_\alpha$ can only assume positive values 
here; this implies that no gluon will be inserted to the right of either
quark\footnote{An exception must be made when $k=n$, when the first gluon 
is inserted in the only possible place, i.e.~onto the then-empty first 
(second) quark line in the secondary R-flow (L-flow).\label{ftn:zero}} in 
$\bsigma$. This constraint is necessary in order to preserve the cyclicity
properties of gluon-only flows. Because of it, one observes that, for any 
given values of \mbox{$i_2\ldots i_k$}, the sum over $i_1$ in 
eq.~(\ref{bAdefgg}) features $n-1$ terms, which are exactly those 
that enter into the dual Ward identity. Therefore
\beqn
\ampOSn(\bsigma)&=&0\,,\phantom{aaa}
\;\;\;\;\;\;\;\;\;\;\;\;\;\;\;\;\forall\,k>0\,,
\label{DWI1}
\\
\ampOSn(\bsigma)&=&\ampCSn(\bsigma)\,,\phantom{aaa}
\;\;\;\;\;\;k=0\,,
\label{DWI2}
\eeqn
and eq.~(\ref{GGAflowsum}) simplifies to read
\beq
\ket{\ampn}=
\frac{1}{2^{n/2}}\sum_{\bsigma\in P_n^\prime}\ampCSn(\bsigma)
\!\!\!\sum_{\setijn}
\!\!\Lambda\left(\{ij\},\bsigma\right)\ket{i_1j_1,\ldots i_nj_n}\,.
\label{GGflowsum}
\eeq
This is the anticipated, and expected, result: when the sum over
flows is carried out, all of the contributions due to $U(1)$ gluons
(which are non-zero vectors that live in the 
\mbox{$\spaceij^{(p)}\setminus\spacea^{(p)}$} subspaces, with $p=1\ldots n$) 
add to zero. One must bear this fact in mind when claiming that, for
gluon-only amplitudes, the fundamental and the flow representation are
strictly equivalent.

One can finally use eq.~(\ref{GGflowsum}) to arrive at the result,
analogous to that of eq.~(\ref{matelfin}), for the matrix elements
\beq
\ampsqn=\frac{1}{2^n}\,
\sum_{\bsigma,\bsigmap\in P_n^\prime}
\ampCSn(\bsigmap)^\star\,\ampCSn(\bsigma)\,
N^{\nloops{\loopset(\bsigmap,\bsigma)}}\,,
\phantom{aa}
\label{matelfingg}
\eeq
which again is as expected. Although the notation of eq.~(\ref{matelfingg}) 
bears memory of the usage of secondary flows, in the absence of $U(1)$ gluons, 
and in view of the definition of colour loops adopted here, there is no 
difference between the original flows and the secondary ones in 
eq.~(\ref{matelfingg}).

\section{Soft limits\label{sec:soft}}
We consider here the tree-level matrix elements relevant to processes 
that feature at least one gluon, in the limit in which one of such
gluons is soft, while all of the other partons are hard and well
separated from each other.

These soft limits have been studied in detail in ref.~\cite{Frixione:2011kh};
we shall show below how the formulae of that paper assume an extremely
simple form with a geometrical character by exploiting what has been
done in sect.~\ref{sec:tree}. The results of ref.~\cite{Frixione:2011kh}
were derived with the goal of employing them in the FKS subtraction
formalism~\cite{Frixione:1995ms,Frixione:1997np} 
for the computation of NLO-accurate cross sections
and, to that end, were given in three different forms, namely: 
at fixed colour configurations, and at fixed colour flows either at the 
real-emission or at the Born level. Since the emphasis of the current
paper is on PSMCs, working at fixed colour configurations is of no
interest. Fixing the colour flows at the real-emission level is also
problematic. Such kinematics configurations are those one obtains 
after the first emission in a PSMC; therefore, information gathered
from the matrix elements at that level cannot easily be used by the PSMC 
to generate those very same emissions. In keeping with its Markovian
structure, a shower will generate emissions in its $k^{th}$ step by
using information available at the \mbox{$(k-1)^{th}$} step. The
most convenient option, then, is to work at fixed Born-level
flows, which is what we are going to do in the present section.

Consistently with the previous statement, when referring to processes
that feature $n$ gluons we understand those gluon to be at the Born level.
In particular, a quark-only case is that of the soft limit of a matrix 
element relevant to those quarks plus an additional gluon that becomes 
soft.

\subsection{Quark-only matrix elements\label{sec:softqrk}}
In this case, the quantity of interest is that in \Feq{3.86} 
(with $n=0$), which we report here for convenience
\beqn
\ampsqQGRSz(\gammap,\gamma)&=&\half\gs^2
\sum_{k,l=-2q}^{-1}\eik{k}{l}
\ampsqQGBggz_{kl}(\gammap,\gamma)\,.
\label{MQGsoftdef2}
\eeqn
The r.h.s.~of this equation features the usual eikonal factors,
\beq\label{eq:eik_def}
\eik{k}{l} \equiv \frac{(p_k\cdot p_l)}{(p\cdot p_k)(p\cdot p_l)}
\eeq
where $p$ is the four-momentum of the emitted soft gluon,
$p_k$ and $p_l$ are those of the corresponding
quarks or antiquarks, and $\ampsqQGBggz_{kl}(\gammap,\gamma)$ is the
colour-linked matrix element for the given closed flow. The latter
have been presented in \Feq{3.115}; again we report here their
expressions
\beqn
\ampsqQGBggz_{kl}(\gammap,\gamma)&=&-2\,\ampCSQGBggz(\gammap)^\star\Big[
C\left(I_+(\igammap(k))\,\gammap,\,I_+(\igamma(l))\,\gamma\right)
\left(1-\delta_{\qb\ident_k}\right)
\left(1-\delta_{\qb\ident_l}\right)
\nonumber\\*&&\phantom{-2\,\ampCSQGBggz(\gammap)}\!\!
-C\left(I_+(\igammap(k))\,\gammap,\,I_-(\igamma(l))\,\gamma\right)
\left(1-\delta_{\qb\ident_k}\right)
\left(1-\delta_{q\ident_l}\right)
\nonumber\\*&&\phantom{-2\,\ampCSQGBggz(\gammap)}\!\!
-C\left(I_-(\igammap(k))\,\gammap,\,I_+(\igamma(l))\,\gamma\right)
\left(1-\delta_{q\ident_k}\right)
\left(1-\delta_{\qb\ident_l}\right)
\nonumber\\*&&\phantom{-2\,\ampCSQGBggz(\gammap)}\!\!
+C\left(I_-(\igammap(k))\,\gammap,\,I_-(\igamma(l))\,\gamma\right)
\left(1-\delta_{q\ident_k}\right)
\left(1-\delta_{q\ident_l}\right)
\Big]
\nonumber\\*&&\!\!\times
\ampCSQGBggz(\gamma)\,.
\label{MQGkldef2sc}
\eeqn
Fuller details can be found in ref.~\cite{Frixione:2011kh}. It turns
out that the explicit computation of eq.~(\ref{MQGsoftdef2}) leads to
the following simple result:
\beqn
\ampsqQGRSz(\gammap,\gamma)&=&2\gs^2\,
\ampCSQGBggz(\gammap)^\star\ampCSQGBggz(\gamma)\,
N^{-\rho(\gamma)-\rho(\gammap)}\!N^{\nloops{\loopset(\gammap,\gamma)}}
\nonumber\\*&&\phantom{aaa}\times
\CF\!\!\sum_{\ell\in\loopset(\gammap,\gamma)}
\sum_{k,l\in\ell}^{k<l}(-1)^{\delta_\ell(k,l)}\eik{k}{l}\,.
\label{qqsoftres}
\eeqn
The proof of eq.~(\ref{qqsoftres}) proceeds as follows.
Start by considering eq.~(\ref{MQGkldef2sc}): of the four terms on
the r.h.s., only one may not be equal to zero, owing to the fact that both
$k$ and $l$ are either a quark or an antiquark, and the conclusion follows
from the presence of the Kronecker $\delta$'s. Thus, a single colour
factor is relevant, which we write as follows (the notation is symbolic: 
the appropriate choices of the signs where $\pm$ appear are given in 
eq.~(\ref{MQGkldef2sc}))
\beq
\pm C\left(I_\pm(\igammap(k))\,\gammap,\,I_\pm(\igamma(l))\,\gamma\right)\,.
\label{softCqq}
\eeq
The general form of eq.~(\ref{softCqq}) can be obtained from 
eq.~(\ref{Lambdagq}). Taking footnote~\ref{ft:empty} into account, and 
denoting by $a$ the colour of the only gluon present in the process, there 
are only two possibilities for the colour factor of eq.~(\ref{softCqq}), 
namely
\beqn
&&\pm C\left(I_\pm(\igammap(k))\,\gammap,\,I_\pm(\igamma(l))\,\gamma\right)
\;\ni\;\sum_a{\rm Tr}\left(\lambda^a\right){\rm Tr}\left(\lambda^a\right)
\,=\,0\,,
\label{softCqqz}
\\
&&\pm C\left(I_\pm(\igammap(k))\,\gammap,\,I_\pm(\igamma(l))\,\gamma\right)
\;\ni\;\sum_a{\rm Tr}\left(\lambda^a\lambda^a\right)
\,=\,\CF{\rm Tr}\left(I\right)\,.\phantom{aaaa}
\label{softCqqnz}
\eeqn
The case of eq.~(\ref{softCqqz}) (eq.~(\ref{softCqqnz})) occurs when the 
operators $I_\pm$ insert the gluon onto open colour lines which do not 
(do) belong to the same closed colour line\footnote{A closed colour
line is obtained by merging open colour lines, after reversing those 
that belong to the L-flow: starting from one given such line in the R-flow, 
one attaches to its end the colour line in the L-flow whose antiquark label 
is identical to that of the line in the R-flow; the procedure is iterated 
until the quark label in the colour line of the last L-flow coincides with 
that of the first line considered in the R-flow.}. It is easy to see that 
this happens when the partons $k$ and $l$ do not (do) belong to the same
Born-level colour loop. This explains the two summations that appear
on the r.h.s.~of eq.~(\ref{qqsoftres}). By considering the only
non-trivial case, eq.~(\ref{softCqqnz}), one sees that the rightmost
side is equivalent to removing the gluon from the closed colour line
where it appears (with the factor $\CF$ accounting for it); the remaining
trace of the identity is the contribution of that closed colour line stripped 
of the gluon (which now contains only quarks and antiquarks, and thus
coincides with a colour loop) to the overall colour factor. Then, since 
the other closed colour lines have not been affected by the $I_\pm$ operators, 
and are therefore identical to their Born-level colour-loop counterparts, 
eq.~(\ref{softCqqnz}) can be written more precisely as
follows
\beqn
\pm C\left(I_\pm(\igammap(k))\,\gammap,\,I_\pm(\igamma(l))\,\gamma\right)
&=&\pm\CF\,C\left(\gammap,\gamma\right)
\nonumber\\
&=&\pm\CF\,N^{-\rho(\gamma)-\rho(\gammap)}\!
N^{\nloops{\loopset(\gammap,\gamma)}}\,,
\label{softCqqnz2}
\eeqn
having repeated the procedure that led us to eq.~(\ref{treeQres}).
Equation~(\ref{softCqqnz2}) essentially completes the proof of 
eq.~(\ref{qqsoftres}), bar the sign in front of each eikonal. 
In order to establish that, one observes that the sign in front of each 
colour-flow matrix element on the r.h.s.~of eq.~(\ref{MQGkldef2sc}) 
(taking the overall minus sign into account) is equal to $-1$ if $k$ 
and $l$ are both quarks or both antiquarks, and equal to $+1$ if one 
of $k$ and $l$ is a quark and the other an antiquark. According to 
item {\rm ii.} in sect.~\ref{sec:loops} the former case is that of $k$ 
and $l$ being both R-elements or L-elements, while the latter case is 
that of one R-element and one L-element. As such, in the former (latter) 
case $k$ and $l$ are separated by an odd (even) number of partons. 
By putting all this together, one obtains the factor 
$(-1)^{\delta_\ell(k,l)}$ on the r.h.s.~of eq.~(\ref{qqsoftres}).
Finally, the factor $1/2$ in eq.~(\ref{MQGsoftdef2}) is necessary
to avoid double counting, the summands on the r.h.s.~being symmetric
when \mbox{$k\leftrightarrow l$}; such a symmetry is removed in
the sum on the r.h.s.~of eq.~(\ref{qqsoftres}) (owing to the 
\mbox{$k<l$} constraint), and therefore no factor $1/2$ is present
there. This concludes the proof.

There are two remarkable things about eq.~(\ref{qqsoftres}). Firstly,
it is an entirely geometrical formula; no colour algebra is required,
and its expression is fully determined by the structure of the colour
loops. Secondly, such loops are Born-level quantities, whereas one would 
expect that soft radiation pattern be determined by \mbox{$(n+1)$}-body
matrix elements. In fact, there is no contradiction between these two
statements: colour-wise, the present case is so tightly constrained
(by the Kronecker $\delta$'s in eq.~(\ref{MQGkldef2sc}), and by the
structure that leads to eqs.~(\ref{softCqqz}) and~(\ref{softCqqnz})) as
to give a one-to-one correspondence between Born- and real-emission-level
quantities.

\subsection{Quark-gluon matrix elements\label{sec:softqg}}
The calculation of the soft limit of interest here is trivial if
one exploits the results of sects.~\ref{sec:treeqg} and~\ref{sec:softqrk}.
The essence of the procedure of sect.~\ref{sec:treeqg} is to write 
the matrix elements for a quark-gluon process in terms of secondary 
(i.e., quark-only) flows. However, the procedure is iterative, and can 
be stopped after dealing with any number of gluons. This is not of 
particular interest when all partons are hard and well separated from
each other, but in the present context it is helpful, since one can 
get rid of all gluons except that which becomes soft. By doing so,
one is left with a sum of $q$-quark-antiquark, one-gluon contributions, 
each of which has a soft limit given by eq.~(\ref{qqsoftres}). By putting all
together one obtains
\beqn
\ampsqQGRS(\gammap,\gamma)&=&2\gs^2\,
\ampCSQGBgg(\gammap)^\star\ampCSQGBgg(\gamma)\,
N^{-\rho(\gamma)-\rho(\gammap)}
\nonumber\\*&&\phantom{aa}\times\,
\CF\sum_{p=0}^n\sum_{s_p\in\Sset{n}{p}}
\frac{(-1)^{p}}{2^n}\,N^{-p}\,
N^{\nloops{\loopset(\bgammap_{\cancel{s_p}},\bgamma_{\cancel{s_p}})}}
\nonumber\\*&&\phantom{aa}\times\!
\sum_{\ell\in\loopset(\bgammap_{\cancel{s_p}},\bgamma_{\cancel{s_p}})}
\sum_{k,l\in\ell}^{k<l}(-1)^{\delta_\ell(k,l)}\eik{k}{l}\,.
\label{qgsoftres2}
\eeqn
Equation~(\ref{qgsoftres2}) allows one to sum over flows, by following 
again the same procedure as in sect.~\ref{sec:treeqg}. Defining
\beq
\ampsqQGRS=\sum_{\gamma,\gammap\in\flowBgg}\ampsqQGRS(\gammap,\gamma)
\eeq
from eqs.~(\ref{matelfin}) and~(\ref{qgsoftres2}) one obtains
\beqn
\ampsqQGRS&=&2\gs^2\,
\CF\sum_{p=0}^n\sum_{s_p\in\Sset{n}{p}}
\frac{(-1)^{p}}{2^n}\,N^{-p}\,
\nonumber
\\*&&\phantom{aa}\times\!\!\!\!\!\!
\sum_{\bgamma,\bgammap\in\flowBggnp^{(s_p)}}
\!\!\!
\ampOSQGBgg(\bgammap)^\star\,\ampOSQGBgg(\bgamma)\,
N^{-\rho(\bgamma)-\rho(\bgammap)}
N^{\nloops{\loopset(\bgammap,\bgamma)}}\,
\nonumber\\*&&\phantom{aa}\times\!
\sum_{\ell\in\loopset(\bgammap,\bgamma)}
\sum_{k,l\in\ell}^{k<l}(-1)^{\delta_\ell(k,l)}\eik{k}{l}\,.
\label{qgsoftsum}
\eeqn
For future use, we denote the sum in the last line of eq.~(\ref{qgsoftsum}) 
as follows:
\beq
\eiksum(\bgammap,\bgamma)=
\sum_{\ell\in\loopset(\bgammap,\bgamma)}
\sum_{k,l\in\ell}^{k<l}(-1)^{\delta_\ell(k,l)}\eik{k}{l}\,,
\label{eiksum}
\eeq
and call it the {\em radiation pattern} associated with the
(secondary) closed flow $(\bgammap,\bgamma)$. 

The similarity between eqs.~(\ref{qgsoftres2}) and~(\ref{qgsoftsum})
on the one hand, and eq.~(\ref{qqsoftres}) on the other, is
remarkable, in view of the fact that, at variance with the case of
a single final-state gluon, matrix elements that feature at least
two gluons have a more involved colour structure, which prevents one
from establishing in a straightforward manner a one-to-one correspondence
between Born- and real-emission-level quantities. As is shown here, this
can be done by employing secondary flows.

Needless to say, eq.~(\ref{qgsoftsum}) can be cast in the 
standard\footnote{With only possible differences of notation,
eq.~(\ref{softsum}) holds for the quark- and gluon-only cases
as well, which we refrain from writing explicitly.} form for soft 
matrix elements summed over flows (see e.g.~\Feq{3.33}), namely
\beq
\ampsqQGRS=\half\,\gs^2\sum_{k,l=-2q}^n\eik{k}{l}\ampsqQGBgg_{kl}\,,
\label{softsum}
\eeq
where use is made of the colour-linked Borns, $\ampsqQGBgg_{kl}$,
whose definition can be found in \Feq{3.35}.
These quantities satisfy the completeness relations
(see e.g.~eq.~(3.14) of ref.~\cite{Frederix:2018nkq})
\beq
\sum_{\stackrel{l\ne k}{l=-2q}}^n\ampsqQGBgg_{kl}=
2C(\ident_k)\ampsqQGBgg\,,
\label{CLBcompl}
\eeq
which, as is shown in ref.~\cite{Frixione:2011kh} using the language of
the present paper, stem from the conservation of colour. In
eq.~(\ref{CLBcompl}) by $C(\ident_k)$ we have denoted the Casimir factor
(equal to $\CF$ for quarks and to $\CA$ for gluons) of a parton with
identity $\ident_k$.

\subsection{Gluon-only matrix elements\label{sec:softgg}}
The case of gluon-only matrix elements can be dealt with as
has been done in sect.~\ref{sec:softqg}, thanks to the equivalence
established in sect.~\ref{sec:treeglu} between gluon-only colour flows
and suitably constructed quark-gluon ones, eqs.~(\ref{qRdef2}) 
and~(\ref{qLdef2}); in particular, the properties in 
eqs.~(\ref{lseq})--(\ref{distparity}) are crucial.
Thus
\beqn
\ampsqnpoS(\sigmap,\sigma)&=&2\gs^2\,
\ampCSn(\sigmap)^\star\ampCSn(\sigma)\,
\nonumber\\*&&\phantom{aa}\times\,
\CF\sum_{p=0}^n\sum_{s_p\in\Sset{n}{p}}
\frac{(-1)^{p}}{2^n}\,N^{-p}\,
N^{\nloops{\loopset(\bsigmapqLarg{\cancel{s_p}},\bsigmaqRarg{\cancel{s_p}})}}
\nonumber\\*&&\phantom{aa}\times\!
\sum_{\ell\in\loopset{\bsigmapqLarg{\cancel{s_p}},\bsigmaqRarg{\cancel{s_p}}}}
\sum_{k,l\in\ell}^{0<k<l}(-1)^{\delta_\ell(k,l)}\eik{k}{l}\,.
\label{ggsoftres}
\eeqn
Note that, at variance with what happens in eq.~(\ref{qgsoftres2}),
the innermost sum on the r.h.s.~of eq.~(\ref{ggsoftres}) constrains
the indices $k$ and $l$ to be larger than zero, which in turn implies
that no quarks or antiquarks will appear in the eikonals. This is an
obvious requirement, since the quarks and antiquarks introduced
in eqs.~(\ref{qRdef2}) and~(\ref{qLdef2}) are unphysical.

With a procedure analogous to that which leads to eq.~(\ref{matelfingg}),
from eq.~(\ref{ggsoftres}) one can also obtain the soft matrix element 
summed over flows
\beqn
\ampsqnpoS&=&
\sum_{\sigma,\sigmap\in P_n^\prime}\ampsqnpoS(\sigmap,\sigma)
\nonumber\\*&=&
\frac{2\gs^2\,\CF}{2^n}\,
\sum_{\bsigma,\bsigmap\in P_n^\prime}
N^{\nloops{\loopset(\bsigmap,\bsigma)}}\,
\ampCSn(\bsigmap)^\star\,\ampCSn(\bsigma)
\nonumber\\*&&\phantom{aaa}\times\!
\sum_{\ell\in\loopset(\bsigmap,\bsigma)}
\sum_{k,l\in\ell}^{0<k<l}(-1)^{\delta_\ell(k,l)}\eik{k}{l}\,,
\label{ggsoftsum}
\eeqn
which results from significant simplifications that stem from employing
the dual Ward identities of eq.~(\ref{DWI1}), and eq.~(\ref{DWI2}).

\subsection{General considerations\label{sec:softcomm}}
Equations~(\ref{qqsoftres}), (\ref{qgsoftres2}), and~(\ref{ggsoftres}), 
or their flow-summed counterparts (eqs.~(\ref{qgsoftsum}) 
and~(\ref{ggsoftsum}), the quark-only case being trivially obtained
from eq.~(\ref{qqsoftres})) tell one that processes that feature
only quarks at the Born level are essentially maximally complicated
as far as soft-radiation patterns are concerned. Gluons are either
treated as equivalent to a quark-antiquark pair, or they just decouple
completely, consistently with their interpretation as $U(1)$ colourless
objects.

These results also suggest a natural generalization of the concept of 
colour partner we are used to in the context of current PSMCs: given 
a parton (understood here as either a quark, or an antiquark, or one of
the two colour indices of an $SU(N)$ gluon), its colour partners are all 
of the partons that belong to the colour loop to which that parton 
also belongs. This idea indeed encompasses the standard one, which 
is relevant to leading-$N$ contributions, where $\gammap=\gamma$ or
$\sigmap=\sigma$, and no $U(1)$ gluons in the case of quark-gluon and
gluon-only processes. In such a situation, loops are trivial, and it
is easy to see that one ends up with quark and antiquarks having one
colour partner, and gluons having two of them (one for the colour, and
one for the anticolour), as usual. In general, however, a parton can 
have up to $(n-1)$ colour partners (i.e.~up to $2(n-1)$ for a gluon,
if its colour and anticolour are simultaneously accounted for), 
corresponding to a more involved radiation pattern than that relevant 
to a leading-$N$ picture.

It should be clear that, having chosen a parton, its colour partners
are those that enter all of the eikonals that also feature the former one.
From the results presented before, one sees that eikonals contribute
to the cross section with either sign, and therefore cancellations
must be expected. Therefore, in a refinement of the previous
paragraph, one is entitled to consider as colour partners only those
partons that enter the eikonals that survive algebraic simplifications.

Some such simplifications can be understood in a fairly general manner.
Consider in particular the case of a gluon, and its role in colour loops,
as detailed in item iv.~in sect.~\ref{sec:loops}. Suppose that the two
colour indices of a given gluon, which we denote by $g^\star$, appear 
in the same loop. It follows that all of the eikonals that feature $g^\star$
will cancel in the physical cross section. This is because those two indices, 
per item iv., will correspond to one L-element and to one R-element. Any 
other parton $k$ in the loop will therefore result in two eikonals,
both equal to \mbox{$[k,g^\star]$} but with opposite signs in front --
this is because the two distances $\delta_\ell(k,g^\star)$ when computed
with $g^\star$ regarded as an L- or an R-element will have opposite
parity. A corollary of this is that when a closed colour flow in
a quark-gluon process corresponds to a single colour loop, then no
eikonal that contains at least one gluon contributes to the cross section.

\subsection{Examples}
\subsubsection[$\bar q q\to gg$]{\boldmath  $\bar q q\to gg$}\label{sec:qqgg}
The set of primary flows was given in eq~(\ref{Bflowex1}). 
The sets of indices of $U(1)$ gluons, eq.~(\ref{Ssetdef}), are
\beq
\Sset{2}{0}=\Big\{\{\}\Big\}\equiv\Big\{s_0\Big\}\,,
\;\;\;\;
\Sset{2}{1}=\Big\{\big\{1\big\}\,,\big\{2\big\}\Big\}\equiv
\Big\{s_{1,1}\,,s_{1,2}\Big\}\,,
\;\;\;\;
\Sset{2}{2}=\Big\{\big\{1,2\big\}\Big\}\equiv\Big\{s_2\Big\}\,.
\label{Ssetex1}
\eeq
The sets of secondary flows, eq.~(\ref{Fnksets}), are
\beqn
\Sflowexot^{(s_0)}&=&\Big\{\big(\Qa;\bo,\bt;\Qb\big),
\big(\Qa;\bt,\bo;\Qb\big)\Big\}\equiv
\Big\{\bgamma_{0,1},\bgamma_{0,2}\Big\}\,,
\label{Sflow1ex1}
\\
\Sflowexoo^{(s_{1,1})}&=&\Big\{\big(\Qa;\bt;\Qb\big)\Big\}\equiv
\Big\{\bgamma_{1,1}\Big\}\,,
\label{Sflow2ex1}
\\
\Sflowexoo^{(s_{1,2})}&=&\Big\{\big(\Qa;\bo;\Qb\big)\Big\}\equiv
\Big\{\bgamma_{1,2}\Big\}\,,
\label{Sflow3ex1}
\\
\Sflowexoz^{(s_2)}&=&\Big\{\big(\Qa;\Qb\big)\Big\}\equiv
\Big\{\bgamma_{2,1}\Big\}\,.
\label{Sflow4ex1}
\eeqn
Table~\ref{tab:loopsets_qqgg} reports the various quantities that 
appear in eq.~(\ref{qgsoftsum}), for each non-trivial secondary 
closed flow -- number of $U(1)$ gluons ($p$, first column), 
$\rho$ parameters for the L- and R-flows (second and third columns), 
set of colour loops (fourth column, eq.~(\ref{loopsetdef})), number 
of colour loops (fifth column), and colour factors (sixth column; these
are defined as the products of all the $N$ monomials in 
eq.~(\ref{qgsoftsum}), times the factor $(-)^p$).
\begin{table}[h!]
  \begin{center}    
    \begin{tabular}{|c|cccccc|}
      \hline
      $(\bgammap,\bgamma)$ & $p$ & $\rho(\bgammap)$ &  $\rho(\bgamma)$
      & $\loopset (\bgammap,\bgamma)$ &
           $\nloops{\loopset(\bgammap,\bgamma)}$ & Col.fac.\\
      \hline
 $(\bgamma_{0,1},\bgamma_{0,1})$ & 0 & 0 & 0 & 
 $\Big\{\big(\Qa,\bo\big),\big(\bo,\bt\big),\big(\bt,\Qb\big)\Big\}$ & 3
                                  & $N^3$ \\
 $(\bgamma_{0,1},\bgamma_{0,2})$ & 0 & 0 & 0 & 
 $\Big\{\big(\Qa,\bt,\bo,\Qb,\bt,\bo\big)\Big\}$ & 1
                                  & $N$ \\
$(\bgamma_{0,2},\bgamma_{0,2})$ & 0 & 0 & 0 & 
 $\Big\{\big(\Qa,\bt\big),\big(\bt,\bo\big),\big(\bo,\Qb\big)\Big\}$ & 3
                                  & $N^3$ \\
$(\bgamma_{1,1},\bgamma_{1,1})$ & 1 & 0 & 0 & 
 $\Big\{\big(\Qa,\bt\big),\big(\bt,\Qb\big)\Big\}$ & 2
                                  & $-N$ \\
$(\bgamma_{1,2},\bgamma_{1,2})$ & 1 & 0 & 0 & 
 $\Big\{\big(\Qa,\bo\big),\big(\bo,\Qb\big)\Big\}$ & 2
                                  & $-N$ \\
$(\bgamma_{2,1},\bgamma_{2,1})$ & 2 & 0 & 0 & 
 $\Big\{\big(\Qa,\Qb\big)\Big\}$ & 1
                                  & $N^{-1}$ \\
     \hline
    \end{tabular}
  \end{center}
  \caption{
    \label{tab:loopsets_qqgg}Quantities relevant to eq.~(\ref{qgsoftsum})
for $\bar q q\to gg$, for each non-trivial closed flow. See the text
for details.}
\end{table}

The radiation patterns (see eq.~(\ref{eiksum})) associated with the
secondary closed flows of table~\ref{tab:loopsets_qqgg} read as follows:
\beqn
\eiksum(\bgamma_{0,1},\bgamma_{0,1})&=&
\eik{\Qa}{\bo}+\eik{\bo}{\bt}+\eik{\bt}{\Qb}\,,
\label{eikf1ex1}
\\
\eiksum(\bgamma_{0,1},\bgamma_{0,2})&=&
\eik{\Qa}{\Qb}\,,
\label{eikf2ex1}
\\
\eiksum(\bgamma_{0,2},\bgamma_{0,2})&=&
\eik{\Qa}{\bt}+\eik{\bt}{\bo}+\eik{\bo}{\Qb}\,,
\label{eikf3ex1}
\eeqn
\beqn
\eiksum(\bgamma_{1,1},\bgamma_{1,1})&=&
\eik{\Qa}{\bt}+\eik{\bt}{\Qb}\,,
\label{eikf4ex1}
\\
\eiksum(\bgamma_{1,2},\bgamma_{1,2})&=&
\eik{\Qa}{\bo}+\eik{\bo}{\Qb}\,,
\label{eikf5ex1}
\\
\eiksum(\bgamma_{2,1},\bgamma_{2,1})&=&
\eik{\Qa}{\Qb}\,.
\label{eikf6ex1}
\eeqn
Notice that (\ref{eikf2ex1}) illustrates the fact, mentioned in
sect.~\ref{sec:softcomm}, that gluons appearing twice in any loop do not
contribute to the associated radiation pattern.
It trivially follows from eq.~(\ref{Bflowex1}) that there are two
independent dual amplitudes
\beq
\ampCSQGexo(\gamma_1)\,,\;\;\;\;
\ampCSQGexo(\gamma_2)\,.
\eeq
Thus, the amplitudes associated with secondary flows are the following
(from eq.~(\ref{bAdef}), and the definition of the $I_+$ operator, 
eq.~(\ref{Ipdef})):
\beqn
\ampOSQGexo(\bgamma_{0,1})&=&\ampCSQGexo(\gamma_1)\,,
\label{bamp1ex1}
\\
\ampOSQGexo(\bgamma_{0,2})&=&\ampCSQGexo(\gamma_2)\,,
\label{bamp2ex1}
\\
\ampOSQGexo(\bgamma_{1,1})&=&\ampCSQGexo(\gamma_1)+\ampCSQGexo(\gamma_2)\,,
\label{bamp3ex1}
\\
\ampOSQGexo(\bgamma_{1,2})&=&\ampCSQGexo(\gamma_1)+\ampCSQGexo(\gamma_2)\,,
\label{bamp4ex1}
\\
\ampOSQGexo(\bgamma_{2,1})&=&\ampCSQGexo(\gamma_1)+\ampCSQGexo(\gamma_2)\,.
\label{bamp5ex1}
\eeqn
This leads to four quantities that enter eqs.~(\ref{matelfin}) (tree level)
and~(\ref{qgsoftsum}) (soft limit). Note that amplitudes associated with
different sets of $U(1)$ gluons do not interfere with each other.
By direct computation, defining $s=(p_{-1}+p_{-2})^2$, $t=(p_{-1}+p_1)^2$,
$u=(p_{-1}+p_2)^2$,
\beqn
M_1&\equiv&
\ampOSQGexo(\bgamma_{0,1})^\star\,\ampOSQGexo(\bgamma_{0,1})=
8\gs^4\left(\frac{u}{t}-2\,\frac{u^2}{s^2}\right)\,,
\label{bMat1ex1}
\\
M_2&\equiv&
2\Re\left(\ampOSQGexo(\bgamma_{0,1})^\star\,
\ampOSQGexo(\bgamma_{0,2})\right)=
16\,\gs^4\frac{t^2+u^2}{s^2}\,,
\label{bMat2ex1}
\\
M_3&\equiv&
\ampOSQGexo(\bgamma_{0,2})^\star\,\ampOSQGexo(\bgamma_{0,2})=
8 \gs^4\left(\frac{t}{u}-2\,\frac{t^2}{s^2}\right)\,,
\label{bMat3ex1}
\\
M_4&\equiv&
\left(\ampCSQGexo(\gamma_1)+\ampCSQGexo(\gamma_2)\right)^\star
\left(\ampCSQGexo(\gamma_1)+\ampCSQGexo(\gamma_2)\right)
\nonumber
\\*&=&
M_1+M_2+M_3=8 \gs^4\left(\frac{t}{u}+\frac{u}{t}\right)\,.
\label{bMat4ex1}
\eeqn
With this, eq.~(\ref{matelfin}) leads to (note that $\rho(\bgamma_\alpha)=0$
for any $\alpha$)
\beq
4\ampsqQGexo=N^3\left(M_1+M_3\right)+N M_2
-\frac{2}{N}\,N^2\,M_4
+\frac{1}{N^2}\,N\,M_4\,,
\label{Mtreeex1}
\eeq
with the first two, the third, and the fourth term on the r.h.s.~stemming
from the contributions associated with zero, one, and two $U(1)$ gluons,
respectively. Likewise, from eq.~(\ref{qgsoftsum}) we obtain
\beqn
\frac{2\ampsqQGSexo}{\gs^2\CF}&=&N^3\eiksum(\bgamma_{0,1},\bgamma_{0,1})M_1+
N^3\eiksum(\bgamma_{0,2},\bgamma_{0,2})M_3+
N\eiksum(\bgamma_{0,1},\bgamma_{0,2})M_2
\nonumber
\\*&-&
\frac{1}{N}\,N^2\Big(\eiksum(\bgamma_{1,1},\bgamma_{1,1})+
\eiksum(\bgamma_{1,2},\bgamma_{1,2})\Big)M_4
\nonumber
\\*&+&
\frac{1}{N^2}\,N
\eiksum(\bgamma_{2,1},\bgamma_{2,1})M_4\,,
\label{Msoftex1}
\eeqn
with the first, second, and third line on the r.h.s.~stemming from the 
contributions associated with zero, one, and two $U(1)$ gluons,
respectively.

In terms of the colour-linked Borns introduced in eq.~(\ref{softsum}),
in this case denoted by $\ampsqQGexo_{kl}$, we have
\beq
\ampsqQGSexo=\half\,\gs^2\sum_{k,l=-2}^2\eik{k}{l}\ampsqQGexo_{kl}\,,
\label{softsum22}
\eeq
and keeping in mind that \mbox{$\ampsqQGexo_{kl}=\ampsqQGexo_{lk}$},
we obtain
\beqn
\label{M22clB1}
\ampsqQGexo_{-2-1}&=&\frac{\gs^4\CF}{32}\frac{1}{N}
\left(M_2+\frac{1}{N^2}\,M_4\right)\,,
\\
\ampsqQGexo_{-21}&=&\frac{\gs^4\CF}{32}\left(NM_3-\frac{1}{N}\,M_4\right)\,,
\\
\ampsqQGexo_{-22}&=&\frac{\gs^4\CF}{32}\left(NM_1-\frac{1}{N}\,M_4\right)\,,
\\
\ampsqQGexo_{-11}&=&\frac{\gs^4\CF}{32}\left(NM_1-\frac{1}{N}\,M_4\right)\,,
\\
\ampsqQGexo_{-12}&=&\frac{\gs^4\CF}{32}\left(NM_3-\frac{1}{N}\,M_4\right)\,,
\\
\ampsqQGexo_{12}&=&\frac{\gs^4\CF}{32}N\Big(M_1+M_3\Big)\,.
\label{M22clB6}
\eeqn
As expected, these expressions satisfy the completeness relations (\ref{CLBcompl})
\beq
\sum_{\stackrel{l\ne k}{l=-2}}^2\ampsqQGexo_{kl}=
2C(\ident_k)\ampsqQGexo\,.
\label{CLBcompl22}
\eeq
 
\subsubsection[$\bar q q\to q'\bar q'g$]{\boldmath  $\bar q q\to
  q'\bar q'g$}\label{sec:qqqqg}
The process is $\bar q_{-1}\,q_{_-3}\to q'_{-2}\bar q'_{-4}g_1$, with $q=2$ and
$n=1$.  The primary flows are
\beqn 
\gamma_1 &=& \big(-1;1;-3\bigr) \big(-2;-4\bigr)\,, \\
\gamma_2 &=& \big(-1;1;-4\bigr) \big(-2;-3\bigr)\,, \\
\gamma_3 &=& \big(-2;1;-3\bigr) \big(-1;-4\bigr)\,, \\
\gamma_4 &=& \big(-2;1;-4\bigr) \big(-1;-3\bigr)\,.
\eeqn 
The secondary flows are
\beqn 
\bgamma_{0,1} &=& \big(-1;\bo ;-3\bigr) \big(-2;-4\bigr)\,, \\
\bgamma_{0,2} &=& \big(-1;\bo ;-4\bigr) \big(-2;-3\bigr)\,, \\
\bgamma_{0,3} &=& \big(-2;\bo ;-3\bigr) \big(-1;-4\bigr)\,, \\
\bgamma_{0,4} &=& \big(-2;\bo ;-4\bigr) \big(-1;-3\bigr)\,, \\
\bgamma_{1,1} &=& \big(-1;-3\bigr) \big(-2;-4\bigr)\,, \\
\bgamma_{1,2} &=& \big(-1;-4\bigr) \big(-2;-3\bigr)\,.
\eeqn 
The sets of associated colour loops are given in
table~\ref{tab:loopsets}, together with the computation of their
colour factors according to eq.~(\ref{qgsoftsum}).
\begin{table}[h!]
  \begin{center}    
    \begin{tabular}{|c|cccccc|}
      \hline
      $(\bgammap,\bgamma)$ & $p$ & $\rho(\bgammap)$ &  $\rho(\bgamma)$
      & $\loopset (\bgammap,\bgamma)$ &
           $\nloops{\loopset(\bgammap,\bgamma)}$ & Col.fac.\\
      \hline
 $(\bgamma_{0,1},\bgamma_{0,1})$ & 0 & 1 & 1 & 
 $\Big\{\big(-1,\bo\big),\big(\bo,-3\big),\big(-2,-4\big)\Big\}$ & 3 
                                  & $N$ \\
 $(\bgamma_{0,1},\bgamma_{0,2})$ & 0 & 1 & 0 & 
 $\Big\{\big(-1,\bo\big),\big(\bo,-4,-2,-3\big)\Big\}$ & 2 
                                  & $N$ \\
$(\bgamma_{0,1},\bgamma_{0,3})$ & 0 & 1 & 0 & 
 $\Big\{\big(-2,\bo,-1,-4\big),\big(\bo,-3\big)\Big\}$ & 2 
                                  & $N$ \\
$(\bgamma_{0,1},\bgamma_{0,4})$ & 0 & 1 & 1 & 
 $\Big\{\big(-2,\bo,-1,-3,\bo,-4\big)\Big\}$ & 1 
                                  & $N^{-1}$ \\
$(\bgamma_{0,2},\bgamma_{0,2})$ & 0 & 0 & 0 & 
 $\Big\{\big(-1,\bo\big),\big(\bo,-4\big),\big(-2,-3\big)\Big\}$ & 3 
                                  & $N^3$ \\
$(\bgamma_{0,2},\bgamma_{0,3})$ & 0 & 0 & 0 & 
 $\Big\{\big(-2,\bo,-1,-4,\bo,-3\big)\Big\}$ & 1 
                                  & $N$ \\
$(\bgamma_{0,2},\bgamma_{0,4})$ & 0 & 0 & 1 & 
 $\Big\{\big(-2,\bo,-1,-3\big),\big(\bo,-4\big)\Big\}$ & 2
                                  & $N$ \\
 $(\bgamma_{0,3},\bgamma_{0,3})$ & 0 & 0 & 0 & 
 $\Big\{\big(-2,\bo\big),\big(\bo,-3\big),\big(-1,-4\big)\Big\}$ & 3 
                                  & $N^3$ \\
$(\bgamma_{0,3},\bgamma_{0,4})$ & 0 & 0 & 1 & 
 $\Big\{\big(-2,\bo\big),\big(\bo,-4,-1,-3\big)\Big\}$ & 2 
                                  & $N$ \\
$(\bgamma_{0,4},\bgamma_{0,4})$ & 0 & 1 & 1 & 
 $\Big\{\big(-2,\bo\big),\big(\bo,-4\big),\big(-1,-3\big)\Big\}$ & 3 
                                  & $N$ \\
$(\bgamma_{1,1},\bgamma_{1,1})$ & 1 & 1 & 1 & 
 $\Big\{\big(-1,-3\big),\big(-2,-4\big)\Big\}$ & 2 
                                  & $-N^{-1}$ \\
$(\bgamma_{1,1},\bgamma_{1,2})$ & 1 & 1 & 0 & 
 $\Big\{\big(-1,-4,-2,-3\big)\Big\}$ & 1 
                                  & $-N^{-1}$ \\
$(\bgamma_{1,2},\bgamma_{1,2})$ & 1 & 0 & 0 & 
 $\Big\{\big(-1,-4\big),\big(-2,-3\big)\Big\}$ & 2 
                                  & $-N$ \\
     \hline
    \end{tabular}
  \end{center}
  \caption{
    \label{tab:loopsets}As in table~\ref{tab:loopsets_qqgg}, for
$\bar q q\to q'\bar q' g$.}
\end{table}

Defining
\beq 
S_{ij}\equiv \frac{s_{i,j}}{s_{i,1}\,s_{j,1}}\,,
\eeq
where $s_{i,j}=(p_i+p_j)^2$, and
\beq
R =8\gs^4
\frac{s^2_{-1,-2}+s^2_{-1,-4}+s^2_{-2,-3}+s^2_{-3,-4}}
{s_{-1,-3}\,s_{-2,-4}}\,,
\eeq
we find by explicit calculation, summing over helicities,
\beqn
\label{eq:qqqqg11}
M_1=
\ampCS^{(4;1)}(\gamma_1)^\star\ampCS^{(4;1)}(\gamma_1)&=& S_{-1,-3}\,R\,, 
\\
M_2=
2\Re\left(\ampCS^{(4;1)}(\gamma_1)^\star\ampCS^{(4;1)}(\gamma_2)\right)
&=&\left(S_{-3,-4}-S_{-1,-3}-S_{-1,-4}\right)R\,,
\\
M_3=
2\Re\left(\ampCS^{(4;1)}(\gamma_1)^\star\ampCS^{(4;1)}(\gamma_3)\right)
&=& \left(S_{-1,-2}-S_{-1,-3}-S_{-2,-3}\right)R\,,
\\
M_4=
2\Re\left(\ampCS^{(4;1)}(\gamma_1)^\star\ampCS^{(4;1)}(\gamma_4)\right)
&=&\left(S_{-1,-4}+S_{-2,-3}-S_{-1,-2}-S_{-3,-4}\right)R\,, 
\\
M_5=
\ampCS^{(4;1)}(\gamma_2)^\star\ampCS^{(4;1)}(\gamma_2)&=&S_{-1,-4}\,R \,,
\\ 
M_6=
2\Re\left(\ampCS^{(4;1)}(\gamma_2)^\star\ampCS^{(4;1)}(\gamma_3)\right)
&=&\left(S_{-1,-3}+S_{-2,-4}-S_{-1,-2}-S_{-3,-4}\right)R\,,
\eeqn
\beqn
M_7=
2\Re\left(\ampCS^{(4;1)}(\gamma_2)^\star\ampCS^{(4;1)}(\gamma_4)\right)
&=&\left(S_{-1,-2}-S_{-1,-4}-S_{-2,-4}\right)R\,,
\\
M_8=
\ampCS^{(4;1)}(\gamma_3)^\star\ampCS^{(4;1)}(\gamma_3) &=& S_{-2,-3}\,R\,, 
\\
M_9=
2\Re\left(\ampCS^{(4;1)}(\gamma_3)^\star\ampCS^{(4;1)}(\gamma_4)\right)
&=&\left(S_{-3,-4}-S_{-2,-3}-S_{-2,-4}\right)R\,, 
\\
\label{eq:qqqqg44}
M_{10}=
\ampCS^{(4;1)}(\gamma_4)^\star\ampCS^{(4;1)}(\gamma_4)&=&S_{-2,-4}\,R \,.
\eeqn
Note that only six of these are independent, owing to the property of
the dual amplitudes
\beq
\sum_{i=1}^4\ampCS^{(4;1)}(\gamma_i)=0\,.
\eeq
The products of secondary flow amplitudes follow from
eq.~(\ref{bAdef}), which gives in this case
\beqn
\ampOS^{(4;1)}(\bgamma_{0,i})&=&\ampCS^{(4;1)}(\gamma_i)\,,
\phantom{+\ampCS^{(4;1)}(\gamma_4)\,,aaaa}
1\le i\le 4\,,
\label{bamp1ex2}
\\
\ampOS^{(4;1)}(\bgamma_{1,1})&=&
\ampCS^{(4;1)}(\gamma_1)+\ampCS^{(4;1)}(\gamma_4)\,,
\label{bamp2ex2}
\\
\ampOS^{(4;1)}(\bgamma_{1,2})&=&
\ampCS^{(4;1)}(\gamma_2)+\ampCS^{(4;1)}(\gamma_3)\,.
\label{bamp3ex2}
\eeqn
The radiation patterns associated with the secondary closed flows are
\beqn
\eiksum(\bgamma_{0,1},\bgamma_{0,1})&=&
\eik{-1}{\bo}+\eik{\bo}{-3}+\eik{-2}{-4}\,, 
\label{eikf11ex2}
\\
\eiksum(\bgamma_{0,1},\bgamma_{0,2})&=&
\eik{-1}{\bo}+\eik{\bo}{-4}-\eik{\bo}{-2}+\eik{\bo}{-3}+\eik{-4}{-2}\nonumber\\
&&-\eik{-4}{-3}+\eik{-2}{-3}\,, 
\label{eikf12ex2}
\\
\eiksum(\bgamma_{0,1},\bgamma_{0,3})&=&
\eik{-2}{\bo}-\eik{-2}{-1}+\eik{-2}{-4}+\eik{\bo}{-1}-\eik{\bo}{-4}\nonumber\\
&&+\eik{-1}{-4}+\eik{\bo}{-3}\,, 
\label{eikf13ex2}
\\
\eiksum(\bgamma_{0,1},\bgamma_{0,4})&=&
\eik{-1}{-3}+\eik{-1}{-4}-\eik{-2}{-1}+\eik{-2}{-3}\nonumber\\
&&+\eik{-2}{-4}-\eik{-3}{-4}\,, 
\label{eikf14ex2}
\\
\eiksum(\bgamma_{0,2},\bgamma_{0,2})&=&
\eik{-1}{\bo}+\eik{\bo}{-4}+\eik{-2}{-3}\,, 
\label{eikf22ex2}
\\
\eiksum(\bgamma_{0,2},\bgamma_{0,3})&=&
-\eik{-2}{-1}+\eik{-2}{-4}+\eik{-2}{-3}+\eik{-1}{-4}\nonumber\\
&&+\eik{-1}{-3}-\eik{-4}{-3}\,, 
\label{eikf23ex2}
\\
\eiksum(\bgamma_{0,2},\bgamma_{0,4})&=&
\eik{-2}{\bo}-\eik{-2}{-1}+\eik{-2}{-3}+\eik{\bo}{-1}-\eik{\bo}{-3}\nonumber\\ 
&&+\eik{-1}{-3}+\eik{\bo}{-4}\,, 
\label{eikf24ex2}
\\
\eiksum(\bgamma_{0,3},\bgamma_{0,3})&=&
\eik{-2}{\bo}+\eik{\bo}{-3}+\eik{-1}{-4}\,,
\label{eikf33ex2}
\\
\eiksum(\bgamma_{0,3},\bgamma_{0,4})&=&
\eik{-2}{\bo}+\eik{\bo}{-4}-\eik{\bo}{-1}+\eik{\bo}{-3}+\eik{-4}{-1}\nonumber\\
&&-\eik{-4}{-3}+\eik{-1}{-3}\,, 
\label{eikf34ex2}
\\
\eiksum(\bgamma_{0,4},\bgamma_{0,4})&=&
\eik{-2}{\bo}+\eik{\bo}{-4}+\eik{-1}{-3}\,,
\label{eikf44ex2}
\eeqn
\beqn
\eiksum(\bgamma_{1,1},\bgamma_{1,1})&=&
\eik{-1}{-3}+\eik{-2}{-4}\,, 
\label{eik1111ex2}
\\
\eiksum(\bgamma_{1,1},\bgamma_{1,2})&=&
\eik{-1}{-4}-\eik{-1}{-2}+\eik{-1}{-3}+\eik{-4}{-2}\nonumber\\
&&-\eik{-4}{-3}+\eik{-2}{-3}\,, 
\label{eik1112ex2}
\\
\eiksum(\bgamma_{1,2},\bgamma_{1,2})&=&
\eik{-1}{-4}+\eik{-2}{-3}\,.
\label{eik1212ex2}
\eeqn
Here again, eqs.~(\ref{eikf14ex2}) and (\ref{eikf23ex2}) reflect the
fact that a gluon appearing twice in the same loop does not contribute
to the associated radiation pattern.
Using eqs.~(\ref{eq:qqqqg11})--(\ref{eq:qqqqg44}), 
eqs.~(\ref{bamp1ex2})--(\ref{bamp3ex2}), and the $\rho$ parameters
and colour factors of table~\ref{tab:loopsets}, from eq.~(\ref{matelfin}) 
one obtains
\beq
\ampsqQGext=\CF\left[N^2\left(M_5+M_8\right)+
M_1+M_2+M_3+M_7+M_9+M_{10}\right]\,,
\eeq
for the tree-level matrix element of the current process. Furthermore,
by writing the flow-summed soft limit in the standard form as was done
in eq.~(\ref{softsum}), namely
\beq
\ampsqQGSext=\half\,\gs^2\sum_{k,l=-4}^1\eik{k}{l}\ampsqQGext_{kl}\,,
\label{softsum2}
\eeq
and using eqs.~(\ref{eikf11ex2})--(\ref{eik1212ex2}) in eq.~(\ref{qgsoftsum}), 
we arrive at the following expressions for the colour-linked Borns:
\beqn
\label{M41clB1}
\ampsqQGext_{-4,-3}&=&\CF\Big[-N\Big(M_2+M_6+M_9\Big)
+\frac{1}{N}\Big(M_2+M_3-M_4+M_7+M_9\Big)\Big]\,,
\\
\ampsqQGext_{-4,-2}&=&\CF\Big[N\Big(M_1+M_2+M_3+M_6\Big)
\\*&&\phantom{\CF\Big[}
-\frac{1}{N}\Big(M_1+M_2+M_3+M_7+M_9+M_{10}\Big)\Big]\,,
\\
\ampsqQGext_{-4,-1}&=&\CF\Big[N^3 M_8+N\Big(M_3-M_5-M_8+M_9\Big)
\\*&&\phantom{\CF\Big[}
-\frac{1}{N}\Big(M_2+M_3-M_4+M_7+M_9\Big)\Big]\,,
\\
\ampsqQGext_{-4,1}&=&\CF N\Big[N^2 M_5+M_2-M_3+M_7+M_9+M_{10}\Big]\,,
\\
\ampsqQGext_{-3,-2}&=&\CF\Big[N^3 M_5+N\Big(M_2-M_5+M_7-M_8\Big)
\\*&&\phantom{\CF\Big[}
-\frac{1}{N}\Big(M_2+M_3-M_4+M_7+M_9\Big)\Big]\,,
\\
\ampsqQGext_{-3,-1}&=&\CF\Big[N\Big(M_6+M_7+M_9+M_{10}\Big)
\\*&&\phantom{\CF\Big[}
-\frac{1}{N}\Big(M_1+M_2+M_3+M_7+M_9+M_{10}\Big)\Big]\,,
\\
\ampsqQGext_{-3,1}&=&\CF N\Big[N^2 M_8+M_1+M_2+M_3-M_7+M_9\Big]\,,
\\
\ampsqQGext_{-2,-1}&=&\CF\Big[-N\Big(M_3+M_6+M_7\Big)+
\frac{1}{N}\Big(M_2+M_3-M_4+M_7+M_9\Big)\Big]\,,\phantom{aa}
\\
\ampsqQGext_{-2,1}&=&\CF N\Big[N^2 M_8-M_2+M_3+M_7+M_9+M_{10}\Big]\,,
\\
\ampsqQGext_{-1,1}&=&\CF N\Big[N^2 M_5+M_1+M_2+M_3+M_7-M_9\Big]\,.
\label{M41clB10}
\eeqn
Again, one can verify that these satisfy the completeness
relations (\ref{CLBcompl})
\beq
\sum_{\stackrel{l\ne k}{l=-4}}^1\ampsqQGext_{kl}=
2C(\ident_k)\ampsqQGext\,.
\label{CLBcompl2}
\eeq

\section{Monte Carlo simulation\label{sec:MCs}}
In a parton shower Monte Carlo, each parton of a primary hard
subprocess has the potential to generate a shower, with initial
conditions to be determined by the kinematics and colour structure of
the hard process.  In the case of a dipole shower, each parton is
paired with another to form a dipole which sets the initial scale of
the shower according to its invariant mass.  In a single-parton
shower, the partner is again involved in setting the initial scale
of evolution.  For example, in an angular-ordered shower it
determines the maximum angle of emission in
the frame of shower evolution.  In the leading-colour approximation,
the parton and its partner form a colour singlet, so that the end
products of the shower from a dipole, or from the angular-ordered
showers of a parton and its partner, can hadronize without involving
the products of showers initiated by other hard partons.

In order to reproduce the leading-order soft gluon emission pattern
from an $m$-parton hard process, the dipoles or parton pairs
have to be associated with
eikonal terms in the soft limit  of the $(m+1)$-parton process.  In
the leading-colour approximation, considering all partons as outgoing,
every quark can be associated with an antiquark or the anticolour of a
gluon, every antiquark with a quark or the colour of a gluon, and the
remaining  colours and anticolours of gluons with other gluons.

As we have seen, at subleading colour there are eikonal terms that
pair quarks with quarks
and antiquarks with antiquarks.  There may also be a term where a
given gluon is paired with a quark and a term where it is paired
with another quark, and similarly for pairing with two antiquarks.
We can still associate dipoles or parton pairs with
such terms, and use them to generate showers, but the final products
cannot form colour singlets without communicating with other showers.
However, since colour is conserved overall, it is always possible to
hadronize after some colour reconnection between showers.

In this section we discuss some strategies that could be adopted to
include subleading colour contributions from the hard subprocess.
Which of these strategies may prove to be the most practical or
efficient probably depends on the process under consideration and
remains for further investigation.  A related but separate issue is 
the inclusion of subleading colour contributions inside parton showers, 
studied for example in refs.~\cite{Nagy:2012bt,Nagy:2015hwa,Hamilton:2020rcu,
DeAngelis:2020rvq,AngelesMartinez:2018cfz,Isaacson:2018zdi}. We envisage 
that the use of colour flows in matrix elements may prove useful in this 
case as well, since it has the advantage that it leads to an easy accounting 
of the number of colours, and thus renders it straightforward to see which 
contributions need to be taken into account to achieve any desired level 
of accuracy w.r.t.~$N$.

\subsection{Single-parton  showers}
In a single-parton shower, the eikonal pairings are used to set the
evolution scales but the showering of partons is otherwise
independent.  To achieve this the dipoles associated with eikonal
terms have to be split in some way between the two partons involved.
As an example, we consider here the splitting that leads to
angular-ordered shower evolution.  For simplicity we adopt the
original formulation used in the Herwig6 event
generator~\cite{Corcella:2000bw}, rather than the more sophisticated
treatments in Herwig++~\cite{Bahr:2008pv} and
Herwig7~\cite{Bellm:2017bvx,Bellm:2019zci}, which are equivalent for
our purposes.
  
\subsubsection{Angular-ordered shower}\label{sec:angord}
For an angular-ordered parton shower, one makes the following
substitution for the eikonal (\ref{eq:eik_def}):
 \beq\label{eq:ang_ord}
 \eik{k}{l}  = \frac 1{\omega^2}\frac{\xi_{kl}}{\xi_{k}\xi_{l}}
 \;\;\;\Longrightarrow\;\;\; \frac
 1{\omega^2}\left[\frac{\Theta(\xi_{kl}-\xi_{k})}{\xi_{k}}
   +\frac{\Theta(\xi_{kl}-\xi_{l})}{\xi_{l}}\right]\,,
 \eeq
 where $\omega$ is the energy of the emitted soft gluon,
 $\xi_{kl}=1-\cos\theta_{kl}$ where $\theta_{kl}$ is
 the angle between the three-momenta of partons $k$ and $l$;
 $\xi_k=1-\cos\theta_k$ where $\theta_k$ is the angle
 between the three-momenta of the soft gluon and parton $k$, and
 similarly for $\xi_l$.  This
 approximation follows from separating the collinear singularities of
 the eikonal and azimuthally averaging each term.  The averaging
 destroys Lorentz invariance and so one has to perform it in the same
 frame as that used by the parton shower.  The subsequent evolution of
 each shower is then ordered with respect to the relevant angular
 variable, $\xi_k$ or $\xi_l$.  This ordering takes into account
 the coherence of emission from the ensemble of partons
 in the shower at angles smaller than the current value.

 Making the substitution (\ref{eq:ang_ord}) in eq.~(\ref{softsum}), we have
\beq
\ampsqQGRS\;\;\;\Longrightarrow\;\;\; 
\frac{\gs^2}{\omega^2}\sum_{k,l}\ampsqQGBgg_{kl}
\frac{\Theta(\xi_{kl}-\xi_{k})}{\xi_{k}}\,.
\eeq
 For any given $k$,  using the phase-space element for soft gluon emission,
 \beq
 \frac{d^3p}{2(2\pi)^3\omega} =
 \frac{\omega}{2(2\pi)^3}d\omega\,d\xi_{k}d\phi_k\,,
 \eeq
 and integrating over $\phi_{k}$, the azimuthal angle of emission with respect to
 parton $k$, the differential distribution of soft gluon emission becomes
\beq\label{eq:ang_ord_dist}
\frac{\as}{2\pi}\sum_{k,l} \ampsqQGBgg_{kl}\,\Theta(\xi_{kl}-\xi_{k})
\frac{d\omega}{\omega}\frac{d\xi_{k}}{\xi_{k}}.
\eeq
The emission of non-soft collinear gluons can be included by
replacing $d\omega/\omega$ by $P(z)\,dz$,  where $P(z)$ is the
appropriate collinear splitting function of $z=1-\omega/p^0_k$.

In a leading-colour treatment~\cite{Odagiri:1998ep}, one first
generates a kinematic configuration of the Born process using
its full matrix element and then chooses among the leading-$N$ colour
flows (if there are more than one) according to their relative
contributions.  In the radiation pattern of a leading-$N$ flow, all
eikonal terms have coefficient $+1$, each quark or antiquark
appears in only one eikonal term, and each gluon
appears in two.  Therefore, if $k$ is a quark or antiquark, there is a
unique colour partner $l$ in eq.~(\ref{eq:ang_ord_dist}), which
specifies the angular region of radiation from $k$, $\xi_k<\xi_{kl}$.
On the other hand, if $k$ is a gluon it has two partners, say $l$ and
$m$ such that  $\xi_{km}<\xi_{kl}$, and it radiates with half intensity
when  $\xi_{km}<\xi_k<\xi_{kl}$ and full intensity when
$\xi_k<\xi_{km}$.

These radiation patterns, together with subsequent
multiple emissions, can be generated by a parton showering process
controlled by a Sudakov form factor
\beq\label{eq:ang_ord_sud}
\Delta_k(\xi_{k}) =
\exp\left[-C(\ident_k)\frac{\as}{2\pi}\int_{\xi_0}^{\xi_{k}}
  \frac{d\xi}{\xi}\int P(z)\,dz\right],
\eeq
where $C(\ident_k)$ is the relevant Casimir factor (see
eq.~(\ref{CLBcompl})) and $\xi_0$ is a cutoff below which emissions
are not resolved.  When parton $k$ has a single colour partner $l$,
the probability of no emission between the
maximum angular scale $\xi_{kl}$ and some lower scale $\xi_{k}$ is
given by $\Delta_k(\xi_{kl}) /\Delta_k(\xi_{k})$.  In the above case
of a gluon $k$ with partners $l$ and $m$ such that  $\xi_{kl}>\xi_{km}$,
the probability of no emission between $\xi_{kl}$ and $\xi_k>\xi_{km}$
is $\sqrt{\Delta_k(\xi_{kl}) /\Delta_k(\xi_{k})}$ while that between
$\xi_{km}$ and $\xi_k<\xi_{km}$ is $\Delta_k(\xi_{km})
/\Delta_k(\xi_{k})$.  In other words, the leading-colour
no-emission probability for a given colour flow is controlled by
\beq\label{eq:ang_sud_pLC}
\left[\Delta_k(\xi_{k})\right]^{p_{\rm LC}(\xi_{k})}
  \eeq
  where
  \beq\label{eq:sud_pxi_LC}
  p_{\rm LC}(\xi_{k})= \frac{\sum_l \Theta(\xi_{kl}-\xi_{k})}{\sum_l 1},
\eeq
the sums being over all colour partners $l$ of $k$ in the chosen
leading-$N$ colour flow.

Equations~(\ref{matelfin}) and (\ref{qgsoftsum}) suggest an obvious
extension of (\ref{eq:sud_pxi_LC}) to include subleading colour.
After selecting a closed flow $(\bgammap,\bgamma)$ according to its
contribution to the Born matrix element in (\ref{matelfin}), the
evolution of parton $k$ in that flow is controlled by
\beq\label{eq:ang_sud_p}
\left[\Delta_k(\xi_{k})\right]^{p_{(\bgammap,\bgamma)}(\xi_{k})}
  \eeq
  where
 \beq\label{eq:sud_pxi}
  p_{(\bgammap,\bgamma)}(\xi_{k})=
  \frac{\sum_{\ell\in\loopset(\bgammap,\bgamma)}\sum_l (-1)^{\delta_\ell(k,l)}
    \Theta(\xi_{kl}-\xi_{k})}
  {\sum_{\ell\in\loopset(\bgammap,\bgamma)}\sum_l (-1)^{\delta_\ell(k,l)}}.
  \eeq
Consider for example the process $\bar q q\to gg$.  We first choose
one of the six flows in table~\ref{tab:loopsets_qqgg} according to
their contributions to the Born matrix element in eq.~(\ref{Mtreeex1}).
Note that the flows $(\bgamma_{1,1}, \bgamma_{1,1})$  and
$(\bgamma_{1,2}, \bgamma_{1,2})$ make negative contributions and would have 
to be treated as counter-events or with a resampling method such as that 
of refs.~\cite{Olsson:2019wvr,Andersen:2020sjs,Nachman:2020fff}. Showering of
partons that contribute to the radiation pattern of the chosen flow,
according to eqs.~(\ref{eikf1ex1})-(\ref{eikf6ex1}), then proceeds as
in leading colour.  However, if any of the subleading colour flows is
chosen, then one or both of the gluons do not contribute and the
power (\ref{eq:sud_pxi}) is not defined for them.  These are either
gluons that appear twice in the same colour loop (in
$(\bgamma_{0,1},\bgamma_{0,2})$) or else (in
$(\bgamma_{i,j},\bgamma_{i,j})$ with $i=1,2$) the $U(1)$ gluons
discussed in sect.~\ref{sec:treeqg}.  They do not shower
and can be treated in a manner similar to the self-connected
gluons to be discussed in sect.~\ref{sec:scg}.  

For our other example process $\bar q q\to q'\bar q' g$, the situation
is more complicated.  The radiation
patterns of the 13 contributing closed flows in table~\ref{tab:loopsets} are
given in (\ref{eikf11ex2})-(\ref{eik1212ex2}).  Here some of the
eikonal terms in subleading flows have negative coefficients,
leading to negative contributions in eq.~(\ref{eq:sud_pxi}). Note however
that the denominator is always $+1$ for quarks and antiquarks, and
either 2 or 0 for gluons.  In the latter case the gluon is
non-showering as before.  In the former, the gluon may have either
two partners with coefficients $+1$ or four partners, three with
$+1$ and one with $-1$.  Similarly a quark or antiquark may have one
partner or three. Consider for example the flow $(\bgamma_{0,1},
\bgamma_{0,4})$ (which is maximally colour-suppressed).  According
to eq.~(\ref{eikf14ex2}) the gluon is non-showering, while each quark
or antiquark has two partners  with coefficient $+1$ and one with
coefficient $-1$.  For example, quark $-1$ has antiquark
partners $-3$ and $-4$ with coefficients $+1$ and quark partner $-2$
with coefficient $-1$ (this shows in a specific case the implications of the
generalization beyond leading colour of colour connections/colour partners, 
which we discussed in sect.~\ref{sec:softcomm}).  
There are six possible angular orderings
\beqn
\xi_{-1,-2}>\xi_{-1,-3}>\xi_{-1,-4} \;\;\;&\Longrightarrow&\;\;\; (-1,0,1)\,,\\
 \xi_{-1,-2}>\xi_{-1,-4}>\xi_{-1,-3}\;\;\;&\Longrightarrow&\;\;\; (-1,0,1)\,,\\
 \xi_{-1,-3}>\xi_{-1,-2}>\xi_{-1,-4}\;\;\;&\Longrightarrow&\;\;\; (1,0,1)\,,\\
\xi_{-1,-3}>\xi_{-1,-4}>\xi_{-1,-2} \;\;\;&\Longrightarrow&\;\;\; (1,2,1)\,,\\
 \xi_{-1,-4}>\xi_{-1,-2}>\xi_{-1,-3}\;\;\;&\Longrightarrow&\;\;\; (1,0,1)\,,\\
 \xi_{-1,-4}>\xi_{-1,-3}>\xi_{-1,-2}\;\;\;&\Longrightarrow&\;\;\; (1,2,1)\,. 
 \eeqn
  The last entry in each case shows the corresponding sequence of
  values of the power (\ref{eq:sud_pxi}), as $\xi_{-1}$ evolves from
  the highest to the lowest values. Note that the power is always 1 at
  the lowest angles, but can be 2 (higher emission), 0 (no emission) or
  even --1 at higher values.  Negative powers, corresponding to a 
 non-negative-definite Sudakov exponent, have already been considered
 in the literature (see e.g.~refs.~\cite{Platzer:2011dq,Hoeche:2011fd,
 Lonnblad:2012hz}), and can be dealt with for example by using a veto and 
 weighting technique.

The incidence of negative flow contributions and/or negative powers
(\ref{eq:sud_pxi}) could lead to low Monte Carlo efficiency, which
might be improved by not preselecting a flow, thus allowing all flows
to contribute to shower evolution.  This corresponds to using
(\ref{eq:ang_ord_dist}) directly to define the evolution power
  \beq\label{eq:sud_psum}
  p_{\rm SUM}(\xi_{k})= \frac{\sum_l
    \ampsqQGBgg_{kl}\,\Theta(\xi_{kl}-\xi_{k})}{\sum_l\ampsqQGBgg_{kl}}
= \frac{\sum_l
  \ampsqQGBgg_{kl}\,\Theta(\xi_{kl}-\xi_{k})}{2C(\ident_k)\ampsqQGBgg}
\eeq
in place of (\ref{eq:sud_pxi}), where we have used the completeness
relation (\ref{CLBcompl}).  The colour-linked Borns
$\ampsqQGBgg_{kl}$ are given explicitly in
(\ref{M22clB1})-(\ref{M22clB6}) for $\bar q q\to gg$ and in
(\ref{M41clB1})-(\ref{M41clB10}) for $\bar q q\to q'\bar q' g$.  Then
since the leading-colour flows contribute to all stages of showering
there will be on average a lower incidence of negative powers.
Statistically, at the parton level the results should be equivalent to
evolving flows separately. On the other hand, all colour information
is lost by the end of the shower, which could be problematic for
hadronization.

\subsection{Dipole showers}
A dipole Monte Carlo implementation including subleading
colour will in general involve contributions from dipole
configurations of all possible types, connecting quark-antiquark,
quark-quark, quark-gluon, gluon-gluon and their conjugates.
With each configuration we
can associate a graph specifying the dipole connections.  The theory
of such graphs is linked to that of flows discussed above, as
elaborated in the following subsection.

\subsubsection{Dipole graphs\label{sec:dipgr}}
For a given process that features $q$ quark-antiquark pairs (all
considered as outgoing) and $n$ gluons
let
\beqn
q_s&=&\Big\{-2q,-2q+1,\ldots -1\Big\}\,,
\\
g_s&=&\Big\{1,2,\ldots n\Big\}\,,
\eeqn
be the sets of the quark/antiquark and gluon labels, respectively,
in keeping with eqs.~(\ref{qbars})--(\ref{glus}). We define a
dipole graph to be a set of $q+n$ pairs, as follows:
\beqn
\Gamma&=&\Big\{\big(\Gamma_{1,1},\Gamma_{1,2}\big),\ldots
\big(\Gamma_{q+n,1},\Gamma_{q+n,2}\big)\;\Big|\;
\Gamma_{\alpha,\beta}\in q_s\cup g_s\,,
\;\forall\alpha\;\;\Gamma_{\alpha,1}\ne \Gamma_{\alpha,2}\,,
\nonumber
\\*&&\phantom{aaaaaa}
\forall\,k\in q_s~~\exists!~(\alpha,\beta)~~{\rm s.t.}~~
k=\Gamma_{\alpha,\beta}~~{\rm and}
\nonumber
\\*&&\phantom{aaaaaa}
\forall\,k\in g_s~~\exists!~(\alpha_1,\beta_1)\,,
(\alpha_2,\beta_2)~~{\rm s.t.}~~
k=\Gamma_{\alpha_1,\beta_1}=\Gamma_{\alpha_2,\beta_2}
\Big\}\,.
\label{graphdef}
\eeqn
In other words: a dipole graph is a set of pairs of non-identical parton 
labels; each quark and antiquark (gluon) label will appear in it exactly 
once (twice). Each pair that belongs to a dipole graph is a dipole, and
its two elements are the dipole ends. The order in which the pairs 
enter a dipole graph is irrelevant from a physics viewpoint: they can be 
re-shuffled as is convenient. We denote by
\beq
\dipset_{2q;n}=\Big\{\Gamma\;\Big|\;
\Gamma\big({\rm eq}.~(\text{\protect\ref{graphdef}})\big)\Big\}\,,
\label{dipset}
\eeq
the set of all dipole graphs relevant to $q$ quark lines and $n$ gluons.
By analogy with eq.~(\ref{eiksum}), we call the quantity
\beq
\dipeik\left(\Gamma\right)=
\sum_{\alpha=1}^{q+n}\eik{\Gamma_{\alpha,1}}{\Gamma_{\alpha,2}}\,,
\label{dipeikdef}
\eeq
the radiation pattern associated with the dipole graph $\Gamma$.

Any dipole graph admits a representation by means of 
a square \mbox{$(2q+n)\times (2q+n)$} matrix\footnote{In keeping with the 
labeling of ref.~\cite{Frixione:2011kh} (see sect.~\ref{sec:basic}),
parton labels equal to zero must be ignored.}, defined as follows:
\beq
m(\Gamma)=\big(m(\Gamma)_{ij}\big)_{i,j=-2q}^{n}\,,\;\;\;
m(\Gamma)_{ij}=\sum_{\alpha=1}^{q+n}\Big[
\delta(i,\Gamma_{\alpha,1})\delta(j,\Gamma_{\alpha,2})+
\delta(j,\Gamma_{\alpha,1})\delta(i,\Gamma_{\alpha,2})\Big]\,.
\label{matdip}
\eeq
That is, each dipole in $\Gamma$ gives a unit contribution
to the matrix element whose indices coincides with the dipole ends,
both in the given and in the reverse order. Owing to eq.~(\ref{graphdef}), 
the matrix defined in eq.~(\ref{matdip}) has the following properties:
{\em a)} it is symmetric; {\em b)} it is traceless; {\em c)} its elements
are equal to either $0$, $1$, or $2$; {\em d)} each of $2q$ of its rows 
sums to one, and each of the remaining $n$ rows sums to two.
By construction
\beq
\dipeik\left(\Gamma\right)=\half\sum_{i,j=-2q}^{n}m(\Gamma)_{ij}
\eik{i}{j}\,.
\label{dipeikvmat}
\eeq
We stress that eqs.~(\ref{dipeikdef})--(\ref{dipeikvmat}) are invariant
under reordering of the dipoles that enter the given dipole graph.

The counting of distinct matrices of the type defined in eq.~(\ref{matdip})
tells us the number of possible dipole configurations for $q$
quark-antiquark pairs and $n$ gluons, $|\dipset_{2q;n}|$.
It is convenient to define the generating function
\beq
{\cal F}(x,y) = \sum_{q,n}  |\dipset_{2q;n}|
\frac{x^q}{q!}\frac{y^n}{n!}\,,
\eeq
so that
\beq
|\dipset_{2q;n}| =\left.\frac{\partial^{q+n}{\cal F}}{\partial x^q\partial y^n}\right|_{x,y=0}. 
\eeq
For a process involving only $q$ quark-antiquark pairs and no gluons,
the dipole matrix is a
$2q\times 2q$  symmetric, traceless matrix with elements $0$ or $1$
and each row summing to $1$.  It is readily seen that the number of
distinct  matrices of this type is $(2q-1)!!$, so that the generating
function is
\beq
{\cal F}(x,0) = \sum_q  \frac{(2q-1)!!}{q!} x^q = \frac 1{\sqrt{1-2x}}.
\eeq
In the case of purely gluonic processes, $q=0$, the corresponding
matrices have all rows and columns summing to 2.   The counting of
such matrices is a known problem, whose solution is given by the 
generating function~\cite{oeisA002137}
\beq
{\cal F}(0,y) = \frac{{\rm e}^{-y/2+y^2/4}}{\sqrt{1-y}}.
\eeq
For processes that involve both quarks and gluons,
we can suppose that $n-k$ of the gluons are connected amongst themselves
while $k$ are inserted onto one or more of the quark lines.  All the
permutations of the latter are distinct, and the 
number of ways of distributing them is the number of ways of
locating $q-1$ subset boundaries amongst $k+q-1$ objects,
so the number of distinct dipole graphs is\footnote{Note therefore
that this is analogous to the counting problem that leads one
to eq.~(\ref{dimBflow}).}
\beq
\label{eq:Sqn}
|\dipset_{2q;n}| =\sum_{k=0}^n\binomial{n}{k}\binomial{k+q-1}{q-1}k!\,
|\dipset_{0;n-k}|\, |\dipset_{2q;0}|\,,
\eeq
which yields the generating function
\beqn\label{eq:Sxy}
{\cal F}(x,y) &=& {\cal F}(0,y)\,  {\cal F}(x/[1-y],0)\nonumber\\
  &=&\frac{{\rm e}^{-y/2+y^2/4}}{\sqrt{1-2x-y}}.
\eeqn
This gives the numbers shown in table~\ref{tab:dipgraphs}.

\begin{table}[h!]
  \begin{center}    
    \begin{tabular}{|c|ccccccc|}
      \hline
      $q\diagdown n$ & 0 & 1 & 2 & 3 & 4 & 5 & 6 \\
      \hline
       0 & 1 & 0 & 1 & 1 & 6 & 22 & 130 \\
       1 & 1 & 1  & 3 & 10 & 46 & 252 & 1642 \\
       2 & 3 & 6  & 21 & 93 & 510 & 3306 & 24762 \\
       3 & 15 & 45 & 195  & 1050 & 6750 & 50280 & 425490  \\
       4 & 105 & 420 & 2205 & 13965  & 103110 & 867510 & 8183490 \\
      \hline
    \end{tabular}
  \end{center}
  \caption{
    \label{tab:dipgraphs}Number of distinct dipole graphs for
    $q$ quark-antiquark pairs and $n$ gluons.}
\end{table}

\subsubsection{Self-connected gluons\label{sec:scg}}
At leading order in colour, and also at subleading order for processes
involving only quarks or only gluons, the eikonal combinations that
appear in the differential cross section for emission of an additional
soft gluon can be represented by a weighted sum of dipole graphs
involving all the partons of the primary process.  In processes
involving both quarks and gluons, however, at subleading order
in colour, it is in general necessary to include graphs with fewer 
participating gluons than in the primary process. In essence, this is
due to the presence of gluons both of whose colour labels enter the
same colour loop; as is shown in sect.~\ref{sec:softcomm}, such gluons
(denoted by $g^\star$ there) cannot contribute to radiation patterns. Further 
discussions on this point will be given in sect.~\ref{sec:ccfvsdp} and
appendix~\ref{sec:flvsdg}.

Although such non-participating gluons are not associated with eikonal terms 
in the soft gluon cross section, they cannot be altogether ignored in a Monte 
Carlo implementation, since to do so would violate momentum conservation.
Instead they can be treated as having self-connected colour and
anticolour.   Such a self-connected gluon $i$ is associated with a null 
eikonal $[i,i]$, which cannot radiate: it propagates without showering and
must be hadronized through momentum transfer with other showers
in the same event.  

To make connection with the theory of dipole graphs,
for any \mbox{$0\le p\le n$} and \mbox{$s_p\in\Sset{n}{p}$}
as in eq.~(\ref{skdef}), we define
\beq
\dipset_{2q;n-p,s_p}=\Big\{\Gamma\;\Big|\;
\Gamma\big({\rm eq}.~(\text{\protect\ref{graphdef}})\,,
g_s\,\longrightarrow\,g_s\!\smallsetminus\!s_p\big)\Big\}
\label{dipsetp}
\eeq
to be the set of $q$-quark, $(n-p)$-gluon dipole graphs, where the
gluon labels do not assume the values in the set $s_p$. The gluons
whose labels belong to the set $s_p$ are understood to be
self-connected gluons.  Note that
\beq
p=0\;\;\;\Longrightarrow\;\;\;s_0=\big\{\big\}
\;\;\;\Longrightarrow\;\;\;\dipset_{2q;n,s_0}\equiv\dipset_{2q;n}\,.
\eeq
For a process involving $q$ quark-antiquark pairs and $n$ gluons, the
number of relevant dipole graphs where we allow any number of self-connected
gluons is then (see eq.~(\ref{dimSnk}))
\beq
\sum_{p=0}^n\left|\dipset_{2q;n-p,s_p}\right|= 
\sum_{p=0}^n \binomial{n}{p} \left|\dipset_{2q;n-p}\right|\,.
\label{countSCg}
\eeq
If we denote the generating function for this by $\widetilde{\cal
  F}(x,y)$, then
\beq
\widetilde{\cal F}(x,y) = {\rm e}^y {\cal F}(x,y)
=\frac{{\rm e}^{y/2+y^2/4}}{\sqrt{1-2x-y}},
\eeq
which gives the numbers shown in table~\ref{tab:dipgraphsU1}.
\begin{table}[h!]
  \begin{center}    
    \begin{tabular}{|c|ccccccc|}
      \hline
      $q\diagdown n$ & 0 & 1 & 2 & 3 & 4 & 5 & 6 \\
      \hline
       0 & 1 & 1 & 2 & 5 & 17 & 73 & 388 \\
       1 & 1 & 2  & 6 & 23 & 109 & 618 & 4096 \\
       2 & 3 & 9  & 36 & 177 & 1035 & 7029 & 54462 \\
       3 & 15 & 60 & 300  & 1785 & 12315 & 96720 & 852630  \\
       4 & 105 & 525 & 3150 & 21945  & 173985 & 1546965 & 15250200 \\
      \hline
    \end{tabular}
  \end{center}
  \caption{
    \label{tab:dipgraphsU1}Number of distinct dipole graphs for
    $q$ quark-antiquark pairs and $n$ gluons, allowing any number of 
    self-connected gluons (see eq.~(\ref{countSCg})).}
\end{table}

\subsection{Examples\label{sec:dipex}}
\subsubsection[$\bar q q\to gg$]{\boldmath  $\bar q q\to gg$}\label{sec:MCqqgg}
\begin{figure}[t!]
\begin{center}
\includegraphics[scale=0.15]{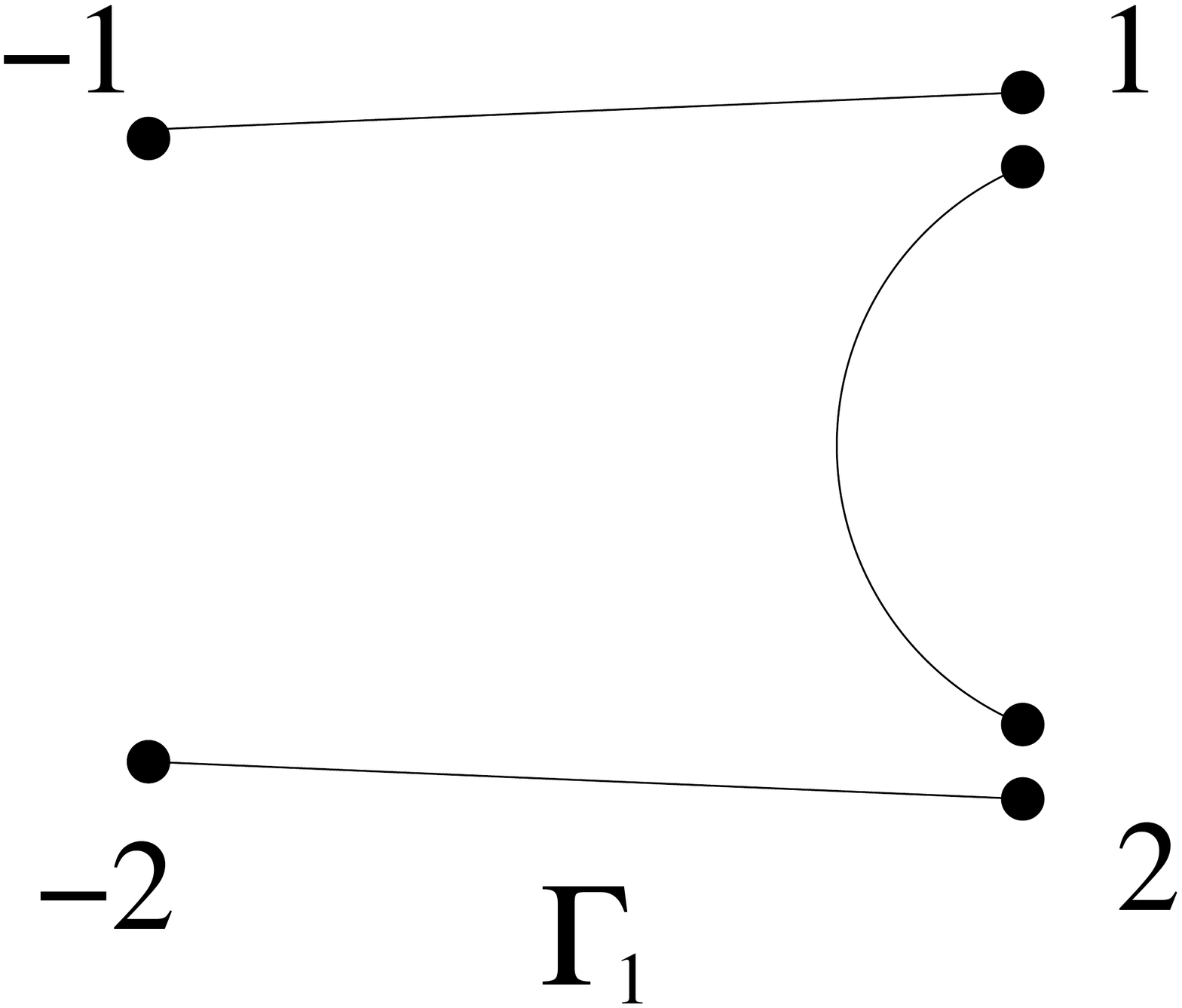}\hspace{5mm}
\includegraphics[scale=0.15]{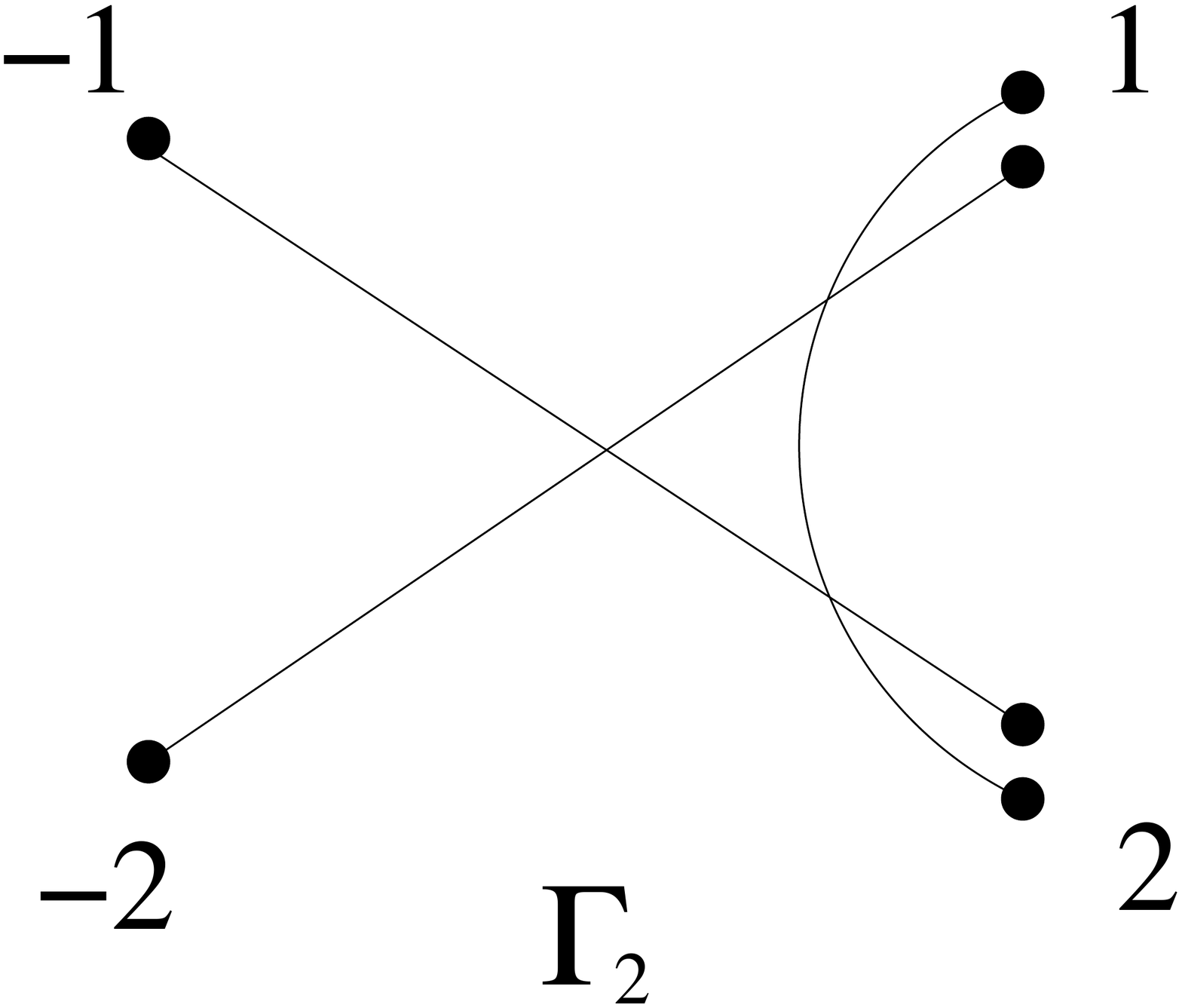}\hspace{5mm}
\includegraphics[scale=0.15]{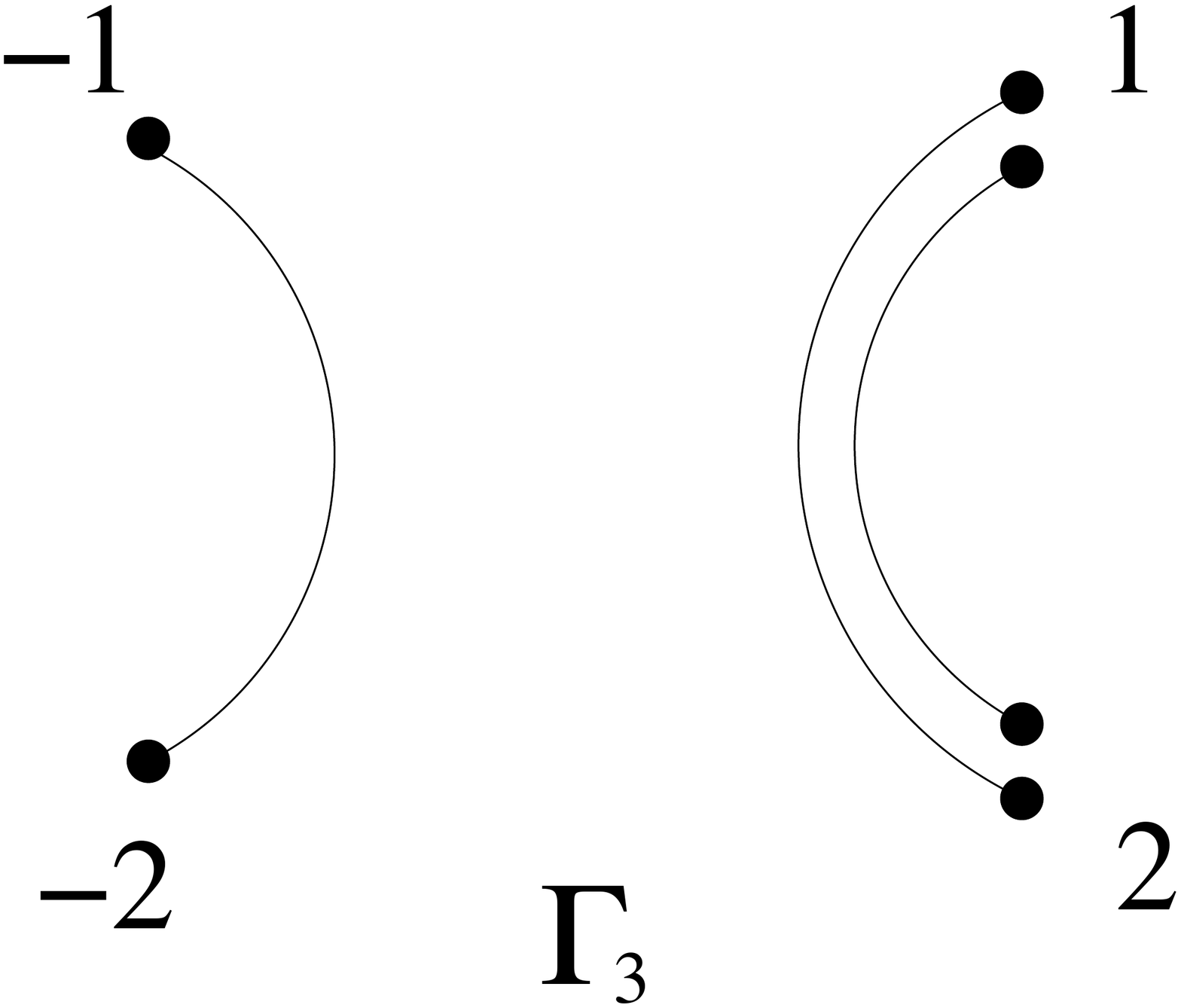}\\
\includegraphics[scale=0.15]{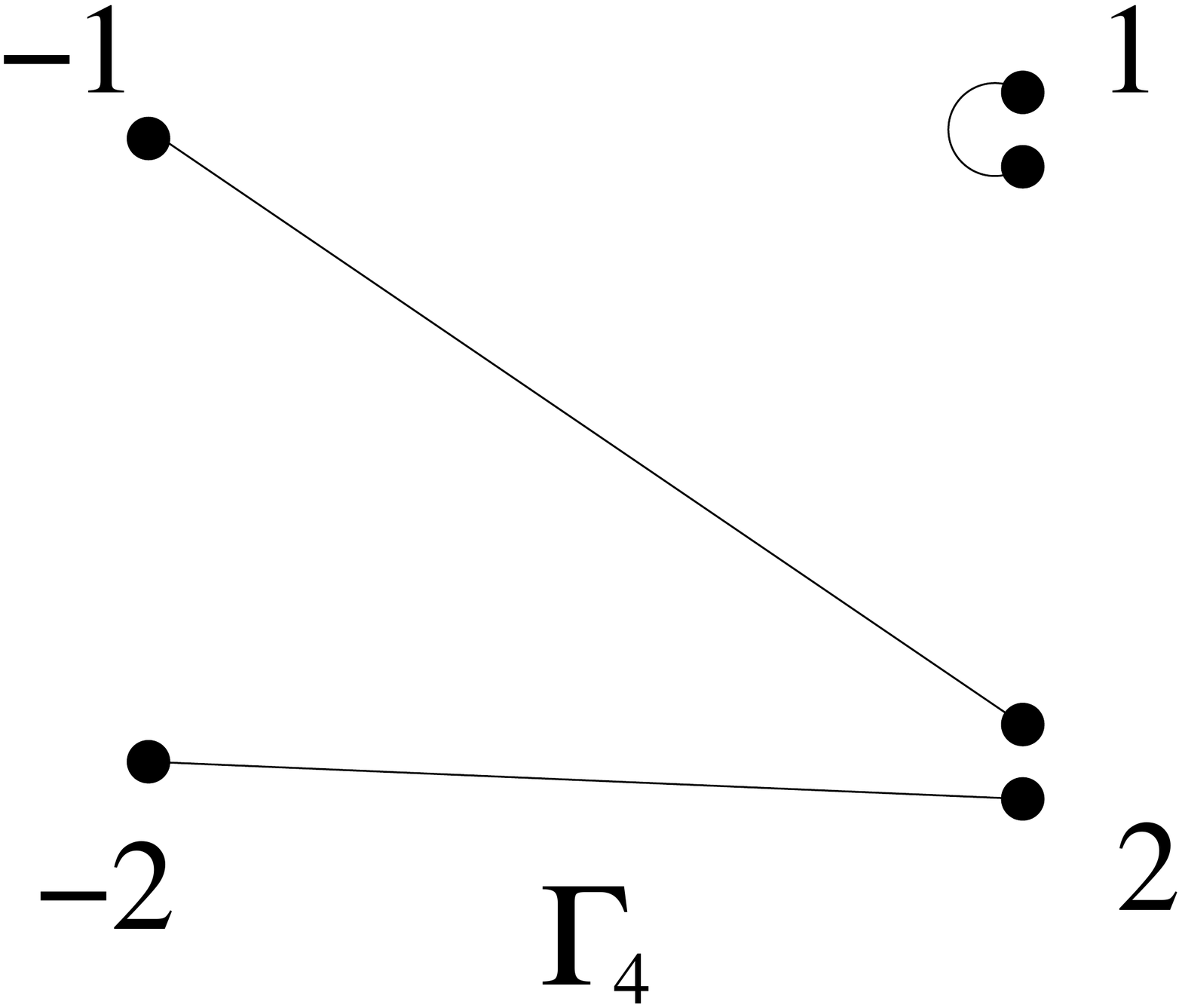}\hspace{5mm}
\includegraphics[scale=0.15]{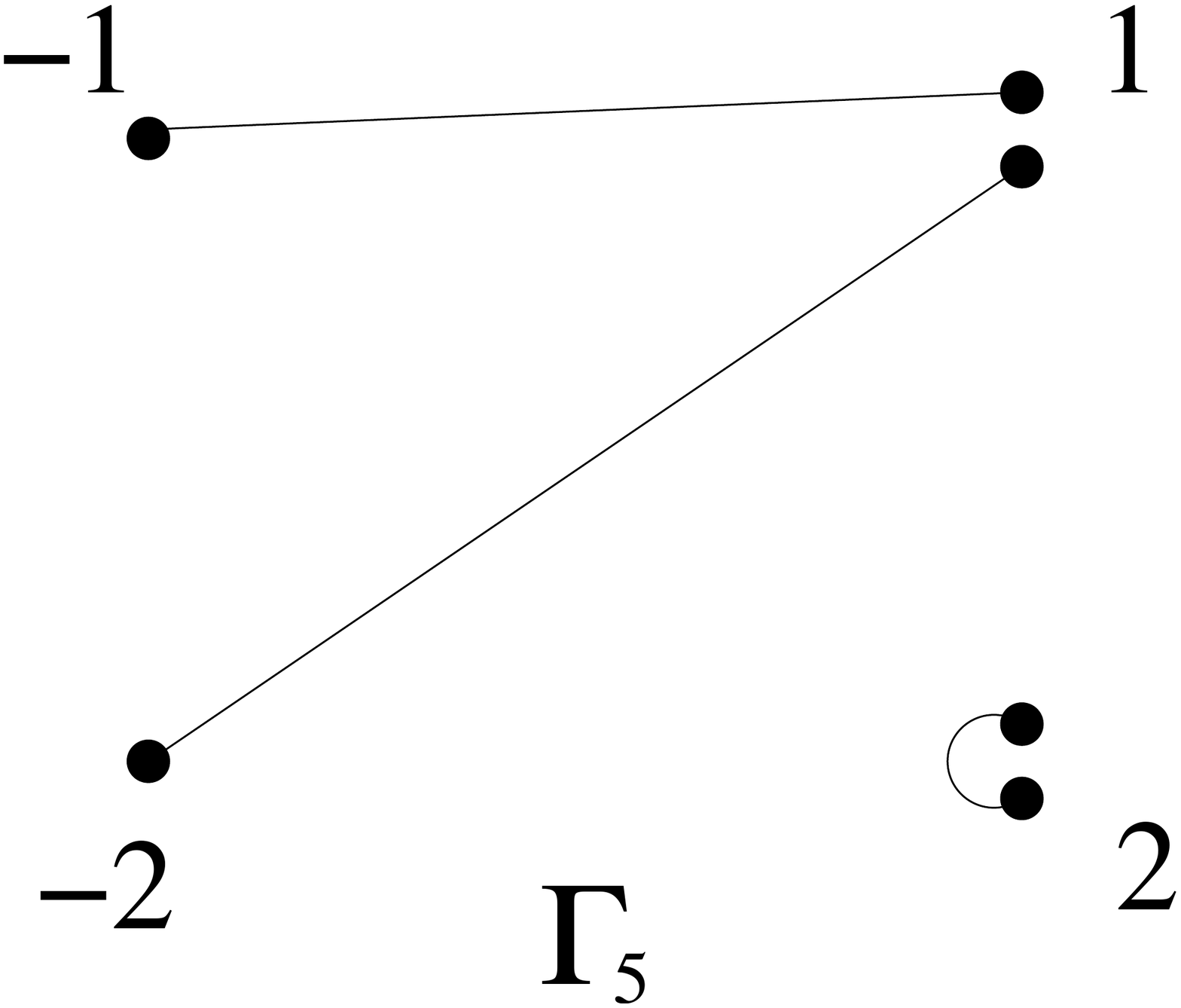}\hspace{5mm}
\includegraphics[scale=0.15]{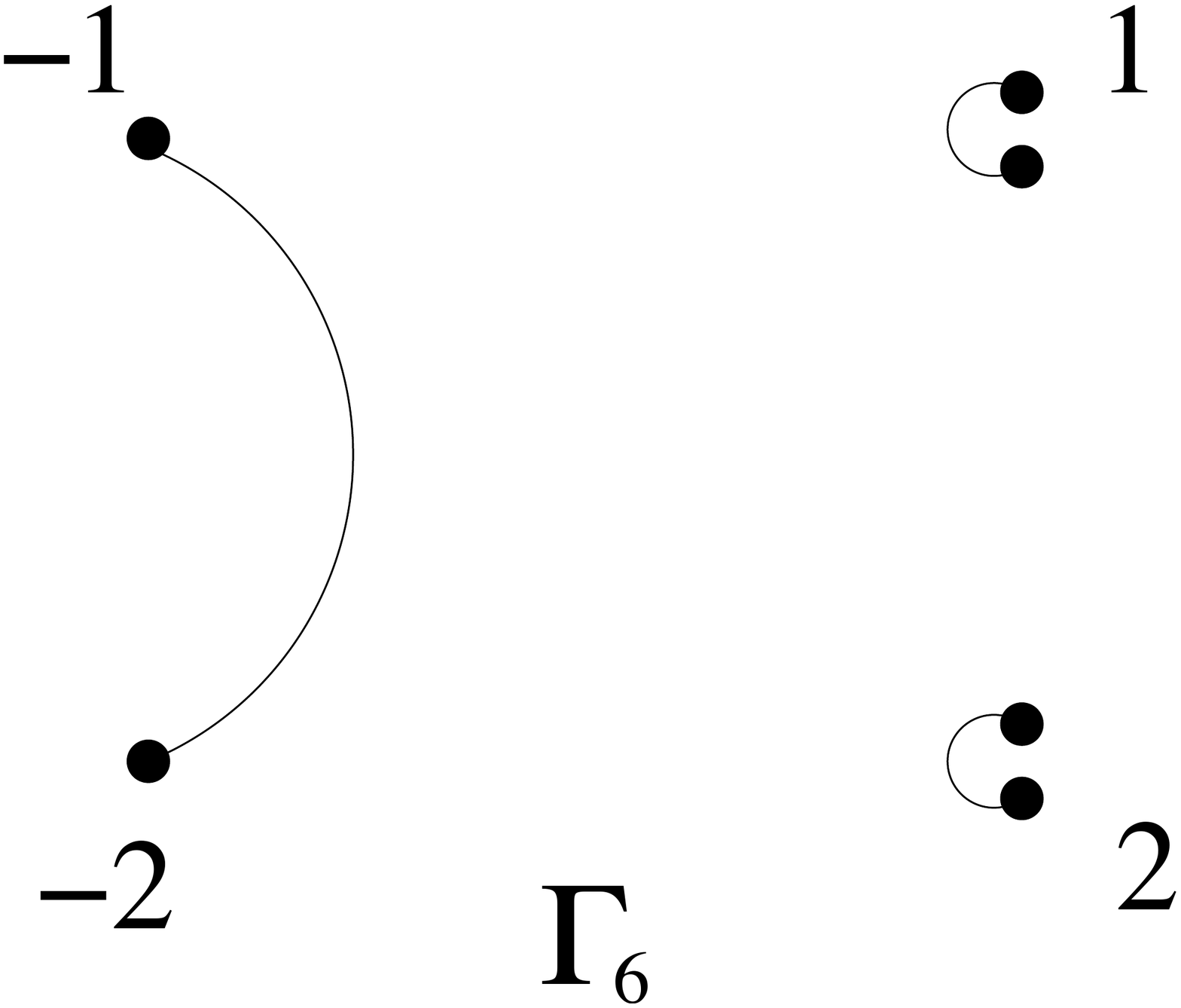}
\caption{\label{fig:qqgg_dipoles}
Dipole graphs for  $q\bar q \to gg$. In the cases of $\Gamma_4$, 
$\Gamma_5$, and $\Gamma_6$ the corresponding
self-connected gluons are also depicted.
}
\end{center}
\end{figure}

Consider again the process $\bar q q\to gg$.  
According to tables~\ref{tab:dipgraphs} and \ref{tab:dipgraphsU1}
there are six relevant dipole graphs, $\Gamma_1\ldots \Gamma_6$,
of which three involve self-connected
gluons, corresponding to the six dipole configurations shown in
Fig.~\ref{fig:qqgg_dipoles}.
The corresponding radiation patterns are
\beqn\label{eq:qqgg_dipeik1}
\dipeik\left(\Gamma_1\right) &=& \eik{-1}{1}+\eik{-2}{2}+\eik{1}{2}\,,\\
\dipeik\left(\Gamma_2\right) &=& \eik{-1}{2}+\eik{-2}{1}+\eik{1}{2}\,,\\
\dipeik\left(\Gamma_3\right) &=& \eik{-1}{-2}+2\eik{1}{2}\,,\\
\dipeik\left(\Gamma_4\right) &=&  \eik{-1}{2}+\eik{-2}{2}\,,\\
\dipeik\left(\Gamma_5\right) &=&  \eik{-1}{1}+\eik{-2}{1}\,,\\
\label{eq:qqgg_dipeik6}
\dipeik\left(\Gamma_6\right) &=&  \eik{-1}{-2}\,.
\eeqn
 Note that these are not independent, since 
 \beq 
 \dipeik\left(\Gamma_1\right)+\dipeik\left(\Gamma_2\right)
 +\dipeik\left(\Gamma_6\right) = \dipeik\left(\Gamma_3\right)
 +\dipeik\left(\Gamma_4\right) +\dipeik\left(\Gamma_5\right).
\label{qqggeikid}
\eeq
The closed-flow radiation patterns (\ref{eikf1ex1})-(\ref{eikf6ex1}) 
can be expressed in terms of the above dipole radiation patterns 
as follows:
\beqn
\eiksum(\bgamma_{0,1},\bgamma_{0,1})&=&
 \dipeik\left(\Gamma_1\right)\,,
\label{eikf1ex1dip}
\\
\eiksum(\bgamma_{0,1},\bgamma_{0,2})&=&
 \dipeik\left(\Gamma_6\right)\,,
\label{eikf2ex1dip}
\\
\eiksum(\bgamma_{0,2},\bgamma_{0,2})&=&
 \dipeik\left(\Gamma_2\right)\,,
\label{eikf3ex1dip}
\eeqn
\beqn
\eiksum(\bgamma_{1,1},\bgamma_{1,1})&=&
 \dipeik\left(\Gamma_4\right)\,,
\label{eikf4ex1dip}
\\
\eiksum(\bgamma_{1,2},\bgamma_{1,2})&=&
 \dipeik\left(\Gamma_5\right)\,,
\label{eikf5ex1dip}
\\
\eiksum(\bgamma_{2,1},\bgamma_{2,1})&=&
 \dipeik\left(\Gamma_6\right)\,.
\label{eikf6ex1dip}
\eeqn
Thus, for the soft radiation matrix element (\ref{Msoftex1}) we now have
\beqn
\frac{2\ampsqQGSexo}{\gs^2\CF}&=&N^3 \dipeik\left(\Gamma_1\right) M_1+
N^3 \dipeik\left(\Gamma_2\right) M_3+
N \dipeik\left(\Gamma_6\right) M_2
\nonumber
\\*&-&
\frac{1}{N}\,N^2\left[\dipeik\left(\Gamma_4\right) +
\dipeik\left(\Gamma_5\right) \right]M_4
+
\frac{1}{N^2}\,N
\dipeik\left(\Gamma_6\right) M_4\,.
\label{Dsoftex1}
\eeqn
Using (\ref{qqggeikid}) we may, for example, eliminate
$\dipeik\left(\Gamma_6\right)$ which involves two
self-connected gluons.

\subsubsection[$\bar q q\to q'\bar q'g$]{\boldmath  $\bar q q\to
  q'\bar q'g$}\label{sec:MCqqqqg}
\begin{figure}[t!]
\begin{center}
\includegraphics[scale=0.15]{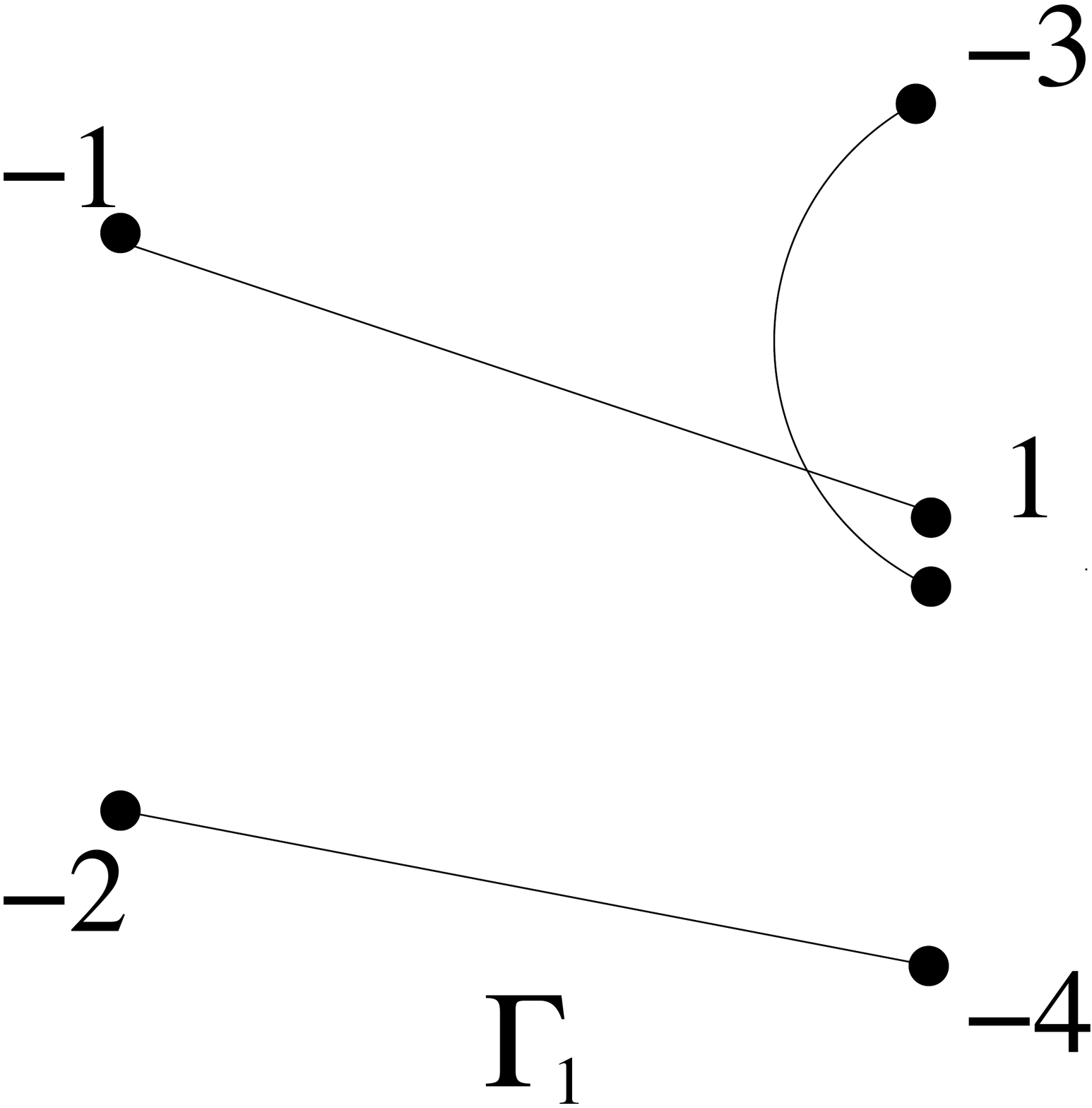}\hspace{5mm}
\includegraphics[scale=0.15]{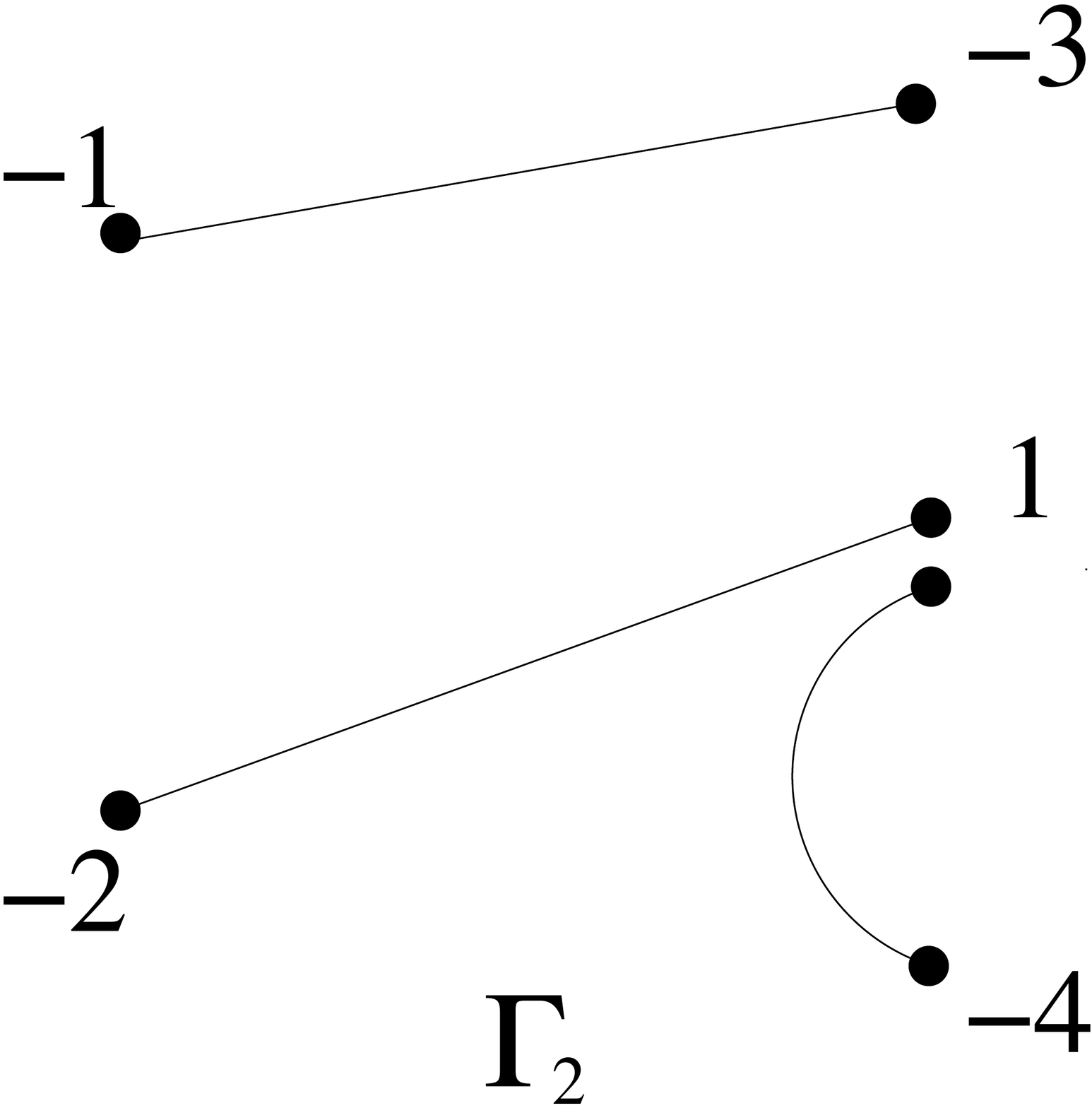}\hspace{5mm}
\includegraphics[scale=0.15]{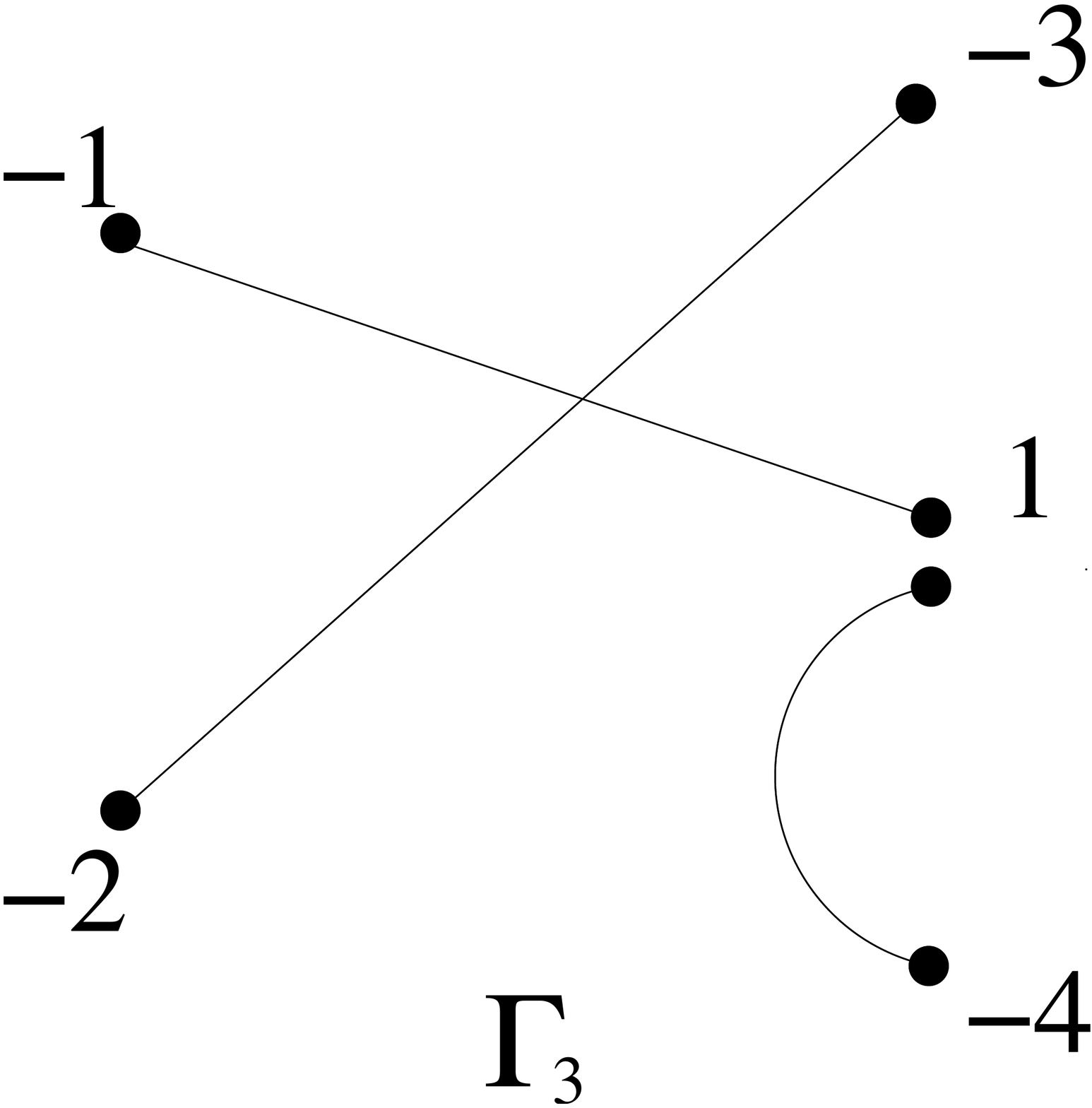}\\
\includegraphics[scale=0.15]{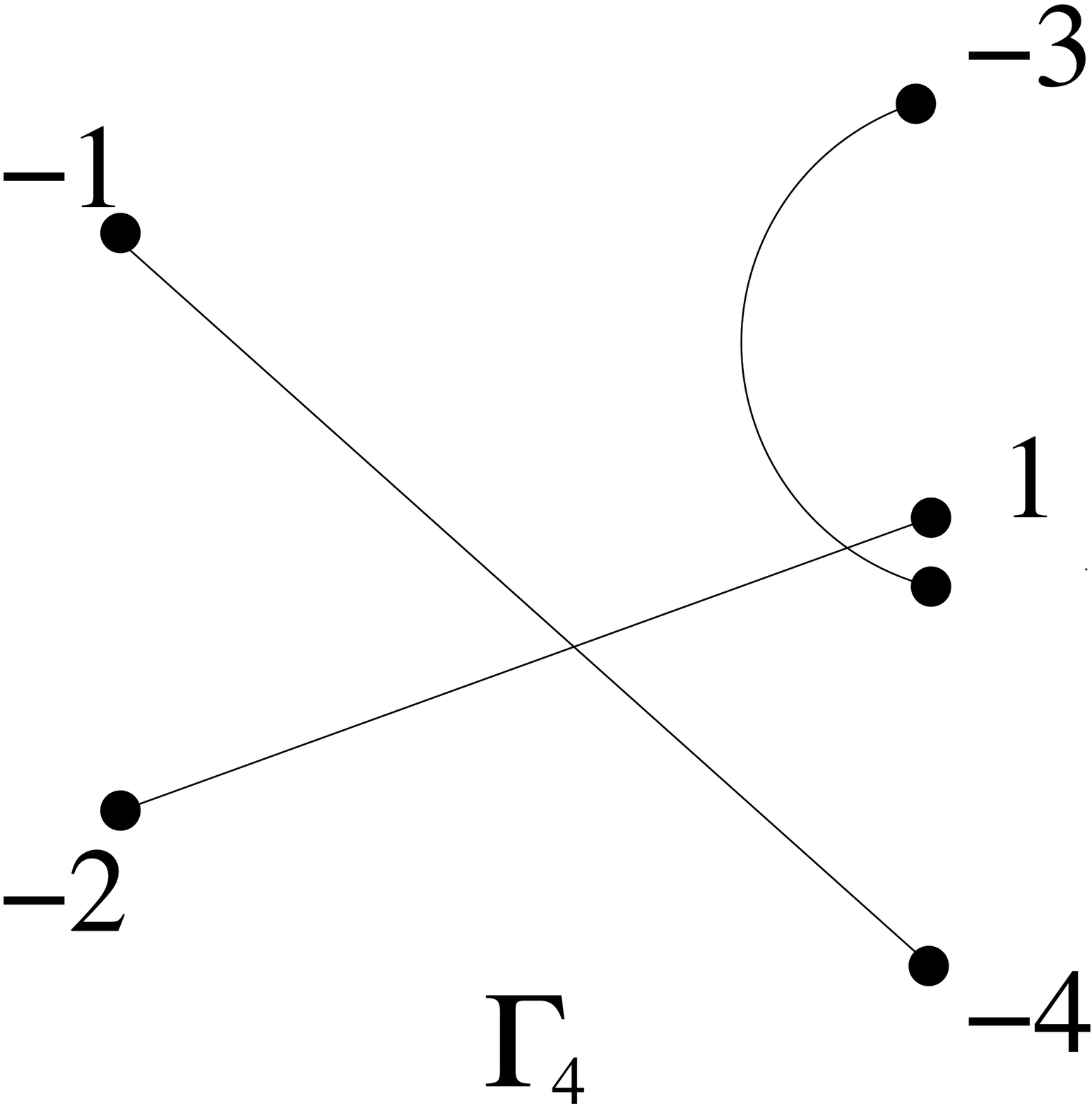}\hspace{5mm}
\includegraphics[scale=0.15]{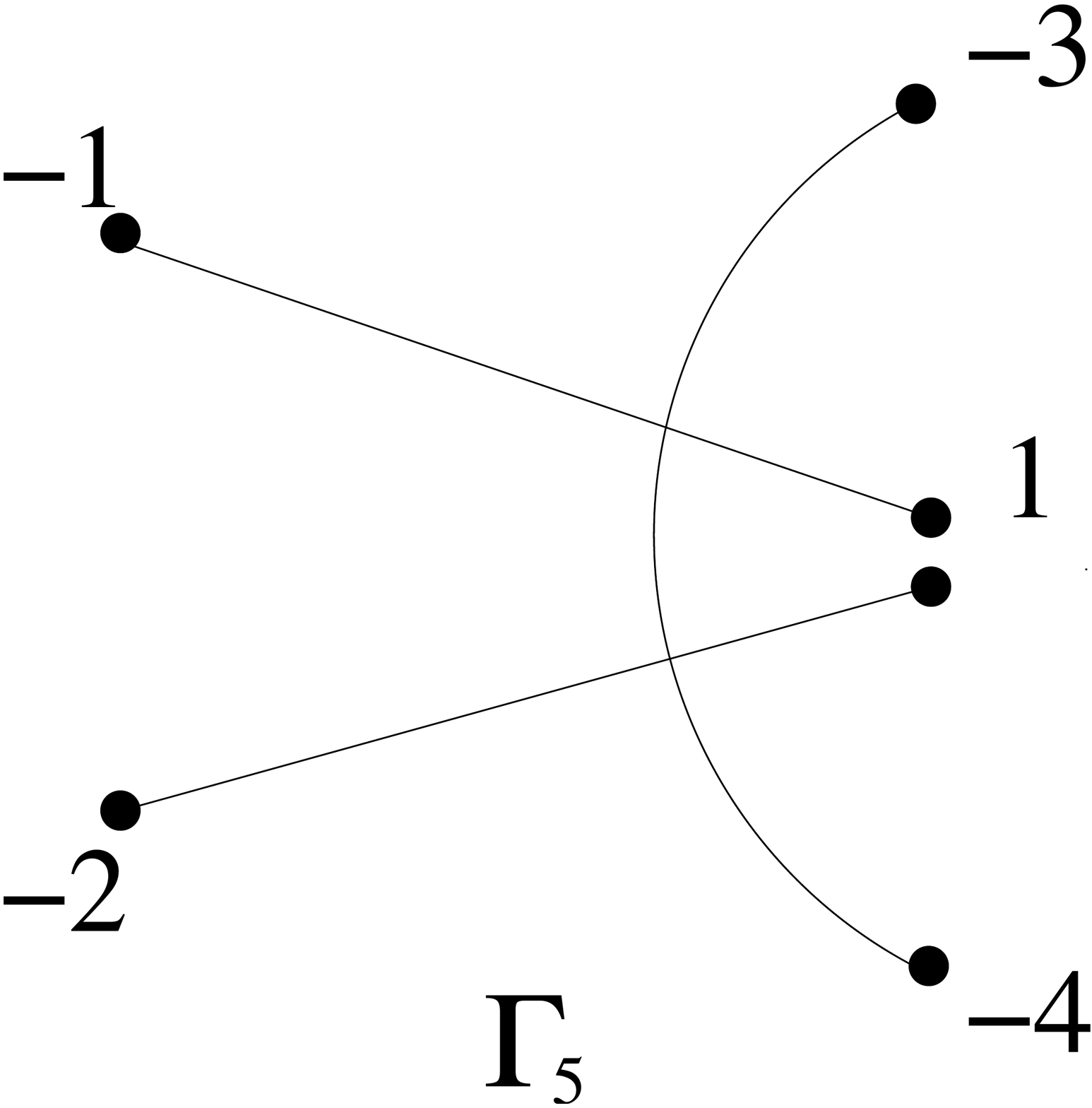}\hspace{5mm}
\includegraphics[scale=0.15]{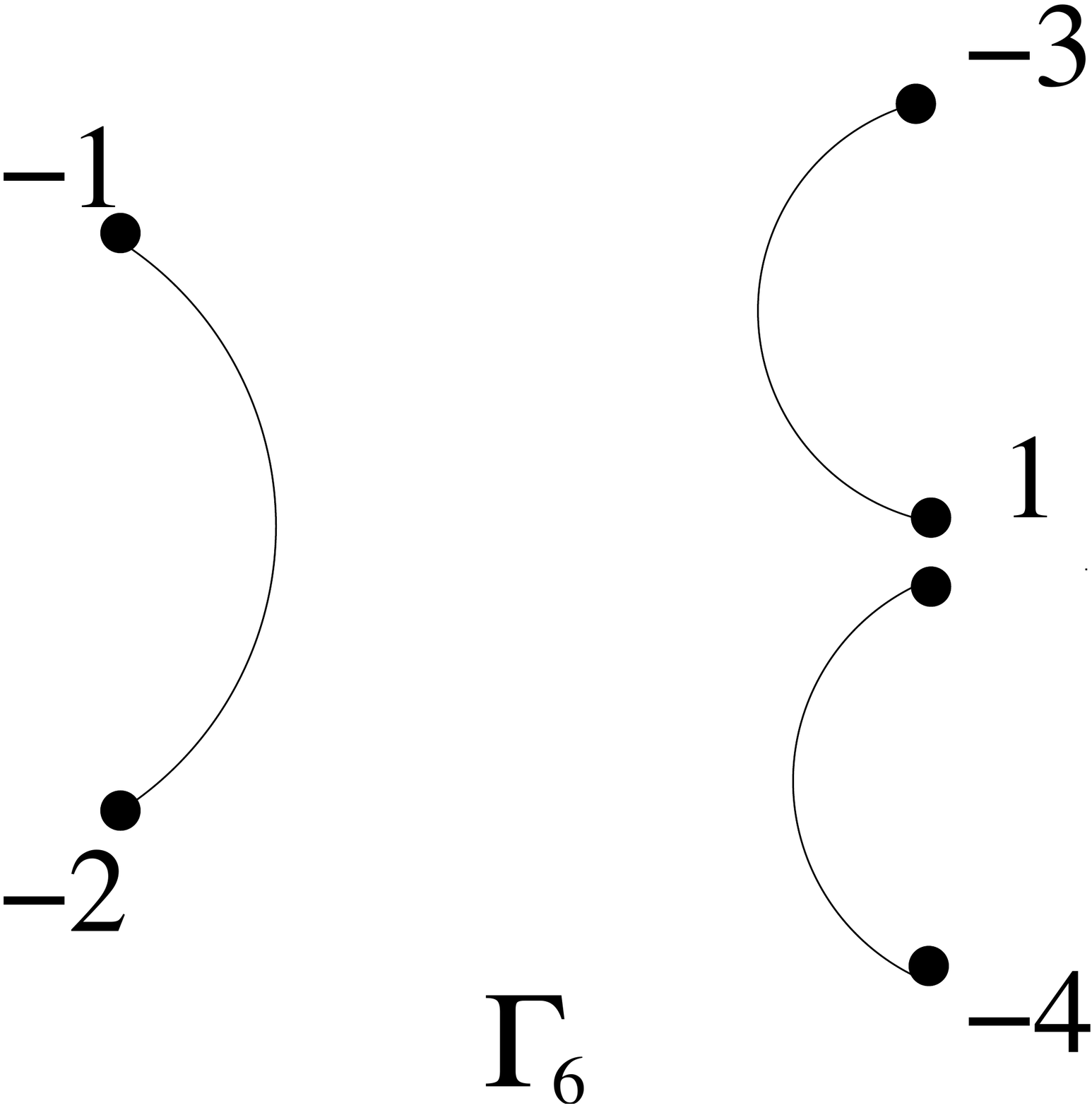}\\
\includegraphics[scale=0.15]{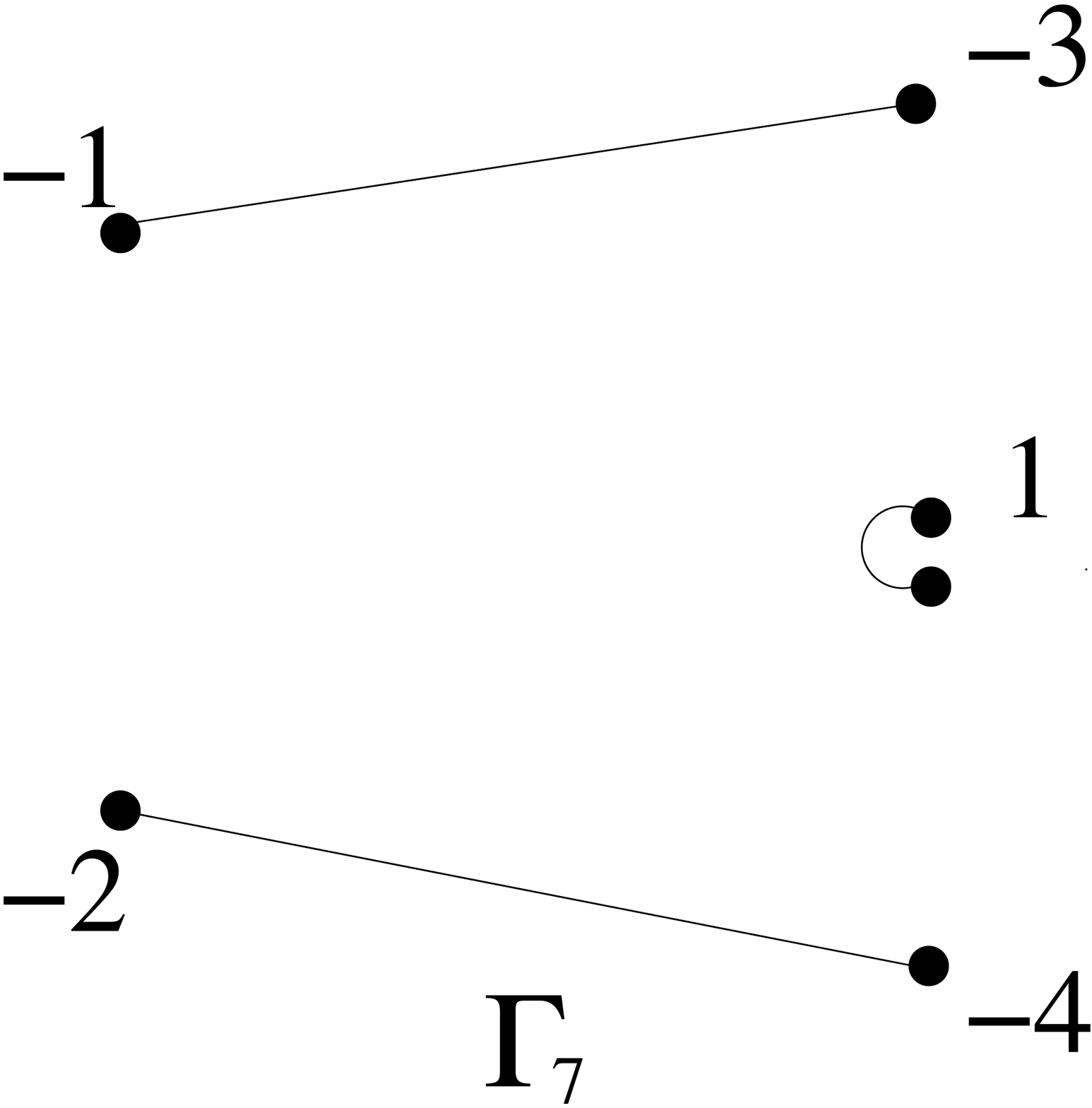}\hspace{5mm}
\includegraphics[scale=0.15]{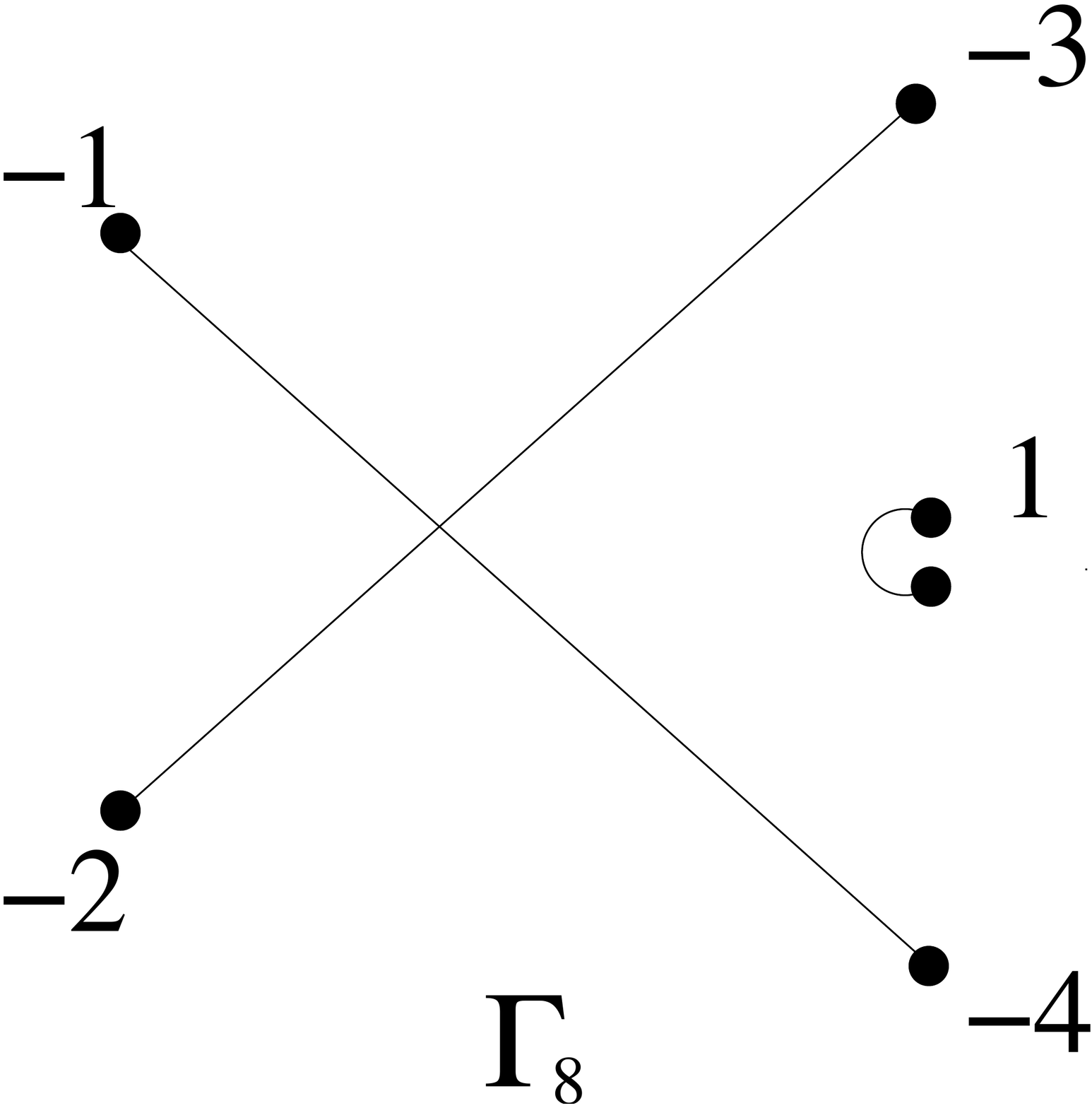}\hspace{5mm}
\includegraphics[scale=0.15]{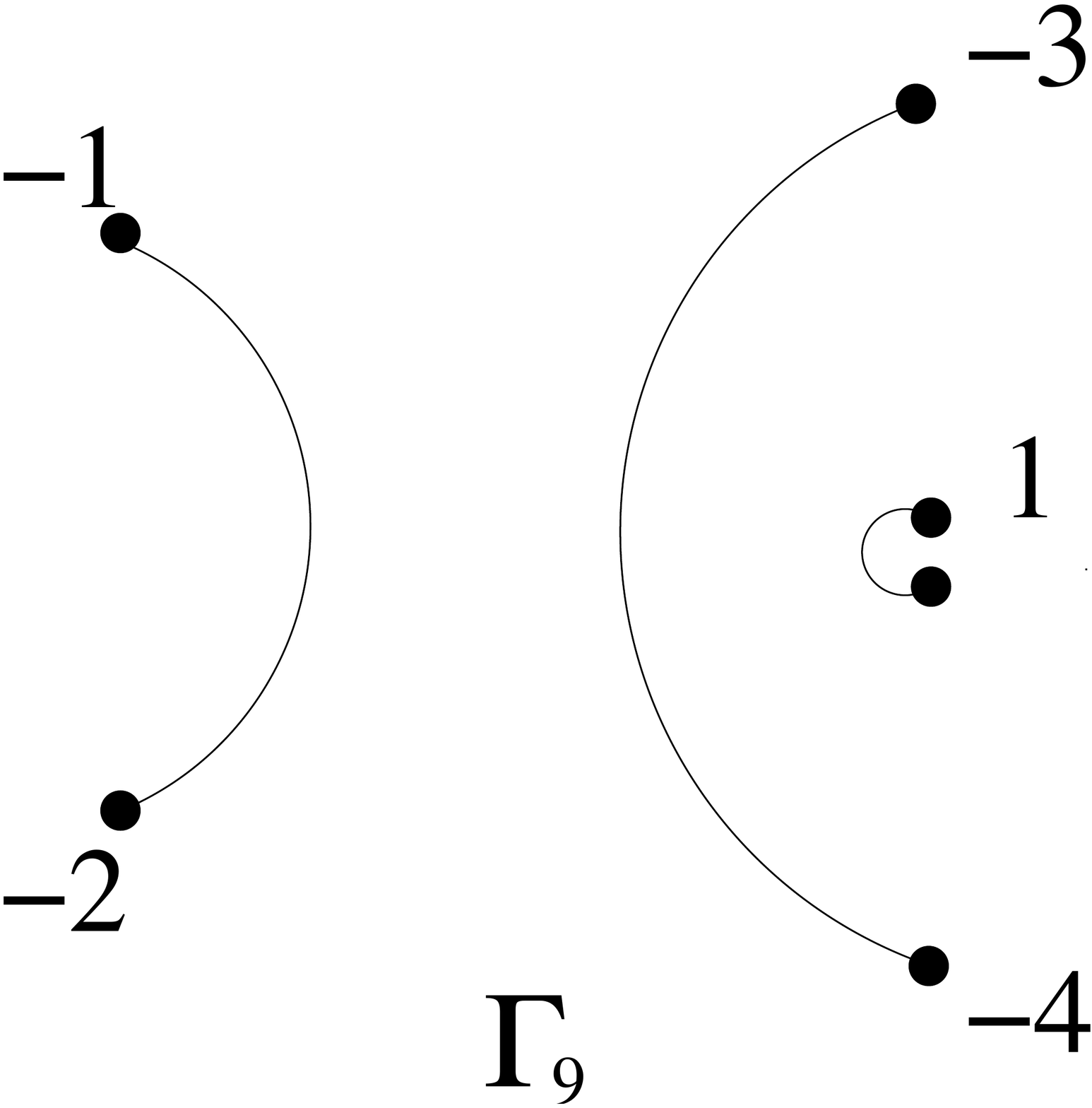}
\caption{\label{fig:qqqqg_dipoles}
Dipole graphs for  $q\bar q \to q'\bar q'g$. In the cases of $\Gamma_7$, 
$\Gamma_8$, and $\Gamma_9$ the self-connected gluon is also depicted.
}
\end{center}
\end{figure}

There are 9 relevant dipole graphs, shown in
 fig.~\ref{fig:qqqqg_dipoles}, with associated radiation patterns
\beqn\label{eq:qqqqg_eikonals}
 \dipeik\left(\Gamma_1\right)&=& \eik{-1}{1}+\eik{1}{-3}+\eik{-2}{-4}\,,
\\
 \dipeik\left(\Gamma_2\right)&=& \eik{-1}{-3}+\eik{-2}{1}+\eik{1}{-4}\,,
\\
 \dipeik\left(\Gamma_3\right)&=& \eik{-1}{1}+\eik{1}{-4}+\eik{-2}{-3}\,,
\\
 \dipeik\left(\Gamma_4\right)&=& \eik{-1}{-4}+\eik{-2}{1}+\eik{1}{-3}\,,
\eeqn
\beqn
 \dipeik\left(\Gamma_5\right)&=& \eik{-1}{1}+\eik{1}{-2}+\eik{-3}{-4}\,,
\\
 \dipeik\left(\Gamma_6\right)&=& \eik{-1}{-2}+\eik{-3}{1}+\eik{1}{-4}\,,
\\
 \dipeik\left(\Gamma_7\right)&=& \eik{-1}{-3}+\eik{-2}{-4}\,,
\\
 \dipeik\left(\Gamma_8\right)&=& \eik{-1}{-4}+\eik{-2}{-3}\,,
\\
 \dipeik\left(\Gamma_9\right)&=& \eik{-1}{-2}+\eik{-3}{-4}\,.
\eeqn
The closed-flow radiation patterns (\ref{eikf11ex2})-(\ref{eik1212ex2}) 
can be expressed in terms of these as follows:
\beqn
\label{eq:qqqqg_dipoles}
 \eiksum(\bgamma_{0,1},\bgamma_{0,1})&=&\dipeik\left(\Gamma_1\right)\,,
\eeqn
\beqn
 \eiksum(\bgamma_{0,1},\bgamma_{0,2})&=&\dipeik\left(\Gamma_1\right)+
 \dipeik\left(\Gamma_3\right)-\dipeik\left(\Gamma_5\right) \,,
\\
 \eiksum(\bgamma_{0,1},\bgamma_{0,3})&=&\dipeik\left(\Gamma_1\right)+
 \dipeik\left(\Gamma_4\right)-\dipeik\left(\Gamma_6\right)\,,
\\
 \eiksum(\bgamma_{0,1},\bgamma_{0,4})&=&\dipeik\left(\Gamma_7\right)+
 \dipeik\left(\Gamma_8\right)-\dipeik\left(\Gamma_9\right) \,,
\\
 \eiksum(\bgamma_{0,2},\bgamma_{0,2})&=&\dipeik\left(\Gamma_3\right) \,,
\\
 \eiksum(\bgamma_{0,2},\bgamma_{0,3})&=&\dipeik\left(\Gamma_7\right)
 +\dipeik\left(\Gamma_8\right)-\dipeik\left(\Gamma_9\right) \,,
\\
 \eiksum(\bgamma_{0,2},\bgamma_{0,4})&=&\dipeik\left(\Gamma_2\right)
 +\dipeik\left(\Gamma_3\right)-\dipeik\left(\Gamma_6\right)\,,
\\
 \eiksum(\bgamma_{0,3},\bgamma_{0,3})&=&\dipeik\left(\Gamma_4\right) \,,
\\
 \eiksum(\bgamma_{0,3},\bgamma_{0,4})&=&\dipeik\left(\Gamma_2\right)+
 \dipeik\left(\Gamma_4\right)-\dipeik\left(\Gamma_5\right)\,,
\\
 \eiksum(\bgamma_{0,4},\bgamma_{0,4})&=&\dipeik\left(\Gamma_2\right)
 \,,
\\
 \eiksum(\bgamma_{1,1},\bgamma_{1,1})&=&\dipeik\left(\Gamma_7\right) \,,
\\
 \eiksum(\bgamma_{1,1},\bgamma_{1,2})&=&\dipeik\left(\Gamma_7\right)+
 \dipeik\left(\Gamma_8\right)-\dipeik\left(\Gamma_9\right) \,,
\\
 \eiksum(\bgamma_{1,2},\bgamma_{1,2})&=&\dipeik\left(\Gamma_8\right) \,.
\eeqn
 We note the following identities:
 \beqn
 \dipeik\left(\Gamma_1\right)+\dipeik\left(\Gamma_2\right) 
 +\dipeik\left(\Gamma_8\right)&=&\dipeik\left(\Gamma_3\right)
 +\dipeik\left(\Gamma_4\right) +\dipeik\left(\Gamma_7\right)\,,\\
\dipeik\left(\Gamma_3\right)+\dipeik\left(\Gamma_4\right) 
+\dipeik\left(\Gamma_9\right)&=&\dipeik\left(\Gamma_5\right)
+\dipeik\left(\Gamma_6\right) +\dipeik\left(\Gamma_8\right)\,,
 \eeqn
 which allow one to eliminate two (but not all) of the dipole graphs with a
 self-connected gluon.
 
\subsection{Closed colour flows versus dipole graphs\label{sec:ccfvsdp}}
In this section, we shall show how the radiation patterns that stem from 
closed colour flows, eq.~(\ref{eiksum}), can be represented in terms of 
analogous quantities, defined as properties of suitably-chosen dipole 
graphs, which involve their radiation patterns of eq.~(\ref{dipeikdef}).

In order to do so, we start by observing that the radiation pattern of 
any closed flow $(\bgammap,\bgamma)$ can be decomposed into the radiation
patterns of the individual colour loops associated with that flow, thus
\beq
\eiksum(\bgammap,\bgamma)=
\sum_{\ell\in\loopset(\bgammap,\bgamma)}\eiksum_\ell(\bgammap,\bgamma)\,,
\;\;\;\;\;\;\;\;
\eiksum_\ell(\bgammap,\bgamma)=
\sum_{k,l\in\ell}^{k<l}(-1)^{\delta_\ell(k,l)}\eik{k}{l}\,.
\label{eiksumL}
\eeq
Furthermore, the reader must bear in mind that the number of elements in a 
colour loop is an even number; half of the elements are L-elements, the other 
half are R-elements (see sect.~\ref{sec:loops}); they are alternating,
beginning with an R-element. We write this as follows:
\beq
\ell\in\loopset(\bgammap,\bgamma)\,,\;\;\;\;\;\;\;\;
\ell=(e_1,e_2,\ldots e_{2m_\ell})\,,
\eeq
so that $2m_\ell$ is the number of elements in the loop, and
\beqn
\eiksum_\ell(\bgammap,\bgamma)&=&
\sum_{i=1}^{2m_\ell-1}\sum_{j=1}^{\ceil*{(2m_\ell-i)/2}}\eik{e_i}{e_{i+2j-1}}-
\sum_{i=1}^{2m_\ell-1}\sum_{j=1}^{\floor*{(2m_\ell-i)/2}}\eik{e_i}{e_{i+2j}}
\label{eiksumLdip}
\\*&=&
\eiksum_\ell^+(\bgammap,\bgamma)-\eiksum_\ell^-(\bgammap,\bgamma)\,.
\label{eiksumLdip2}
\eeqn
There are
\beq
\sum_{k=1}^{2m_\ell-1}\ceil*{\frac{k}{2}}=m_\ell^2
\label{posel}
\eeq
and
\beq
\sum_{k=1}^{2m_\ell-1}\floor*{\frac{k}{2}}=m_\ell(m_\ell-1)
\label{negel}
\eeq
summands in $\eiksum_\ell^+(\bgammap,\bgamma)$ and 
$\eiksum_\ell^-(\bgammap,\bgamma)$, respectively. Note that
\beq
m_\ell^2+m_\ell(m_\ell-1)=\frac{2m_\ell(2m_\ell-1)}{2}\,,
\eeq
is indeed the total number of summands in $\eiksum_\ell(\bgammap,\bgamma)$
(before any algebraic simplifications).

Since the two sums in eq.~(\ref{eiksumLdip2}) are linear combinations
of eikonals with coefficients all equal to one, we see the possible 
emergence of radiation patterns of dipole graphs in eq.~(\ref{eiksumL}). 
More precisely, the argument goes as follows. Equation~(\ref{posel})
suggests organizing the parton labels associated with the
summands in $\eiksum_\ell^+(\bgammap,\bgamma)$ in $m_\ell$ sets,
each composed of $m_\ell$ pairs of labels. Likewise, from eq.~(\ref{negel})
one obtains $m_\ell-1$ sets, each composed of $m_\ell$ pairs of labels,
relevant to $\eiksum_\ell^-(\bgammap,\bgamma)$. These sets of pairs of
labels are candidates to be subsets of suitable dipole graphs.
For this to happen, all quark and antiquark labels that belong to 
the colour loop $\ell$ must appear exactly once in each of such sets,
whereas gluon labels must appear at least once and at most twice.
These properties must be fulfilled for each of the loops in
$\loopset(\bgammap,\bgamma)$, so that when the candidate subsets
are combined
one forms properly-defined dipole graphs. This can happen since
\beq
\sum_\ell m_\ell=q+n\,,
\label{sumellm}
\eeq
that is, the number of elements in the joined candidate subsets is equal
to the number of dipoles in the dipole graphs. The fact that candidate
subsets emerging from different loops have disjoint sets of labels
is trivially true for quarks and antiquarks, but not necessarily
so for gluons (see items i.--iv.~in sect.~\ref{sec:loops}). We point
out that there are
\beq
\prod_\ell m_\ell\,,
\;\;\;\;\;\;\;\;
\prod_\ell \left(m_\ell-1\right)\,,
\label{numofdip}
\eeq
possible ways of joining the candidate subsets that emerge from different
loops, for $\eiksum_\ell^+(\bgammap,\bgamma)$ and 
$\eiksum_\ell^-(\bgammap,\bgamma)$, respectively.

In appendix~\ref{sec:flvsdg} we shall sketch a proof of the equivalence
of colour flows and dipole graphs. Here, we limit ourselves to stating
that the consequence of this equivalence is the existence, for any given 
closed colour flow $(\bgammap,\bgamma)$, of two sets of dipole graphs
\beq
\Big\{\Gamma_a\Big\}_{(\bgammap,\bgamma)}^+\,,\;\;\;\;\;\;\;\;
\Big\{\Gamma_b\Big\}_{(\bgammap,\bgamma)}^-\,,
\label{dipsets}
\eeq
such that
\beqn
\sum_{\ell\in\loopset(\bgammap,\bgamma)}\eiksum_\ell^+(\bgammap,\bgamma)&=&
\sum_a c^+\left(\Gamma_a\right)\,\dipeik\left(\Gamma_a\right)\,,
\label{eikpvsdip}
\\*
\sum_{\ell\in\loopset(\bgammap,\bgamma)}\eiksum_\ell^-(\bgammap,\bgamma)&=&
\sum_b c^-\left(\Gamma_b\right)\,\dipeik\left(\Gamma_b\right)\,,
\label{eikmvsdip}
\eeqn
for suitable {\em integer} coefficients $c^\pm(\Gamma)$ (which, as we
shall show in appendix~\ref{sec:flvsdg}, are typically equal to one). 
We point out that the sets in eq.~(\ref{dipsets}) are not necessarily 
disjoint, and that their choice is not necessarily unique.

While the proof of eqs.~(\ref{eikpvsdip}) and~(\ref{eikmvsdip})
is complicated in general, an exception is that of a diagonal closed 
flow (i.e.~one for which the L- and R-flow coincide, $(\bgamma,\bgamma)$). 
In that case, 
there are \mbox{$q+n$} colour loops\footnote{In order to be definite, here 
we consider the case of zero $U(1)$ gluons. The results can then be applied 
to the case of $p$ $U(1)$ gluons by means of the formal replacement $n\to n-p$.
\label{ft:u1}}, each with two elements; that is, $m_\ell=1$ $\forall\,\ell$, 
and
\beq
\eiksum_\ell^-(\bgamma,\bgamma)=0
\;\;\;\;\Longrightarrow\;\;\;\;
\eiksum_\ell(\bgamma,\bgamma)=\eiksum_\ell^+(\bgamma,\bgamma)\,.
\eeq
Using eqs.~(\ref{qgflow}) and~(\ref{qgflowp})
\beqn
\loopset(\bgamma,\bgamma)&=&
\bigcup_{p=1}^q \Big\{\Big(\Mp,\sigma(t_{p-1}+1)\Big)\,,
\Big(\sigma(t_{p-1}+1),\sigma(t_{p-1}+2)\Big)\,,\ldots
\nonumber
\\*&&\phantom{\bigcup_{p=1}^q \Big\{}\;
\Big(\sigma(t_p-1),\sigma(t_p)\Big)\,,
\Big(\sigma(t_p),\mu(-p-q)\Big)
\Big\}\,.
\label{diagloops}
\eeqn
The set on the r.h.s.~of eq.~(\ref{diagloops}) is a dipole graph
that belongs to $\dipset_{2q;n}$; therefore, the leftmost set in
eq.~(\ref{dipsets}) contains a single element, and the rightmost one
is empty.

When the L- and R-flows do not coincide the sums on the r.h.s.~of
eqs.~(\ref{eikpvsdip}) and~(\ref{eikmvsdip}) are non-trivial.
Without knowing their explicit forms, one can still establish a rather
general property, namely that the leftmost set in eq.~(\ref{dipsets}) is not 
a subset of $\dipset_{2q;n}$, but rather of 
\mbox{$\cup_{p,s_p}\dipset_{2q;n-p,s_p}$},
for suitable values of $p$ and $s_p$. The argument is the following: a 
necessary condition for the following relations to hold true
\beq
\Big\{\Gamma_a\Big\}_{(\bgammap,\bgamma)}^+\;\subseteq\;\dipset_{2q;n}
\label{Gsub}
\eeq
is the absence of self-connected gluons. Indeed, when this is
not the case, of the $m_\ell^2$ summands in the leftmost sum in 
eq.~(\ref{eiksumLdip}) there are $n_{g^\star,\ell}$ which 
are equal to zero, being self-eikonals. Here, by $n_{g^\star,\ell}$ 
we have denoted the number of gluon labels that appear twice in 
the loop $\ell$. Therefore, in general out of $m_\ell^2-n_{g^\star,\ell}$ 
pairs one cannot identify an integer number of non-overlapping sets (the 
candidate subsets of suitable dipole graphs), each with $m_\ell$ pairs of 
labels. The fact that in general $n_{g^\star,\ell}\ne 0$ for some $\ell$ 
implies that eq.~(\ref{Gsub}) does not hold true, 
and must be modified as follows:
\beq
\Big\{\Gamma_a\Big\}_{(\bgammap,\bgamma)}^+\;\subseteq\;
\bigcup_{p,s_p}\dipset_{2q;n-p,s_p}\,.
\label{Gsub2}
\eeq
We point out that this is necessary in spite of the absence of $U(1)$ gluons 
in the flow (or, more precisely, for any given number of $U(1)$ gluons -- 
see footnote~\ref{ft:u1}), which further clarifies the fact that the $p$ 
gluons missing from the sets on the r.h.s.~of eq.~(\ref{Gsub2}) are not 
$U(1)$ gluons, but rather the self-connected ones introduced
in sect.~\ref{sec:scg}. In order to avoid any confusion, when possible
we shall number the self-connected and $U(1)$ gluons by means of labels
$k$ and $p$, respectively.

Thus, for each $0\le k\le n$ and $s_k\in\Sset{n}{k}$, we construct 
the subsets
\beq
\bdipset_{q;n-k,s_k}\subseteq\dipset_{2q;n-k,s_k}
\label{subdipset}
\eeq
so that
\beq
\bigcup_{p=0}^n\bigcup_{s_p\in\Sset{n}{p}}
\bigcup_{\bgamma,\bgammap\in\flowBggnp^{(s_p)}}
\left(\Big\{\Gamma_a\Big\}_{(\bgammap,\bgamma)}^+
\,\bigcup\,
\Big\{\Gamma_b\Big\}_{(\bgammap,\bgamma)}^-\right)\,=\,
\bigcup_{k=0}^n\bigcup_{s_k\in\Sset{n}{k}}
\bdipset_{2q;n-k,s_k}\,.
\eeq
By inserting eqs.~(\ref{eikpvsdip}) and~(\ref{eikmvsdip}) into
eq.~(\ref{qgsoftsum}) we arrive at an alternative expression
for the soft limit of a generic quark-gluon matrix element
\beq
\ampsqQGRS=2\gs^2\,
\CF\sum_{k=0}^n\sum_{s_k\in\Sset{n}{k}}
\sum_{\Gamma\in\bdipset_{2q;n-k,s_k}}\!\!\!\!
\Wampsq^{(2q;n)}\left(\Gamma\right)
\dipeik\left(\Gamma\right)\,,
\label{qgsoftsumdip}
\eeq
where
\beqn
\Wampsq^{(2q;n)}\left(\Gamma\right)&=&
\sum_{p=0}^n\sum_{s_p\in\Sset{n}{p}}
\frac{(-1)^{p}}{2^n}\,N^{-p}\!\!\!
\sum_{\bgamma,\bgammap\in\flowBggnp^{(s_p)}}\!\!\!
\ampOSQGBgg(\bgammap)^\star\,\ampOSQGBgg(\bgamma)\,
\label{subevmats}
\\*&&\times
N^{-\rho(\bgamma)-\rho(\bgammap)}
N^{\nloops{\loopset(\bgammap,\bgamma)}}\,
\nonumber\\*&&\phantom{aaa}\times
\Bigg[
c^+(\Gamma)\,
\delta\!\left(\Gamma\in\Big\{\Gamma_a\Big\}_{(\bgammap,\bgamma)}^+\right)-
c^-(\Gamma)\,
\delta\!\left(\Gamma\in\Big\{\Gamma_b\Big\}_{(\bgammap,\bgamma)}^-\right)
\Bigg]\,.
\nonumber
\eeqn
We point out that the matrix elements defined in eq.~(\ref{subevmats}),
at variance with all of their counterparts used so far, generally
contain contributions associated with different powers of $N$. Because
of this, it is necessary to include the monomials in $N$ into their
definitions, again contrary to what was done so far. 
There is an ultimate  physical reason for that. Namely,
the structure of dipole graphs, although it extends that of the
colour connections employed in current leading-colour PSMCs, is still
based on a picture which takes a single side of the Cutkosky cut into 
account. However, the results of sect.~\ref{sec:soft} show that beyond 
leading colour it is necessary to include simultaneously the information 
on both sides of the cut, which can be done by means of closed colour 
loops. If the more complicated structure of such loops is traded for
the simpler one of the dipole graphs, this entails a set of more 
complicated matrix elements.

By construction, the sets on the l.h.s.~of eq.~(\ref{subdipset}) may
not be unique. However, different choices of $\bdipset_{q,n-k,s_k}$, while 
changing the individual summands on the r.h.s.~of eq.~(\ref{qgsoftsumdip}), 
will leave the sum invariant -- in other words, there are subsets of
dipole graphs whose radiation patterns are not linearly independent
of one another, as we have seen in the examples of sect.~\ref{sec:dipex}. 
With increasing numbers of quarks and gluons, the possibilities 
of choosing different $\bdipset_{q,n-k,s_k}$ also increase.
This flexibility may turn out to be useful, since a suitable choice 
could lead to a reduction of negative weights.

\subsection{Subevents}
In order to implement a dipole shower simulation including
subleading colour, one needs to take account of all relevant dipole graphs for
a given hard process configuration, and then combine the results.  This is
necessary because, as we have seen, contributions from graphs with
self-connected gluons cannot in general be avoided. Simply choosing a 
single one out of the set of dipole graphs according to their relative 
weights would lead to final states where hard hadrons appear that stem 
from non-showering self-connected gluons, and are mostly isolated. 
Instead, those hadrons can be more conveniently represented as components 
of a jet that includes showering of other contributing graphs.  
We call each dipole graph contribution a {\em subevent} of that hard
configuration.  The final state of that event is then the result of
summing all subevents. 

As an example, consider the process $\bar q q\to gg$ studied in
sect.~\ref{sec:MCqqgg}.  For any given Born configuration, one would
generate six subevents by showering and hadronization starting from
the dipole configurations in graphs $\Gamma_1\ldots\Gamma_6$ of
fig.~\ref{fig:qqgg_dipoles}, respectively.  These subevents would then be
weighted by the corresponding coefficients $N^3 M_1\ldots M_4/N$ in
eq.~(\ref{Dsoftex1}), then combined to produce a complete Monte Carlo
event from that Born configuration.  Each final-state object would
be entered into histograms with the weight of the subevent that
generated it.  In this example, all the products of dual amplitudes
$M_1\ldots M_4$, given in eqs.~(\ref{bMat1ex1})-(\ref{bMat4ex1}), are
positive, so (\ref{Dsoftex1}) implies that only the subevents
from graphs $\Gamma_4$ and $\Gamma_5$ have negative weights, while
$\Gamma_3$ does not contribute.  Alternatively, we may use
eq.~(\ref{qqggeikid}) to eliminate  $\Gamma_4$ and $\Gamma_5$ (or
$\Gamma_6$), but then other subevents will get negative weights for certain Born
configurations.  This flexibility in the dipole graph approach may be
useful for optimizing Monte Carlo efficiency.  A possible way to deal
with the remaining negative weights could be a resampling method
such as that of ref.~\cite{Olsson:2019wvr,Andersen:2020sjs,Nachman:2020fff}.

A similar approach would appear necessary in single-parton shower
implementations that treat individual colour flows separately, as in
eq.~(\ref{eq:sud_pxi}).  Flows that do not involve one or more gluons
from the Born process will again give rise to isolated hard hadrons
coming from such non-showering gluons.  If the result of showering
each flow is treated as a subevent from the same Born configuration,
these hadrons become a subleading correction to the hard component of
the associated gluon jet.

\section{The mixed QCD-QED case\label{sec:mix}}
An interesting aspect of the formulation presented in this paper
is that it applies, essentially unchanged, to a mixed-coupling 
scenario, i.e.~one in which perturbative expansions are obtained
as power series in two parameters, both of which are regarded
as ``small''. The foremost candidate is that of QCD+QED, where the
two parameters alluded to before are the coupling constants, $\as$
and $\aem$, of the respective theories; in order to be definite, in
the following we shall understand this case. We point out that while 
$\aem\ll\as$, the existence of a clear hierarchy between the expansion 
parameters is not a necessary condition for this formulation to work.
We also remind the reader that matrix elements are proportional
to $\as^n\aem^m$; at the LO, $n$ and $m$ are constrained to
obey $n+m=k_0$, with $k_0$ a process-dependent integer, while at
the NLO $n+m=k_0+1$, and so forth. The interested reader can find
in ref.~\cite{Frederix:2018nkq} a classification scheme for the various
contributions to a given N$^k$LO accuracy.

In the QCD+QED case, several kinds of results become relevant at the
LO and NLO accuracy for any given combination $\as^n\aem^m$:
\begin{itemize}
\item[{\em a)}] tree-level matrix elements;
\item[{\em b)}] soft-gluon emission patterns;
\item[{\em c)}] soft-photon emission patterns.
\end{itemize}
The statement we make is that the formulae of sect.~\ref{sec:tree}
can be used, with minimal or no changes, to deal with item {\em a)},
and those of sect.~\ref{sec:soft} to deal with item {\em b)}; as far
as item {\em c)} is concerned, its treatment entails a significantly
simplified version of its gluon-emission counterpart.

A detailed discussion of the items above is beyond the scope of this paper.
Therefore, here we shall limit ourselves to presenting a simple
example, which should be sufficient to give the reader an idea of the
procedures to follow in a general case.

\vspace{2mm}
\noindent
$\bullet$ {\em Example.}  Consider the following process:
\beq
0\;\longrightarrow\;u(-1)\,\ub(-3)\,u(-2)\,\ub(-4)\,.
\label{uuuu}
\eeq
Incidentally, the fact that there are two quark-antiquark pairs of
the same flavour allows us to show how to handle in practice such a
case, which has been neglected so far in the interest of a simpler
notation. Following ref.~\cite{Frixione:2011kh}, we start by distinguishing
flavour-line configurations
\beqn
f_A&=&\{-1,-3\}\{-2,-4\}\,,
\label{fAl}
\\
f_B&=&\{-1,-4\}\{-2,-3\}\,,
\label{fBl}
\eeqn
i.e.~$f_A$ ($f_B$) corresponds to the graphs where there is a fermion
line that connects quark $-1$ with antiquark $-3$ ($-4$), and quark $-2$ 
with antiquark $-4$ ($-3$). Conventionally, we identify $f_A$- and $f_B$-type
graphs with $s$- and $t$-channel exchanges, respectively. 

Although the two flavour-line configurations of eqs.~(\ref{fAl})
and~(\ref{fBl}) induce the same colour flows, such flows need to be
distinguished, because their contributions to the matrix elements must
be accounted for separately (also in view of the fact that the associated
colour factors might differ). In the case of gluon exchange we have
\beqn
\gamma_{1,A}&=&(-1,-4)(-2-3)\;\equiv\;\gamma_{2,B}\,,
\label{gaA1eqgaB}
\\
\gamma_{2,A}&=&(-1,-3)(-2-4)\;\equiv\;\gamma_{1,B}\,,
\label{gaA2eqgaB}
\eeqn
whereas in the case of a photon exchange
\beqn
\gamma_{3,A}&\equiv&\gamma_{2,A}\,,
\label{gaA3eqga1}
\\
\gamma_{3,B}&\equiv&\gamma_{1,A}\,.
\label{gaA3eqga2}
\eeqn
According to eq.~(3.32) of ref.~\cite{Frixione:2011kh}
\beqn
&&
\rho(\gamma_{1,A})=0\,,\;\;\;\;\;\;\;\;
\rho(\gamma_{2,A})=1\,,
\label{rhogaA}
\\*&&
\rho(\gamma_{1,B})=0\,,\;\;\;\;\;\;\;\;
\rho(\gamma_{2,B})=1\,,
\label{rhogaB}
\\*&&
\rho(\gamma_{3,A})=0\,,\;\;\;\;\;\;\;\;
\rho(\gamma_{3,B})=0\,.
\label{rhoga3}
\eeqn
In view of eqs.~(\ref{gaA1eqgaB})--(\ref{gaA3eqga2}), the only 
independent colour-loop structures are the following:
\beqn
\loopset\left(\gamma_{1,A},\gamma_{1,A}\right)&=&
\Big\{(-1,-4),(-2,-3)\Big\}\,,
\\*
\loopset\left(\gamma_{2,A},\gamma_{1,A}\right)&=&
\Big\{(-1,-4,-2,-3)\Big\}\,,
\\*
\loopset\left(\gamma_{2,A},\gamma_{2,A}\right)&=&
\Big\{(-1,-3),(-2,-4)\Big\}\,,
\eeqn
whence one derives the independent radiation patterns
\beqn
\eiksum\left(\gamma_{1,A},\gamma_{1,A}\right)&=&
\eik{-1}{-4}+\eik{-2}{-3}\,,
\\*
\eiksum\left(\gamma_{2,A},\gamma_{1,A}\right)&=&
\eik{-1}{-4}-\eik{-1}{-2}+\eik{-1}{-3}
\\*&+&
\eik{-4}{-2}-\eik{-4}{-3}+\eik{-2}{-3}\,,
\\*
\eiksum\left(\gamma_{2,A},\gamma_{2,A}\right)&=&
\eik{-1}{-3}+\eik{-2}{-4}\,.
\eeqn
All of the other combinations are identical to those above, according
to the identifications of eqs.~(\ref{gaA1eqgaB})--(\ref{gaA3eqga2}).
Conversely, as far as the dual amplitudes associated with the colour 
flows are concerned, we have the following:
\beqn
&&
\ampCS^{(4;0)}(\gamma_{1,A})=\amp_s\,,\;\;\;\;\;\;\;\;\phantom{a}\;
\ampCS^{(4;0)}(\gamma_{2,A})=-\amp_s\,,
\\*&&
\ampCS^{(4;0)}(\gamma_{1,B})=-\amp_t\,,\;\;\;\;\;\;\;\;
\ampCS^{(4;0)}(\gamma_{2,B})=\amp_t\,,
\\*&&
\ampCS^{(4;0)}(\gamma_{3,A})=\amp_s\,,\;\;\;\;\;\;\;\;\phantom{a}\;
\ampCS^{(4;0)}(\gamma_{3,B})=-\amp_t\,.
\eeqn
Here, we have denoted by $\amp_s$ and $\amp_t$ the $s$- and $t$-channel
amplitudes, stripped of charge and colour factors (contrary to the 
conventions used so far), so that
\beqn
\ampsq_s&\equiv&\amp_s^\star\,\amp_s=
\frac{t^2+u^2}{s^2}\,,
\\
\ampsq_t&\equiv&\amp_t^\star\,\amp_t=
\frac{s^2+u^2}{t^2}\,,
\\
\ampsq_{st}&\equiv&\Re\left[\amp_s^\star\,\amp_t\right]=
\frac{u^2}{st}\,.
\eeqn
There are three tree-level LO matrix elements, understood to be multiplied
by the factor $\as^n\aem^m$ with $n+m=2$, which we denote by
\beq
\ampsq_{(n,m)}^{(4;0)}\,.
\eeq
In order to compute them, it is sufficient to apply eq.~(\ref{treeQressum}),
with the sums over flows determined by the specific contribution we need to
derive. Explicitly
\beqn
\ampsq_{(2,0)}^{(4;0)}&=&
\sum_{i,j=1}^2\sum_{a,b=A,B}
\ampCS^{(4;0)}(\gamma_{i,a})^\star \,\ampCS^{(4;0)}(\gamma_{j,b})\,
N^{-\rho(\gamma_{i,a})-\rho(\gamma_{j,b})}\,
N^{\nloops{\loopset(\gamma_{i,a},\gamma_{j,b})}}\,,
\label{MQCDsum}
\\*
\ampsq_{(1,1)}^{(4;0)}&=&
\sum_{i=1}^2\sum_{a,b=A,B}
\ampCS^{(4;0)}(\gamma_{i,a})^\star \,\ampCS^{(4;0)}(\gamma_{3,b})\,
N^{-\rho(\gamma_{i,a})-\rho(\gamma_{3,b})}\,
N^{\nloops{\loopset(\gamma_{i,a},\gamma_{3,b})}}+{\rm c.c.}\,,\phantom{aaa}
\label{MQEDQCDsum}
\\*
\ampsq_{(0,2)}^{(4;0)}&=&
\sum_{a,b=A,B}
\ampCS^{(4;0)}(\gamma_{3,a})^\star \,\ampCS^{(4;0)}(\gamma_{3,b})\,
N^{-\rho(\gamma_{3,a})-\rho(\gamma_{3,b})}\,
N^{\nloops{\loopset(\gamma_{3,a},\gamma_{3,b})}}\,,
\label{MQEDsum}
\eeqn
so that
\beqn
\ampsq_{(2,0)}^{(4;0)}&=&
\left(N^2-1\right)\left(\ampsq_s+\ampsq_t+\frac{2}{N}\ampsq_{st}\right)\,,
\label{bornQCDres}
\\*
\ampsq_{(1,1)}^{(4;0)}&=&
-4\left(N^2-1\right)\ampsq_{st}\,,
\label{bornQEDQCDres}
\\*
\ampsq_{(0,2)}^{(4;0)}&=&
N^2\left(\ampsq_s+\ampsq_t-\frac{2}{N}\ampsq_{st}\right)\,,
\label{bornQEDres}
\eeqn
which can be easily verified to agree with the results obtained by means
of standard computational techniques. As far as the soft-gluon limits
are concerned, we need to employ eq.~(\ref{qqsoftres}), with the
colour flows summed as in eqs.~(\ref{MQCDsum})--(\ref{MQEDsum}).
We obtain
\beqn
\ampsq_{(3,0){\rm\sss SOFT}}^{(4;1)}&=&2\CF\Bigg\{
\left[(N^2-1)\left(\eik{-1}{-4}+\eik{-2}{-3}\right)+\Xi\right]\ampsq_s
\label{softQCDres}
\\*&&\phantom{2\CF}
+\left[(N^2-1)\left(\eik{-1}{-3}+\eik{-2}{-4}\right)+\Xi\right]\ampsq_t
\nonumber
\\*&&\phantom{2\CF}
+\frac{2}{N}
\left[(N^2-1)\left(\eik{-1}{-2}+\eik{-3}{-4}\right)+\Xi\right]\ampsq_{st}
\Bigg\}\,,
\nonumber
\\*
\ampsq_{(2,1){\rm\sss SOFT}}^{(4;1)}&=&4\CF\Bigg\{
N\left(\eik{-1}{-4}+\eik{-2}{-3}-\eik{-1}{-2}-\eik{-3}{-4}\right)\ampsq_s
\label{softQEDQCDres}
\\*&&\phantom{2\CF}\!
+N\left(\eik{-1}{-3}+\eik{-2}{-4}-\eik{-1}{-2}-\eik{-3}{-4}\right)\ampsq_t
\nonumber
\\*&&\phantom{a}\!
-\left[N^2\left(\eik{-1}{-3}+\eik{-2}{-4}+\eik{-1}{-4}+\eik{-2}{-3}\right)
+2\Omega\right]\ampsq_{st}
\Bigg\}\,,
\nonumber
\\*
\ampsq_{(1,2){\rm\sss SOFT}}^{(4;1)}&=&2\CF\Bigg\{
N^2\left(\eik{-1}{-3}+\eik{-2}{-4}\right)\ampsq_s
\label{softQEDres}
\\*&&\phantom{2\CF}\!
+N^2\left(\eik{-1}{-4}+\eik{-2}{-3}\right)\ampsq_t
\nonumber
\\*&&\phantom{2\CF}\!
+2N\,\Omega\,\ampsq_{st}
\Bigg\}\,,
\nonumber
\eeqn
having defined
\beqn
\Omega&=&\eik{-1}{-2}+\eik{-3}{-4}-\eik{-1}{-3}
\nonumber
\\*&&
-\eik{-1}{-4}-\eik{-2}{-3}-\eik{-2}{-4}\,,
\label{Omdef}
\\
\Xi&=&\eik{-1}{-2}+\eik{-3}{-4}+\Omega\,.
\label{Xidef}
\eeqn
As for the LO matrix elements, one can verify that the results
of eqs.~(\ref{softQCDres})--(\ref{softQEDres}) coincide with those
obtained with standard techniques.

It is instructive to compare eq.~(\ref{bornQEDQCDres}) with 
eqs.~(\ref{bornQCDres}) and~(\ref{bornQEDres}). In all three cases, 
the leading-$N$ contributions stem from a closed flow in which the 
L- and R-flows coincide (bearing in mind the identifications in 
eqs.~(\ref{gaA1eqgaB})--(\ref{gaA3eqga2})). Such contributions would 
constitute the building blocks for the determination of the shower 
initial conditions in a leading-$N$ PSMC, e.g.~according to the
prescription of ref.~\cite{Odagiri:1998ep}. But while for
eqs.~(\ref{bornQCDres}) and~(\ref{bornQEDres}) the relevant kinematical
quantities are ``squared'' matrix elements (either $\ampsq_s$ or 
$\ampsq_t$), for eq.~(\ref{bornQEDQCDres}) what is relevant
is the interference between the $s$- and $t$-channel amplitudes
($\ampsq_{st}$). This shows that, even at leading $N$, when the
matrix elements that underpin the hard events that initiate the 
QCD showers do not have the maximal power of $\as$ one must consider
objects that are not necessarily positive definite. What is important
is that the nature of such objects is determined by closed colour flows, and 
not by their kinematical characteristics. Direct evidence of this fact can 
be seen in the comparison of eqs.~(\ref{bornQCDres})--(\ref{bornQEDres})
with eqs.~(\ref{softQCDres})--(\ref{softQEDres}): the hierarchy in $N$
of the radiation patterns in the latter equations is reflected in that
of the contributions to the former ones, and the correspondence is
driven by the underlying closed colour flows.

We conclude this section by considering soft-photon emission patterns. 
This case is significantly simpler than its soft-gluon counterpart,
owing to the fact that the charge-linked Borns (i.e.~the matrix elements
that multiply suitable linear combinations of eikonals) are all proportional
to the LO tree-level matrix elements. As far as the linear combinations of 
eikonals are concerned, they are entirely determined by the charges of
external partons that participate in the hard process. Thus, from
eq.~(3.12) of ref.~\cite{Frederix:2018nkq}, one defines the photon
radiation pattern, i.e.~the QED analogue of eq.~(\ref{eiksum})
\beq
\eiksum_{\rm\sss QED}=
\sum_{k<l}e_k\,e_l\,\eik{k}{l}\,,
\label{QEDeiksum}
\eeq
with $e_i$ the charge of parton $i$, the sums are extended to all 
external partons, and, in keeping with the rest of this paper, 
all partons are outgoing. With this, the analogues of 
eqs.~(\ref{softQCDres})--(\ref{softQEDres}) read
\beq
\ampsq_{(n,m+1){\rm\sss SOFT}}^{(4;1)\gamma}=
\left(\frac{2}{3}\right)^2\Omega\,\ampsq_{(n,m)}^{(4;0)}\,,
\eeq
since for the process of eq.~(\ref{uuuu}), the QED radiation pattern
of eq.~(\ref{QEDeiksum}) coincides with eq.~(\ref{Omdef}) up to an overall
factor equal to the charge of the $u$ quark squared.

The fact that on the r.h.s.~of eq.~(\ref{QEDeiksum}) one sums over all
pairs of external partons implies that this radiation pattern is quite
similar to a soft-gluon one relevant to a closed flow with a single
colour loop that includes all the partons of the process. This is 
remarkable, because it puts QED and QCD radiation on the same footing, and
allows one to treat the former in the same way as (one of the cases of)
the latter. However, in general the two patterns are not identical, since
the eikonals in $\eiksum_{\rm\sss QED}$ have coefficients that, in absolute
value, are not all equal to each other, which does not happen in QCD.
While this entails only minor modifications to the angular-ordered
showering approach outlined in sect.~\ref{sec:angord}, it is
potentially more problematic for dipole showering,
since the equivalence of the radiation pattern of eq.~(\ref{QEDeiksum}) 
and dipole graphs must be re-considered.

\section{Conclusions\label{sec:conc}}
In this paper we have investigated the role that colour flows play
in both the calculation of tree-level matrix elements and of their
soft limits, and in the inclusion of effects beyond leading colour
in parton shower Monte Carlos.

For these applications, and in general for any kind of non-approximate
approach, it is appropriate to work with what we have called closed
colour flows (sect.~\ref{sec:basic}), which generalize the colour flows
usually employed in leading-$N$ computations (which are therefore more
appropriately referred to as open colour flows). In essence, closed 
colour flows are based on a picture that features a double copy of the 
colour/anticolour labels of the partons that participate in the process 
(one copy for each side of the Cutkosky cut; each such copy is an open 
colour flow). Leading-$N$ contributions are (possibly a subset of) those 
for which the orderings of the labels in the two open flows that form a 
closed flow are identical to each other.

Colour loops, which we have defined algorithmically in sect.~\ref{sec:loops},
give a representation of closed colour flows that can be conveniently used
in matrix element computations and PSMCs alike. By means of colour loops,
one generalizes the concept of colour connection that is used at the
leading $N$ (see in particular sect.~\ref{sec:softcomm}). 
We have discussed how colour loops can in turn be equivalently 
described in terms of dipole graphs (sect.~\ref{sec:MCs}), which are 
an alternative to the former in PSMC implementations. Interestingly,
the counting of the numbers of dipole graphs gives rise to integer
sequences that, to the best of our knowledge, are presented here
for the first time (sects.~\ref{sec:dipgr} and~\ref{sec:scg}), and
for which we give generating functions.

In the case of purely gluonic processes, the colour loops as defined
in this paper are seen (app.~\ref{sec:gflows}) to correspond to quantities 
emerging from the mathematics of permutations (in particular, the alternating
cycles of a cycle graph). This is interesting, because it could pave the way 
to exploiting results that are increasingly important in other branches of 
science (mainly genome studies -- the curious reader can find a list of 
classical mathematics papers, relevant to biology, for example in chapter~9 
of ref.~\cite{Bona:2012}).

While when working with colour loops each colour or anticolour label
may have as many as $m-1$ colour partners ($m$ being the number of partons
of the process), dipole graphs are more similar to the colour connections 
one is used to at the leading $N$, in that they feature a single partner per 
(anti)quark, and two partners per gluon. Furthermore, while colour loops
require a double copy of the partons (as the closed colour flows from which 
they stem), dipole graphs need a single copy, and thus are topologically 
identical to any set of colour connections employed in existing PSMCs. 
At variance with the latter, however, dipole graphs feature colour-colour 
and anticolour-anticolour connections, in addition to the usual 
colour-anticolour ones. We stress again that, in spite of their topological
differences, colour loops and dipole graphs give equivalent representations
of the same thing. The simpler topological structure of the latter is
compensated by the necessity of working with more complicated objects
at the level of amplitudes squared.

Regardless of their possible use in the context of PSMCs, colour flows give 
an organising principle for the computation of tree-level matrix elements
(sect.~\ref{sec:tree}). We have shown that the treatment of any process 
is in fact no more complicated than that of one featuring only quarks 
and antiquarks, thanks to the introduction of the idea of secondary flows 
(sect.~\ref{sec:treeqg}); these allow one to express the results for 
processes that include gluons in terms of the formulae relevant to the 
no-gluon case.

Although from tree-level matrix elements results it might seem that secondary
flows are a computational device, the results for the soft limits 
(sect.~\ref{sec:soft}) show that they are in fact more fundamental objects, 
in that they control the soft-radiation patterns with a clear hierarchy in $N$.
Furthermore, through secondary flows one recovers a fundamental property
of the leading-$N$ PSMCs, namely the fact that the radiation pattern
of the first emission is entirely controlled by Born-level quantities.
This property, which is trivially true at leading $N$ but not beyond it,
allows one to incorporate subleading-$N$ effects in PSMCs in a Markovian way.

For matrix element computations we often make use of vector spaces (that
represent colour degrees of freedom), which are separable into spaces 
relevant to individual partons (sect.~\ref{sec:vspace}). In the presence
of gluons, we have shown how the so-called fundamental and flow 
representations are associated with vectors in an $(N^2-1)$- and 
$N^2$-dimensional space respectively for each gluon, the former of which 
can be seen as a subspace of the latter, and whose complement in the latter 
is interpreted as representative of the degrees of freedom of $U(1)$ gluons. 

Matrix elements for processes that feature both quarks and gluons can be 
written as the incoherent sum of non-null contributions associated with 
different sets of $U(1)$ gluons, in the same way as for amplitude-level
quantities (see in particular eqs.~(\ref{matelfin}) and~(\ref{qgsoftsum})).
In turn, these sets are associated with non-overlapping sets of secondary 
flows, thereby confirming the physical nature of the latter. $U(1)$ gluons 
do appear also in purely-gluonic processes but, as is well known, they do 
not contribute to the cross section. However, this is true only after 
summing over all possible flows; conversely, if one restricts oneself to 
a subset of the flows, $U(1)$-gluon contributions are generally different
from zero.

In sect.~7 we have discussed possible applications of colour flows to
parton shower Monte Carlo simulations.  In particular, we have
proposed ways in which the subleading colour structure of Born matrix
elements could be included in the matching of hard processes to parton
showers.  One possible approach, based on eq.~(\ref{qgsoftsum}), is to
select closed flows according to their relative contributions to the
Born process and then use the radiation pattern of each flow to
influence the evolution of the associated showers.  This could be done
by using a modified Sudakov factor that reflects the sequence of
scales associated with the colour connections of the evolving partons
in the selected flow.  In the case of an angular-ordered shower, the
eikonal terms in the radiation pattern of the flow define a sequence
of angular regions for the evolution of each participating parton,
the rate of evolution in each region being controlled by the
relevant colour-linked Born matrix elements.

Another possible Monte Carlo approach is more suited to the evolution
of dipole showers, being based on dipole graphs and
eq.~(\ref{qgsoftsumdip}) rather than directly on colour flows.  We
showed that each flow can be represented by a linear combination of
dipoles, which can serve as initiators of dipole showers.

An issue that will affect the implementation of any scheme for
including subleading colour contributions is that these can have
either sign, leading to negative Monte Carlo weights.  There are
established methods for handling such weights~\cite{Olsson:2019wvr,
Andersen:2020sjs,Nachman:2020fff,Platzer:2011dq,Hoeche:2011fd,
Lonnblad:2012hz}; which of them is most suitable in this case 
will probably depend on the process
under study.  A less familiar issue that we have highlighted is the
appearance of self-connected gluons in both the single-parton and
dipole implementations.  Rather than treating these as non-showering
objects that would give rise to isolated hard hadrons, we have
suggested interpreting them as corrections to gluon jets generated by
more leading contributions from the same momentum configuration of the
Born process.  In this approach, each Born configuration would be used
to generate several ``subevents'' corresponding to different flows or
dipole graphs.

The inclusion of subleading colour contributions via subevents, based
on either colour flows or dipole graphs, will require
some modification of hadronization models, since, as we have seen, the
concept of colour partners is generalized to include quark-quark and
antiquark-antiquark pairs, as well as gluons that connect to two
quarks or two antiquarks.  However, since colour is conserved overall,
it is always possible to hadronize after allowing some colour
reconnection among showers within each subevent.

We have also briefly discussed (sect.~\ref{sec:mix}) how the techniques 
developed here in QCD can be applied with minor or no changes to the case 
where two perturbative couplings are relevant. Presently, the most important
application of this is in the context of QCD+QED computations and,
for matrix elements that do not have a maximal power of $\as$,
it affects PSMC simulations even at the leading $N$.

We have usually presented our results both at fixed (Born-level) flows
and after summing over flows. While only flow-summed predictions are physical,
it is not a foregone conclusion that they must be obtained by means of
analytically-summed formulae. In particular, when the number of partons 
involved becomes large, a random sum might be the only viable option, and 
this can be carried out only by using fixed-flow expressions. We point out
that, even in the context of an MC@NLO matching~\cite{Frixione:2002ik},
where a local connection between the short-distance cross sections and the 
PSMC must be established, there is no loss of efficiency, if suitable 
flow-based NLO-subtraction formulae are employed, e.g.~as proposed in
ref.~\cite{Frixione:2011kh}\footnote{A flow-based representation must also
be chosen for the finite virtual contributions, the discussion of which 
was not within the scope of ref.~\cite{Frixione:2011kh}.}. Conversely, a 
random sum behaves in any case poorly from a statistics viewpoint unless one 
is capable of understanding {\em a priori} which flows give the dominant 
contributions in $N$ (or, more in general, some selected $N^k$ contributions), 
and of generating them with high efficiency. While the generation of flows 
is a topic not directly relevant to this paper, since flows are assumed
to be given here, it may become important for the application of our results
to multi-parton processes, and it is well worth pursuing.

\section*{Acknowledgments}
SF is grateful to Fabio Maltoni for collaborating during the early
stages of this work, and for the endless and very entertaining
brainstorming sessions during one of the Covid-19 lockdowns.
Conversations with Stefan Prestel are also gratefully acknowledged.
BRW thanks members of the Cambridge Pheno Working Group for
valuable comments.
This work was partially supported by STFC Consolidated HEP grants
ST/P000681/1 and ST/T000694/1.

\appendix
\section{On gluon-only colour flows and colour factors\label{sec:gflows}}
In this appendix we discuss some properties of colour flows 
relevant to gluon-only processes, and their associated colour
factors. Where necessary, we shall exploit the following property:
\beq
C(\sigmap,\sigma)=C(I,\sigmapmo\circ\sigma)=C(\sigmamo\circ\sigmap,I)
\label{linecompl}
\eeq
of the colour-flow matrix element (see eqs.~(\ref{Lambdashort})
and~(\ref{CFmatdef}))
\beq
C(\sigmap,\sigma)=\sum_{\setan}
{\rm Tr}\Big(\lambda^{a_{\sigmap(1)}}\ldots
\lambda^{a_{\sigmap(n)}}\Big)^\star\,
{\rm Tr}\Big(\lambda^{a_{\sigma(1)}}\ldots\lambda^{a_{\sigma(n)}}\Big)\,.
\label{CFmatgluex}
\eeq
In eq.~(\ref{linecompl}), by the symbol $\circ$ we have denoted the 
composition of permutations, to be applied from right to left, i.e.: 
\beq
\Big(\sigma_A\circ\sigma_B\Big)(i)=\sigma_A\left(\sigma_B(i)\right)\,,
\eeq
for any two permutations $\sigma_A$ and $\sigma_B$.

\subsection{Colour flows\label{sec:appflow}}
A gluon-only closed flow $(\sigmap,\sigma)$ has an interesting property.
Define the permutation
\beq
\varsigma(\sigmap,\sigma)=
\sigmamo\circ\sigmap_{\rightarrow}\circ
\sigmapmo\circ\sigma_{\leftarrow}\,,
\label{eqperm}
\eeq
where the subscripts $\rightarrow$ and $\leftarrow$ indicate that the 
corresponding permutation has to be rotated by one position to the
right and to the left, respectively, i.e.
\beqn
&&\sigma_A=\Big(\sigma_A(1),\sigma_A(2),\ldots\sigma_A(n)\Big)
\\&&\phantom{aaa}
\Longrightarrow\;\;\;\;
\sigma_{A\rightarrow}=\Big(\sigma_A(n),\sigma_A(1),\ldots\sigma_A(n-1)\Big)\,,
\\&&\phantom{aaa\Longrightarrow}
\;\;\;\;\;
\sigma_{A\leftarrow}=\Big(\sigma_A(2),\sigma_A(3),\ldots\sigma_A(1)\Big)\,.
\eeqn
The definition in eq.~(\ref{eqperm}) is motivated by what has been
presented in sect.~\ref{sec:loops}; more precisely, the four permutations
on the r.h.s.~of eq.~(\ref{eqperm}) correspond, from right to left, to
items 2, 3, 4, and 5 in the definition of colour loops. With this,
one obtains
\beq
\nloops{\loopset(\sigmap,\sigma)}\;=\;
\nloops{{\rm cyc}\Big[\varsigma(\sigmap,\sigma)\Big]}\,.
\label{loopvscyc}
\eeq
That is, the number of colour loops in the closed flow $(\sigmap,\sigma)$ 
is equal to the number of ordinary cycles\footnote{Here, the cycles of a 
permutation include explicitly all of its $1-$cycles; these are sometimes
understood in the mathematics literature} of the permutation 
$\varsigma(\sigmap,\sigma)$ defined in eq.~(\ref{eqperm}). 
Equation~(\ref{loopvscyc}) stems from a more fundamental property, 
which is the following. For any colour loop
\beq
\ell=\left(\sigma(i_1),\sigmap(j_1),\sigma(i_2),\ldots\right)\in\loopset\,,
\label{colloop}
\eeq
then
\beq
\exists!\,c\in {\rm cyc}\Big[\varsigma(\sigmap,\sigma)\Big]
\;\;\;\;{\rm s.t.}\;\;\;\;
c=\left(i_1,i_2,\ldots\right)\,.
\eeq
That is: for a given closed flow, each colour loop is in a one-to-one
correspondence with a cycle of the permutation defined in eq.~(\ref{eqperm}).
Furthermore, the labels of such a cycle are mapped onto the labels of
the R-elements of the corresponding colour loop by means of the R-flow
permutation.

By establishing a direct connection between colour loops and ordinary
cycles of permutations (which are a much better studied subject) these
results are helpful in that they give an alternative viewpoint on
colour loops. Note that eq.~(\ref{loopvscyc}) is manifestly invariant
under the operations on the r.h.s.~of eq.~(\ref{linecompl}).

There is a further remarkable fact: the idea of colour loops, 
and in particular their construction given in steps 0--7 of 
sect.~\ref{sec:loops}, is essentially identical to that of 
{\em alternating cycles}, which (as far as we know) has been introduced 
in ref.~\cite{Bafna:1998} in the context of the mathematics of permutations, 
and which is nowadays a powerful tool in the study of genome re-arrangements; 
a good summary can be found e.g.~in chap.~9 of ref.~\cite{Bona:2012}.
The essence is that, given a permutation $p$, one constructs a 
double-edged graph, denoted by $G(p)$ and called the {\em cycle graph}
of $p$, which connects the various elements of $p$. Such connections
are represented as ordered sets of elements, called the alternating
cycles of $p$, whose number is usually denoted by $c(G(p))$. The
interested reader can easily verify that the rules to construct
a cycle graph and its alternating cycles are very closed related
to the rules given in sect.~\ref{sec:loops}. An immediate consequence
is therefore\footnote{Theorems on permutations usually include cyclic ones.
Since any non-cyclic $n$-element permutation (say, $\sigma$) is in 
one-to-one correspondence with an $(n-1)$-element permutations (say, $p$), 
a relabelling (e.g.~\mbox{$\sigma(i)=p(i-1)$}, with \mbox{$2\le i\le n$}), 
plus the insertion of a conventional fixed point (e.g.~\mbox{$\sigma(1)=1$}), 
allows one to straightforwardly apply the results of such theorems to colour 
flows. This is understood in eq.~(\ref{loopid}), and is the reason for the 
$n-1$ argument on the r.h.s.~of eq.~(\ref{countn}).\label{ft:cyc}}
\beq
\loopset\left(\sigma,I\right)=c\left(G(p)\right)\,.
\label{loopid}
\eeq
It is intuitively clear that this equation leads to eq.~(\ref{loopvscyc});
in fact, we shall show below that the analogy is even stricter.

The advantage of being able to establish a direct connection between
colour loops and alternating cycles lies in the ability to exploit
the growing (thanks to the interest in biology) number of mathematics 
results relevant to the latter. A first example is that of counting.
From definition 9.12 of ref.~\cite{Bona:2012}:
\begin{itemize}
\item
The number of $n$-element permutations with $k$ alternating cycles is
the Hultman number~\cite{Hultman:1999} $\hult{n}{k}$.
\end{itemize}
Therefore, thanks to eq.~(\ref{loopid}), we have
\beq
\sum_{\sigma\in P_n^\prime}
1\mydot\delta\Big(k,\nloops{\loopset(\sigma,I)}\Big)=
\hult{n-1}{k}\,,
\;\;\;\;\;\;\;\;
1\le k\le n\,.
\label{countn}
\eeq
In other words, the number of closed colour flows that correspond to
a given number ($k$) of $n$-gluon colour loops is equal to the Hultman 
number, computed at \mbox{$(n-1,k)$}. Fortunately, Hultman numbers are known 
in a closed form~\cite{Doignon:2007}
\beqn
\hult{n}{k}&=&\stirlingS1{n+2}{k}\left/\binomial{n+2}{2}\right. \,,
\;\;\;\;\;\;\;
n-k~{\rm odd}\,,
\label{hult1}
\\
\hult{n}{k}&=&0\,,
\phantom{\stirlingS1{n+2}{k}\left/\binomial{n+2}{2}\right.}
\;\;\;\;\;\;\,
n-k~{\rm even}\,,
\label{hult2}
\eeqn
where $[n\; k]$ is the unsigned Stirling number of the first kind, which
counts the number of $n$-element permutations with $k$ ordinary cycles. 
Note that from eq.~(\ref{countn}) one must have
\beq
\sum_{k=1}^n \hult{n-1}{k}=(n-1)!\,,
\label{hultsum}
\eeq
which can indeed be found with eqs.~(\ref{hult1}) and~(\ref{hult2}). Note,
also, that eqs.~(\ref{countn}) and~(\ref{hult2}) confirm that the number
of colour loops for any given flow has the same parity as the number of
gluons.

A theorem by Doignon and Labarre~\cite{Doignon:2007} (specifically,
theorem~8 there) is directly relevant to the counting of colour
loops. It reads as follows:
\begin{itemize}
\item
Let\footnote{As is standard in the mathematics of permutations,
$(\ldots)$ is the cycle notation. Furthermore, ref.~\cite{Doignon:2007} 
labels elements starting from zero, as opposed to one as is done here.} 
$\alpha=(0\ldots n-1)$. The mapping
\end{itemize}
\vskip -0.2truecm
\beqn
&&\!\!\!\!\!F:\Big\{\pi\in S_{n-1}|c(G(\pi))=k\Big\}\;\longrightarrow\;
\label{DLth81}
\\*
&&\phantom{aaaa}
\Big\{\sigma\in S(n)|c(\Gamma(\sigma))=k\&\exists\rho\in S(n):
c(\Gamma(\rho))=1\&\alpha=\rho\circ\sigma\Big\}\phantom{aaa}
\nonumber
\\*&&
\!\!\!\!\!F:\;\;\pi\;\longmapsto\;\stackrel{\circ}{\pi}
\label{DLth82}
\eeqn
\begin{itemize}
\item[]
is bijective.
\end{itemize}
As is customary, for the map $F$ one introduces the domain and codomain
(eq.~(\ref{DLth81})) and a symbol (\mbox{$\stackrel{\circ}{\pi}\equiv F(\pi)$})
for its result when acting on a specific element ($\pi$) of the domain
(eq.~(\ref{DLth82})).
As before, $c(G(\pi))$ is the number of alternating cycles of $\pi$,
while $c(\Gamma(\rho))$ is the number of ordinary cycles of $\rho$. $S_{n-1}$ 
is the set of $n$-object permutations; $S(n)$ is the set of factorizations
of a given $n$-cycle (in this case, $\alpha$) into some $n$-cycle
($\rho$) and some permutation ($\sigma$). By taking into account the 
different labelling conventions of ref.~\cite{Doignon:2007} and of the 
present paper, and by observing that
\beq
I_{\leftarrow}=\alpha\left(\text{\protect\cite{Doignon:2007}}\right)\,,
\eeq
the relationships between the quantities of eqs.~(\ref{DLth81}) 
and~(\ref{DLth82}) and those introduced here relevant to $\loopset(\sigma,I)$
are the following (see footnote~\ref{ft:cyc}):
\beqn
\sigma
&\longleftrightarrow&
\pi\left(\text{\protect\cite{Doignon:2007}}\right)\,,
\label{DLth8r1}
\\
\varsigma(\sigma,I)&\longleftrightarrow&
\sigma\,,\stackrel{\circ}{\pi}\left(\text{\protect\cite{Doignon:2007}}\right)\,,
\label{DLth8r2}
\\
\left(\sigma_{\rightarrow}\circ\sigmamo\right)^{-1}
&\longleftrightarrow&
\rho\left(\text{\protect\cite{Doignon:2007}}\right)\,.
\label{DLth8r3}
\eeqn
The bijective nature of the map $F$ and eq.~(\ref{loopid}) give a
formal proof of eq.~(\ref{loopvscyc}) (taking the property of
eq.~(\ref{linecompl}) into account).

We point out that ref.~\cite{Doignon:2007} need only introduce $\rho$,
and not the combination on the l.h.s.~of eq.~(\ref{DLth8r3}), which
in our case stems directly from the construction of colour loops.
The consistency of the two procedures follows from the following fact:
\begin{itemize}
\item
For any given $n$-element permutation $\sigma$, the quantity
$\sigma_{\rightarrow}\circ\sigmamo$ is an $n$-cycle.
\end{itemize}
{\em Proof:} first observe that cycles must be invariant under
a general index relabelling; in other words, label indices by $\sigma(i)$
rather than by $i$. Then\footnote{The $\oplus$ and $\ominus$ symbols
are normal addition and subtraction operators, augmented by a cyclicity
condition: $n\oplus 1=1$ and $1\ominus 1=n$.}
\beq
\sigma_{\rightarrow}\circ\sigmamo(\sigma(i))=\sigma(i\ominus 1)\,,
\eeq
whence
\beq
\left(\sigma_{\rightarrow}\circ\sigmamo\right)^k(\sigma(i))=
\sigma(i\ominus k)\,.
\eeq
Thus
\beq
\left(\sigma_{\rightarrow}\circ\sigmamo\right)^k(\sigma(i))=
\sigma(i)
\;\;\;\;
\Longleftrightarrow
\;\;\;\;
k=n
\;\;\;\;\;\;
\forall\,i\,,
\eeq
which concludes the proof.

We finally quote two theorems relevant to permutations that could be useful 
in dealing with closed colour flows. The first is reported as theorem 
9.9~of ref.~\cite{Bona:2012}, and stems from ref.~\cite{Christie:1996}.
It uses the concept of {\em block interchange distance} ${\rm bid}(\sigma)$
for a given permutation $\sigma$, defined as the smallest number of
block interchanges that sort $\sigma$ (i.e.~transform $\sigma$ into 
the identity). That theorem states that (in the language of this paper)
\beq
{\rm bid}(\sigma)=\frac{n-\nloops{\loopset(\sigma,I)}}{2}\,.
\label{th99}
\eeq
This gives an interesting geometric interpretation to the number of
colour loops which, as is explicit in eq.~(\ref{th99}), decreases
linearly with the block interchange distance\footnote{The block interchange
distance can be defined between two permutations, exploiting the same
property as in eq.~(\ref{linecompl}). Therefore, eq.~(\ref{th99}) holds
in the case of a generic closed colour flow, as it should for consistency.}.

The second theorem  alluded to before is reported as theorem 9.22~of 
ref.~\cite{Bona:2012}. Again using the language of this paper, it states that
\beq
\langle\nloops{\loopset}\rangle =
\left(\left\lfloor\frac{n+1}{2}\right\rfloor\right)^{-1}
+H_1(n-1)\,,
\label{loopH1}
\eeq
with $H_1$ the first harmonic number, and the average on the l.h.s.~is 
computed over all $n$-gluon flows. From eq.~(\ref{countn}) it then 
follows that
\beq
\left(\left\lfloor\frac{n+1}{2}\right\rfloor\right)^{-1}+H_1(n-1) =
\frac{1}{(n-1)!}\,\sum_{k=1}^n k\,\hult{n-1}{k}\,.
\eeq
Furthermore, eq.~(\ref{loopH1}) implies
\beq
\langle\nloops{\loopset}\rangle
\;\;\stackrel{n\to\infty}{\longrightarrow}\;\;
\log n\,.
\label{largen}
\eeq
Since $\max(\nloops{\loopset})=n$, eq.~(\ref{largen}) tells one that the 
larger $n$, the more unlikely that a flat random generation results
in a flow with a large number of colour loops. Conversely, a flat random 
generation should be relatively efficient in producing highly-suppressed 
colour contributions.

\subsection{Colour factors\label{sec:appfact}}
We now turn to the discussion of some features of the colour factors.
Firstly, the colour-flow matrix elements of eq.~(\ref{CFmatgluex})
on the diagonal (i.e.~$\sigmap=\sigma$) can be easily computed
by means of secondary flows, by employing eqs.~(\ref{Cnglueq})
and~(\ref{treeGGres}) since, for any $n$-gluon colour flow $\sigma$,
we have
\beq
\nloops{\loopset(\bsigmaqLarg{\cancel{s_k}},\bsigmaqRarg{\cancel{s_k}})}=
\left\{\begin{array}{ll}
n-k\phantom{aaaaaa}0\le k\le n-1\\
2\phantom{aaaaaaaaaa}k=n
\end{array}\right.\;\;\;\;\;\;\;\;
\forall\,s_k\in\Sset{n}{k}\,.
\eeq
Thus
\beqn
C(\sigma,\sigma)&=&\sum_{k=0}^n\frac{(-1)^k}{2^n}\binomial{n}{k}\,N^{-k}
\left[\Theta(k\le n-1)\,N^{n-k}+\delta_{kn}N^2\right]
\nonumber\\&=&
\frac{1}{2^nN^n}\,(N^2-1)\,\Big((N^2-1)^{n-1}+(-1)^n\Big)\,.
\label{Cdiagres}
\eeqn
The rightmost side of eq.~(\ref{Cdiagres}) exhibits the usual factor
\beq
f_N=N^2-1\,.
\label{fNdef}
\eeq
On top of that, it is easy to show analytically for any $n$ that
\beq
C(\sigma,\sigma)=(N^2-1)\,(N^2-2)\,q(N)/N^n
\;\;\;\;\Longleftrightarrow\;\;\;\;n~{\rm odd}\,,
\label{Nmt}
\eeq
for some polynomial $q(N)$. We stress that while the factorization
of $f_N$ occurs also for off-diagonal elements, this is {\em not} the
case for the factor \mbox{$N^2-2$}.

As is the case for the diagonal colour-flow matrix elements,
for the off-diagonal ones the emergence of the factor $f_N$ is not
immediately apparent, the r.h.s.~of eq.~(\ref{treeGGres}) being a 
linear combinations of monomials $N^p$. It is the contributions of 
the $U(1)$ gluons that add and subtract suitable terms to the leading 
$N^n$ monomial, so that $f_N$ eventually appears. This poses an
interesting question, namely: the matrix elements summed over flows, with 
the colour factors computed with traces of $\lambda$ matrices, manifestly 
factor out $f_N$ because each of the products of two traces of $\lambda$ 
does so. On the other hand, as has been shown in eq.~(\ref{matelfingg}),
such matrix elements can be computed by simply counting colour loops, and
by ignoring $U(1)$-gluon contributions. Therefore, in view of the fact that
it is those contributions that cause the emergence of the $f_N$ factor
{\em prior to summing} over flows, where does $f_N$ come from if 
eq.~(\ref{matelfingg}) is employed?

In order to answer that question, we re-write eq.~(\ref{matelfingg}) more 
compactly as follows:
\beqn
\ampsqn&=&\frac{1}{2^n}\,\sum_{i,j=1}^{(n-1)!}\ampsqn_{ij}\,,
\label{matelfingg1}
\\
\ampsqn_{ij}&=&
N^{\nloops{\loopset_{ij}}}\,
\ampCS_i^{(n)^\star}\ampCSn_j\,,
\label{matelfingg2}
\eeqn
i.e.~we temporarily denote the flows by integer indices. Next, consider 
one of the contributions to eq.~(\ref{matelfingg1}) stemming from summing
over columns after choosing a row -- to be definite, take the $i=1$ row
\beq
X=\frac{1}{\ampCS_1^{(n)^\star}}\sum_{j=1}^{(n-1)!}\ampsqn_{1j}=
\sum_{j=1}^{(n-1)!}N^{\nloops{\loopset_{1j}}}\ampCSn_j\,.
\label{Xdef}
\eeq
In view of eq.~(\ref{countn}) and of the properties of the Hultman
numbers in eqs.~(\ref{hult1}) and~(\ref{hult2}), eq.~(\ref{Xdef})
can be re-cast as follows:
\beq
X={\sum_{k=n}^{k_{\min}}}^{~\bigstar}\,N^k\!\!
\sum_{j=\Shult{n-1}{k+2}+1}^{\Shult{n-1}{k}}\ampCSn_j\,,
\label{Xform}
\eeq
generally after a relabelling of the flows, so that those with the same
number of colour loops are contiguous. The $\bigstar$ symbol restricts the
sum over $k$ to run over values whose parity is that same as that of $n$. 
In eq.~(\ref{Xform}) we have used the sequence $\Shult{n}{k}$ that is 
monotonically increasing for decreasing $k$
\beqn
\Shult{n}{k}&=&\sum_{j=n+1}^k\hult{n}{j}\,,
\;\;\;\;\;\;\;
1\le k\le n+1\,,
\\
\Shult{n}{k}&=&0\,,
\phantom{aaaaaaaaaaaaaaa}
k\ge n+2\,,
\eeqn
and
\beq
k_{\min}=2-{\rm mod}(n,2)\,.
\eeq
Note that, thanks to eqs.~(\ref{hult2}) and~(\ref{hultsum})
\beq
\Shult{n-1}{k_{\min}}=(n-1)!\,,
\eeq
so that the sums on the r.h.s.'s of eqs.~(\ref{Xdef}) and~(\ref{Xform})
indeed contain the same number of terms.
Because of the dual Ward identities, there are many ways to collect
$(n-1)$ amplitudes whose sum is equal to zero. However, it is not guaranteed
that the corresponding colour factors will all be equal to each other. In 
other words, for some $(n-1)$ flows labelled by \mbox{$j_1\ldots j_{n-1}$},
there might not exist a value $k$ such that
\beq
j_1\ldots j_{n-1}\;\in\;\Big[\Shult{n-1}{k+2}+1\,,\Shult{n-1}{k}\Big]\,.
\eeq
Therefore, the dual Ward identity
\beq
\ampCSn_{j_1}+\ldots+\ampCSn_{j_{n-1}}=0
\label{DWI3}
\eeq
cannot be easily exploited in eq.~(\ref{Xform}). On the other hand,
eq.~(\ref{DWI3}) implies
\beq
\sum_{j=1}^{(n-1)!}\ampCSn_j=0\,,
\label{DWIglob}
\eeq
and therefore
\beq
\sum_{j=\Shult{n-1}{k_{\min}+2}+1}^{\Shult{n-1}{k_{\min}}}\ampCSn_j=
-\sum_{k=n}^{k_{\min}+2}\sum_{j=\Shult{n-1}{k+2}+1}^{\Shult{n-1}{k}}
\,\ampCSn_j\,,
\label{DWI4}
\eeq
whence
\beq
X=\sum_{k=n}^{k_{\min}+2}
\left(N^k-N^{k_{\min}}\right)\!\!
\sum_{j=\Shult{n-1}{k+2}+1}^{\Shult{n-1}{k}}\ampCSn_j\,.
\label{Xform2}
\eeq
One finally exploits the fact that, for any $n$ and $k$, one can write
\beqn
k~{\rm even:}\;\;\;\;\;\;
N^k-N^{k_{\min}}&=&N^2\left(N^{2p}-1\right)\,,
\\
k~{\rm odd:}\;\;\;\;\;\;
N^k-N^{k_{\min}}&=&N\left(N^{2p}-1\right)\,.
\eeqn
For any integer $p$
\beq
N^{2p}-1=\left(N^2-1\right)\sum_{m=0}^{p-1}N^{2m}\,,
\label{N2p}
\eeq
which shows that $f_N$ indeed factors out of eq.~(\ref{Xform2}),
as was expected. 

It is also easy to see that eq.~(\ref{DWIglob}) is not only a sufficient
condition for the emergence of $f_N$, but also a necessary one. Indeed,
if the polynomial $X$ of $N$ factors out $f_N$, then $X(N=1)=0$. Thus,
from eq.~(\ref{Xdef})
\beq
X(N=1)=\left.\sum_{j=1}^{(n-1)!}
N^{\nloops{\loopset_{1j}}}\ampCSn_j\right|_{N=1}=
\sum_{j=1}^{(n-1)!}\ampCSn_j=0\,,
\label{XNeq1}
\eeq
which proves the statement above.

These arguments about $f_N$ allow us to point out that, while the fundamental
and the flow representations are equivalent for flow-summed gluon-only
quantities, this is not true in the intermediate steps of any computation.
Whether this has any practical consequences depends on which role such 
intermediate steps play in attaining the final result. An example relevant
to the case discussed here is the following: suppose one performs the
sum over flows by means of Monte Carlo methods. The difference between
the results obtained with the fundamental (where $U(1)$-gluon contributions
are kept non-null) and the flow representation will vanish 
only in the infinite-statistics limit. At finite statistics,
it is likely that the flow representation will induce a larger number
of cancellations among different terms, because this is the only manner
in which the factor $f_N$ can arise. In order to increase the efficiency,
then, it is wise to set up a flow-representation computation in a way
that incorporates the dual Ward identities exactly and in each step of
the procedure.

\section{Equivalence of closed colour flows and dipole 
graphs\label{sec:flvsdg}}
In this appendix, we sketch the proof of the equivalence between closed 
colour flows and dipole graphs that underpins eq.~(\ref{qgsoftsumdip}).
Although we are unable to prove formally this equivalence for any number
of quarks and gluons, we show a possible way to do this. We stress that,
in all the cases we have explicitly dealt with, the equivalence has
been found to hold true. We also show that, should the equivalence
break down for some values of $q$ and $n$, there is an alternative
whereby one still employs eq.~(\ref{qgsoftsumdip}), by means of a
minor extension of the definition of dipole graphs.

We start by observing that the radiation patterns of eq.~(\ref{eiksumLdip}) 
can be represented as a matrix, whose row and column indices are the 
positions of the elements in the colour loop, and whose entries are equal 
to either $+1$ or $-1$, according to the sign in front of the eikonal
associated with those elements; the diagonal elements are set equal
to zero (since they correspond to self eikonals). This matrix being 
symmetric, we limit ourselves to writing its upper right triangle. 
Therefore
\beqn
M\!\left(\eiksum_\ell(\bgammap,\bgamma)\right)=
\begin{blockarray}{ccccccc}
    e_1 & e_2 & e_3 & \cdots & e_{2m_\ell-1} & e_{2m_\ell} & \\
\begin{block}{(cccccc)c}
 0 & +1 & -1 & \ldots & -1 & +1 & e_1 \\
   &  0 & +1 & \ldots & +1 & -1 & e_2 \\
   &    &  0 & \ldots & -1 & +1 & e_3 \\
   &    &    & \ddots & \vdots & \vdots & \cdots \\
   & & & & 0 & +1 & e_{2m_\ell-1}\\
   & & & & & 0 & e_{2m_\ell} \\
\end{block}
\end{blockarray}
\label{eikpattM}
\eeqn
so that
\beq
\eiksum_\ell(\bgammap,\bgamma)=\half
\sum_{i,j=1}^{2m_\ell}M\!\left(\eiksum_\ell(\bgammap,\bgamma)\right)_{ij}
\eik{e_i}{e_j}\,.
\label{eiksummat}
\eeq
Equation~(\ref{eikpattM}) shows that the two sums on the r.h.s.~of 
eq.~(\ref{eiksumLdip}) coincide with those of the matrix elements with 
like signs on the northwest-southeast diagonals.

The basic idea is that of exploiting the matrix representation of
eq.~(\ref{eikpattM}) for $\eiksum^+$ and $\eiksum^-$ separately, 
and to show that it can be written as a sum of matrices that are
representations of dipole graphs, eq.~(\ref{matdip}).

In order to do so, we start from a case that features only quarks
and antiquarks. Therefore, from eq.~(\ref{sumellm})
\beq
2q=\sum_\ell 2m_\ell\,.
\eeq
The \mbox{$(2q)\times (2q)$} matrices\footnote{Henceforth, in order 
to simplify the notation we often understand the closed flow
$(\bgammap,\bgamma)$, which is given and fixed.}
\beq
M\!\left(\eiksum^+\right)=M\!\left(\sum_\ell\eiksum_\ell^+\right)\,,
\;\;\;\;\;\;\;\;
M\!\left(\eiksum^-\right)=M\!\left(\sum_\ell\eiksum_\ell^-\right)\,,
\eeq
are block matrices constructed by starting from the 
\mbox{$(2m_\ell)\times (2m_\ell)$} matrices
\beq
M\!\left(\eiksum_\ell^+\right)\,,
\;\;\;\;\;\;\;\;
M\!\left(\eiksum_\ell^-\right)\,,
\eeq
and by embedding them (as blocks) in a \mbox{$(2q)\times (2q)$} null matrix, 
so that we can formally write
\beq
M\!\left(\sum_\ell\eiksum_\ell^+\right)=
\sum_\ell M\!\left(\eiksum_\ell^+\right)\,,
\;\;\;\;\;\;\;\;
M\!\left(\sum_\ell\eiksum_\ell^-\right)=
\sum_\ell M\!\left(\eiksum_\ell^-\right)\,,
\label{MsumeqsumM}
\eeq
This property stems from item iii. in sect.~\ref{sec:loops}:
for any given quark or antiquark, there exists a single colour
loop that contains its (anti)colour label.
Equation~(\ref{MsumeqsumM}) allows one to work with an individual
$M(\eiksum_\ell^\pm)$ matrix. From eq.~(\ref{eikpattM}), we see that
\beqn
M\!\left(\eiksum_\ell^+\right)_{ij}&=&
\delta\big(1,{\rm mod}(i-j,2)\big)\,,
\\*
M\!\left(\eiksum_\ell^-\right)_{ij}&=&
\delta\big(0,{\rm mod}(i-j,2)\big)\left(1-\delta_{ij}\right)\,,
\label{Mepmres}
\eeqn
so that, in keeping with eqs.~(\ref{eiksumLdip2}) and~(\ref{eiksummat})
\beq
M\!\left(\eiksum_\ell\right)_{ij}=
M\!\left(\eiksum_\ell^+\right)_{ij}-
M\!\left(\eiksum_\ell^-\right)_{ij}\,,
\label{MeqMpMm}
\eeq
and therefore
\beq
\eiksum_\ell^\pm(\bgammap,\bgamma)=\half
\sum_{i,j=1}^{2m_\ell}M\!\left(\eiksum_\ell^\pm(\bgammap,\bgamma)\right)_{ij}
\eik{e_i}{e_j}\,,
\label{eiksummatpm}
\eeq
which are the analogues of eq.~(\ref{eiksummat}).
Let us first consider the case of $\eiksum_\ell^+$. We define 
the following $m_\ell$ matrices:
\beq
M\!\left(\eiksum_{\ell,k}^+\right)_{ij}=
\delta\big(j,1+{\rm mod}(1-k-i,2m_\ell)\big)\,,
\;\;\;\;\;\;
1\le k\le 2m_\ell-1\,,\;\;\;\;\;\;k~{\rm odd}\,.
\label{Mpkdef}
\eeq
These matrices are symmetric, traceless, and have the following
properties:
\beqn
&&\sum_k M\!\left(\eiksum_{\ell,k}^+\right)=
M\!\left(\eiksum_\ell^+\right)\,,
\label{sEpEpk}
\\*
&&\sum_{i=1}^{2m_\ell} M\!\left(\eiksum_{\ell,k}^+\right)_{ij}=
\sum_{j=1}^{2m_\ell} M\!\left(\eiksum_{\ell,k}^+\right)_{ij}=1\,.
\label{srowcolp}
\eeqn
It should be clear, then, that each of the matrices defined
in eq.~(\ref{Mpkdef}) represents what in sect.~\ref{sec:ccfvsdp} 
has been called a candidate subset of a dipole graph. This essentially
concludes the argument, since each of the matrices $M(\eiksum_{\ell,k}^+)$ 
can be transformed into a \mbox{$(2q)\times (2q)$} block matrix as was
done above, so that
\beq
M\!\left(\eiksum^+\right)=
\sum_\ell\sum_k M\!\left(\eiksum_{\ell,k}^+\right)=
\sum_{k_1,\ldots k_{\nloops{\loopset}}}
\sum_\ell M\!\left(\eiksum_{\ell,\{k_\ell\}}^+\right)\,.
\label{pflowvsdip}
\eeq
In the rightmost side of this equation we have exchanged the order
of the summations over the loops and the index that identifies the
individual matrices introduced in eq.~(\ref{Mpkdef}).
In doing so, we need to have one such index for each block; in other words,
a choice of the values of the indices \mbox{$k_1,\ldots k_{\nloops{\loopset}}$}
corresponds to choosing one of the matrices of eq.~(\ref{Mpkdef}) 
for each of the blocks. In this way, for any
given \mbox{$k_1,\ldots k_{\nloops{\loopset}}$} set of values, the matrix
\beq
M\!\left(\eiksum_{\{k_\ell\}}^+\right):=
\sum_\ell M\!\left(\eiksum_{\ell,\{k_\ell\}}^+\right)
\label{Mflowdip}
\eeq
is a matrix representative of a dipole graph. In summary, by replacing
eq.~(\ref{pflowvsdip}) into (the positive-eikonal part of)
eq.~(\ref{eiksummat}) one shows that 
the radiation pattern of $\eiksum^+$ can be written as the sum
of the radiation patterns of the dipoles $\eiksum_{\{k_\ell\}}^+$.
Thus, eq.~(\ref{pflowvsdip}) is the matrix form of eq.~(\ref{eikpvsdip}),
and also proves that all the $c^+$ coefficients in the latter equation
are equal to one.

It is important to stress again that not only is eq.~(\ref{Mpkdef}) not 
expected to be unique, but also that the decomposition of the radiation
pattern of a closed flow into its positive- and negative-definite
components is not mandatory. Therefore, it is in general possible to achieve 
a representation of a closed flow in terms of dipoles, with the number of 
the latter smaller than those in eq.~(\ref{numofdip})\footnote{This will
of course also change the individual values of the $c^\pm$ coefficients.};
we shall give an explicit example of this fact later in this appendix.
However, this is not important here, since what is crucial is
that the above proves that there is {\em at least} a way in which
this representation can be achieved.

Unfortunately, things are not as simple for the radiation patterns
of eikonals with negative signs, eq.~(\ref{Mepmres}), since we have not 
been able to find matrices analogous to those of eq.~(\ref{Mpkdef})
that have all the properties of the latter. We note in particular that, 
when $q$ is an odd number, the analogue of eq.~(\ref{sEpEpk}) cannot be true:
some contributions must have opposite signs, so as to achieve a cancellation.
This is because the radiation pattern $\eiksum^-$ arises necessarily
from quark-quark and antiquark-antiquark eikonals (owing to the properties
of colour loops, sect.~\ref{sec:loops}), but never from quark-antiquark
eikonals. For this to correspond to a set of dipole graphs fulfilling
the analogue of eq.~(\ref{sEpEpk}), each such graph would therefore
contain $q/2$ pairs of quark indices and $q/2$ pairs of antiquark indices,
which is impossible since $q/2$ is not in general an integer number.

Having said that, we observe that the \mbox{$m_\ell-1$} matrices
\beq
M\!\left(\eiksum_{\ell,k}^-\right)_{ij}=
\delta\big(j,1+{\rm mod}(i+k-1,2m_\ell)\big)\,,
\;\;\;\;\;\;
2\le k\le 2m_\ell-2\,,\;\;\;\;\;\;k~{\rm even}\,,\phantom{aaa}
\label{Mmkdef}
\eeq
are traceless and fulfill properties identical to those in 
eqs.~(\ref{sEpEpk}) and~(\ref{srowcolp}), namely
\beqn
&&\sum_k M\!\left(\eiksum_{\ell,k}^-\right)=
M\!\left(\eiksum_\ell^-\right)\,,
\label{sEmEmk}
\\*
&&\sum_{i=1}^{2m_\ell} M\!\left(\eiksum_{\ell,k}^-\right)_{ij}=
\sum_{j=1}^{2m_\ell} M\!\left(\eiksum_{\ell,k}^-\right)_{ij}=1\,,
\label{srowcolm}
\eeqn
but {\em not being symmetric} (except when $k=m_\ell$; thus, this happens
only if $m_\ell$ is even) cannot be representative of (a subset of) any dipole 
graph. However, by means of eq.~(\ref{Mmkdef}) we can construct the following 
set of $\floor*{m_\ell/2}$ matrices:
\beqn
M^\prime\!\left(\eiksum_{\ell,k}^-\right)&=&
\frac{1}{1+\delta_{km_\ell}}\left(
M\!\left(\eiksum_{\ell,k}^-\right)+
M\!\left(\eiksum_{\ell,2m_\ell-k}^-\right)\right)
\nonumber
\\*&&\phantom{aaaaaaaaaaaaaaaaaa}
2\le k\le m_\ell\,,\;\;\;\;\;\;k~{\rm even}\,.
\label{Mpndef}
\eeqn
These matrices are traceless and symmetric, and obey
\beqn
&&\sum_k M^\prime\!\left(\eiksum_{\ell,k}^-\right)=
M^\prime\!\left(\eiksum_\ell^-\right)\,,
\label{psEmEmk}
\\*
&&\sum_{i=1}^{2m_\ell} M^\prime\!\left(\eiksum_{\ell,k}^-\right)_{ij}=
\sum_{j=1}^{2m_\ell} M^\prime\!\left(\eiksum_{\ell,k}^-\right)_{ij}=
\delta_{km_\ell}+2\left(1-\delta_{km_\ell}\right)\,.
\label{psrowcolm}
\eeqn
This implies that they can be used in a dipole-like representation
of the $\eiksum^-$ radiation pattern, provided that one extends the definition
of dipole graphs to include the cases where each quark and each antiquark
label appears in exactly two (as opposed to one) pairs. This corresponds
to allowing, in a PSMC implementation, any quark and any antiquark to 
have two colour partners. 

The procedure described so far cannot be directly applied in the 
presence of gluons, since a gluon may appear in two different loops, 
which renders invalid the construction of the block matrices as was
done thus far. However, one may treat the colour and anticolour labels of 
a gluon as a quark and an antiquark. One then proceeds as above and,
at the very end, merges the two rows and columns relevant to 
each gluon (by summing their contents). The resulting rows 
and columns will sum to two, rather than to one (to four rather than
to two in the case of the $M^\prime(\eiksum_{\ell,k}^-)$ matrices), with the 
symmetry and traceless properties obviously unspoilt. One thus obtains 
again matrices that represent dipole graphs (or generalized dipole
graphs, in the case of $\eiksum^-$) that include gluons, {\em except} 
for the following important detail. By treating, during the construction, 
the colour and anticolour indices of a gluon as independent from each 
other, one obtains self-eikonals which are not on the diagonal (see
e.g.~eq.~(\ref{eikpCL2ex1})). However, such a case is easy
to diagnose, because after merging the rows and columns it results
in an entry equal to two on the diagonal. When this happens, that gluon
has to be removed, and thus interpreted as a self-connected gluon.
After the removal, one is left with a dipole graph of dimension smaller
than the original one, which is what has already been discussed
in sect.~\ref{sec:scg}.

\noindent
$\bullet$ {\em Example.} We consider the case of the closed flow
$(\bgamma_{0,1},\bgamma_{0,2})$ in table~\ref{tab:loopsets_qqgg}.
Equation~(\ref{eikpattM}) reads (there is a single loop, so the
index $\ell$ is omitted)
\beqn
M\!\left(\eiksum(\bgamma_{0,1},\bgamma_{0,2})\right)=
\begin{blockarray}{rrrrrrr}
   \Qa & \bt & \bo & \Qb & \bt & \bo & \\
\begin{block}{(rrrrrr)r}
 0 & +1 & -1 & +1 & -1 & +1 & \Qa \\
   &  0 & +1 & -1 & +1 & -1 & \bt \\
   &    &  0 & +1 & -1 & +1 & \bo \\
   &    &    &  0 & +1 & -1 & \Qb \\
   &    &    &    &  0 & +1 & \bt \\
   &    &    &    &    & 0  & \bo \\
\end{block}
\end{blockarray}
\label{eikMCL2ex1}
\eeqn
and, equivalently
\beqn
\eiksum^+(\bgamma_{0,1},\bgamma_{0,2})&=&
\eik{\Qa}{\bt}+
\eik{\bt}{\bo}+
\eik{\bo}{\Qb}+
\eik{\Qb}{\bt}+
\eik{\bt}{\bo}+
\nonumber
\\*&&
\eik{\Qa}{\Qb}+
\eik{\bt}{\bt}+
\eik{\bo}{\bo}+
\nonumber
\\*&&
\eik{\Qa}{\bo}\,,
\label{eikpCL2ex1}
\\*
\eiksum^-(\bgamma_{0,1},\bgamma_{0,2})&=&
\eik{\Qa}{\bo}+
\eik{\bt}{\Qb}+
\eik{\bo}{\bt}+
\eik{\Qb}{\bo}+
\nonumber
\\*&&
\eik{\Qa}{\bt}+
\eik{\bt}{\bo}\,.
\label{eikmCL2ex1}
\eeqn
The three lines of eq.~(\ref{eikpCL2ex1}) correspond to the three
northwest-southeast diagonals (from left to right) with positive signs,
and the two lines of eq.~(\ref{eikmCL2ex1}) to the two diagonals (from left 
to right) with negative signs, of the matrix in eq.~(\ref{eikMCL2ex1}).
The sum of the r.h.s.'s of eqs.~(\ref{eikpCL2ex1}) and~(\ref{eikmCL2ex1})
is equal to the r.h.s.~of eq.~(\ref{eikf2ex1}), as it should
by construction. In this example, we have
\beq
q=1\,,\;\;\;\;
n=2\,,\;\;\;\;
m=3\,,\;\;\;\;
n_{g^\star}=2\,,
\eeq
and indeed, in keeping with eqs.~(\ref{posel}) and~(\ref{negel}),
\beq
m^2=9\,,\;\;\;\;
m(m-1)=6\,,
\eeq
are the number of terms in eq.~(\ref{eikpCL2ex1}) and the number of 
terms in eq.~(\ref{eikmCL2ex1}), respectively.

Since $n_{g^\star}\ne 0$, we expect the radiation pattern $\eiksum^+$
to be given in terms of dipoles graphs at least one of which will feature
a number of gluons smaller than two (the ``missing'' gluons being
self-interacting ones). Furthermore, we shall see that the radiation pattern 
$\eiksum^-$ can also be expressed in terms of dipole graphs, notwithstanding
the lack of formal proof that this is possible. However, before going
into that, we observe that by applying the merging-line procedure advocated 
above to the each of the gluons\footnote{That is, row/column number 2 and 3
is merged with row/column number 5 and 6, respectively.} that appear 
in the matrix in eq.~(\ref{eikMCL2ex1}) we obtain the following:
\beqn
\begin{blockarray}{rrrrrrr}
   \Qa & \bt & \bo & \Qb & \bt & \bo & \\
\begin{block}{(rrrrrr)r}
 0 & +1 & -1 & +1 & -1 & +1 & \Qa \\
   &  0 & +1 & -1 & +1 & -1 & \bt \\
   &    &  0 & +1 & -1 & +1 & \bo \\
   &    &    &  0 & +1 & -1 & \Qb \\
   &    &    &    &  0 & +1 & \bt \\
   &    &    &    &    & 0  & \bo \\
\end{block}
\end{blockarray}
\;\;\;\longrightarrow\;\;\;
\begin{blockarray}{rcccr}
   \Qa & \bt & \bo & \!\!\!\Qb  & \\
\begin{block}{(rccc)r}
 0 &  0 &  0 &  1 &  \Qa \\
   &  2 &  0 &  0 &  \bt \\
   &    &  2 &  0 &  \bo \\
   &    &    &  0 &  \Qb \\
\end{block}
\end{blockarray}
\,.
\label{eikMCL2ex1mer}
\eeqn
As has been anticipated in the discussion given before, in the matrix
on the r.h.s.~of eq.~(\ref{eikMCL2ex1mer}) we see the presence of two
elements equal to two on the diagonal, which is equivalent to saying
that there are two self-connected gluons. Indeed, it is immediate to
see that this matrix is representative of the dipole graph $\Gamma_6$
of fig.~\ref{fig:qqgg_dipoles}, with the corresponding radiation
pattern of eq.~(\ref{eikf2ex1}).

We now consider the matrices of eq.~(\ref{Mpkdef}), which we construct
as explained before, namely by first treating the colour and anticolour
of each gluon as a quark and an antiquark, and by merging the relevant
lines afterwards. Thus
\beqn
M\!\left(\eiksum_1^+\right)=
\begin{blockarray}{ccccccr}
   \Qa & \bt & \bo & \Qb & \bt & \bo & \\
\begin{block}{(cccccc)r}
 0 &  1 &  0 &  0 &  0 &  0 & \Qa \\
   &  0 &  0 &  0 &  0 &  0 & \bt \\
   &    &  0 &  0 &  0 &  1 & \bo \\
   &    &    &  0 &  1 &  0 & \Qb \\
   &    &    &    &  0 &  0 & \bt \\
   &    &    &    &    & 0  & \bo \\
\end{block}
\end{blockarray}
\;\longrightarrow\;
\begin{blockarray}{ccccr}
   \Qa & \bt & \bo & \Qb  & \\
\begin{block}{(cccc)r}
 0 &  1 &  0 &  0 &  \Qa \\
   &  0 &  0 &  1 &  \bt \\
   &    &  2 &  0 &  \bo \\
   &    &    &  0 &  \Qb \\
\end{block}
\end{blockarray}
\,,\phantom{aaa}
\label{matp1}
\eeqn
\beqn
M\!\left(\eiksum_3^+\right)=
\begin{blockarray}{ccccccr}
   \Qa & \bt & \bo & \Qb & \bt & \bo & \\
\begin{block}{(cccccc)r}
 0 &  0 &  0 &  1 &  0 &  0 & \Qa \\
   &  0 &  1 &  0 &  0 &  0 & \bt \\
   &    &  0 &  0 &  0 &  0 & \bo \\
   &    &    &  0 &  0 &  0 & \Qb \\
   &    &    &    &  0 &  1 & \bt \\
   &    &    &    &    &  0 & \bo \\
\end{block}
\end{blockarray}
\;\longrightarrow\;
\begin{blockarray}{ccccr}
   \Qa & \bt & \bo & \Qb  & \\
\begin{block}{(cccc)r}
 0 &  0 &  0 &  1 &  \Qa \\
   &  0 &  2 &  0 &  \bt \\
   &    &  0 &  0 &  \bo \\
   &    &    &  0 &  \Qb \\
\end{block}
\end{blockarray}
\,,\phantom{aaa}
\label{matp3}
\eeqn
\beqn
M\!\left(\eiksum_5^+\right)=
\begin{blockarray}{ccccccr}
   \Qa & \bt & \bo & \Qb & \bt & \bo & \\
\begin{block}{(cccccc)r}
 0 &  0 &  0 &  0 &  0 &  1 & \Qa \\
   &  0 &  0 &  0 &  1 &  0 & \bt \\
   &    &  0 &  1 &  0 &  0 & \bo \\
   &    &    &  0 &  0 &  0 & \Qb \\
   &    &    &    &  0 &  0 & \bt \\
   &    &    &    &    &  0 & \bo \\
\end{block}
\end{blockarray}
\;\longrightarrow\;
\begin{blockarray}{ccccr}
   \Qa & \bt & \bo & \Qb  & \\
\begin{block}{(cccc)r}
 0 &  0 &  1 &  0 &  \Qa \\
   &  2 &  0 &  0 &  \bt \\
   &    &  0 &  1 &  \bo \\
   &    &    &  0 &  \Qb \\
\end{block}
\end{blockarray}
\,.\phantom{aaa}
\label{matp5}
\eeqn
We see that the rightmost matrices in eqs.~(\ref{matp1}), (\ref{matp3}), 
and~(\ref{matp5}) correspond to the dipole graphs $\Gamma_4$, $\Gamma_3$,
and $\Gamma_5$ of fig.~\ref{fig:qqgg_dipoles}, respectively, with the first 
and the third being associated
with one self-connected gluon (with label $\bo$ and $\bt$ respectively).
This concludes the explicit construction of the r.h.s.~of 
eq.~(\ref{eikpvsdip}); as was expected from eq.~(\ref{Mpkdef}), all 
coefficients $c^+$ that appear in the former equation are equal to one.

We now turn to the radiation pattern $\eiksum^-$. We construct a regular 
dipole representation, thereby avoiding the use of the matrices
in eq.~(\ref{Mmkdef}). Explicitly
\beqn
M\!\left(\eiksum_{b_1}^-\right)=
\begin{blockarray}{ccccccr}
   \Qa & \bt & \bo & \Qb & \bt & \bo & \\
\begin{block}{(cccccc)r}
 0 &  1 &  0 &  0 &  0 &  0 & \Qa \\
   &  0 &  0 &  0 &  0 &  0 & \bt \\
   &    &  0 &  0 &  1 &  0 & \bo \\
   &    &    &  0 &  0 &  1 & \Qb \\
   &    &    &    &  0 &  0 & \bt \\
   &    &    &    &    &  0 & \bo \\
\end{block}
\end{blockarray}
\;\longrightarrow\;
\begin{blockarray}{ccccr}
   \Qa & \bt & \bo & \Qb  & \\
\begin{block}{(cccc)r}
 0 &  1 &  0 &  0 &  \Qa \\
   &  0 &  1 &  0 &  \bt \\
   &    &  0 &  1 &  \bo \\
   &    &    &  0 &  \Qb \\
\end{block}
\end{blockarray}
\,,\phantom{aaa}
\label{matn1}
\eeqn
\beqn
M\!\left(\eiksum_{b_2}^-\right)=
\begin{blockarray}{ccccccr}
   \Qa & \bt & \bo & \Qb & \bt & \bo & \\
\begin{block}{(cccccc)r}
 0 &  0 &  0 &  0 &  1 &  0 & \Qa \\
   &  0 &  0 &  0 &  0 &  1 & \bt \\
   &    &  0 &  1 &  0 &  0 & \bo \\
   &    &    &  0 &  0 &  0 & \Qb \\
   &    &    &    &  0 &  0 & \bt \\
   &    &    &    &    & 0  & \bo \\
\end{block}
\end{blockarray}
\;\longrightarrow\;
\begin{blockarray}{ccccr}
   \Qa & \bt & \bo & \Qb  & \\
\begin{block}{(cccc)r}
 0 &  1 &  0 &  0 &  \Qa \\
   &  0 &  1 &  0 &  \bt \\
   &    &  0 &  1 &  \bo \\
   &    &    &  0 &  \Qb \\
\end{block}
\end{blockarray}
\,,\phantom{aaa}
\label{matn2}
\eeqn
\beqn
M\!\left(\eiksum_{b_3}^-\right)=
\begin{blockarray}{ccccccr}
   \Qa & \bt & \bo & \Qb & \bt & \bo & \\
\begin{block}{(cccccc)r}
 0 &  0 &  1 &  0 &  0 &  0 & \Qa \\
   &  0 &  0 &  1 &  0 &  0 & \bt \\
   &    &  0 &  0 &  0 &  0 & \bo \\
   &    &    &  0 &  0 &  0 & \Qb \\
   &    &    &    &  0 &  1 & \bt \\
   &    &    &    &    &  0 & \bo \\
\end{block}
\end{blockarray}
\;\longrightarrow\;
\begin{blockarray}{ccccr}
   \Qa & \bt & \bo & \Qb  & \\
\begin{block}{(cccc)r}
 0 &  0 &  1 &  0 &  \Qa \\
   &  0 &  1 &  1 &  \bt \\
   &    &  0 &  0 &  \bo \\
   &    &    &  0 &  \Qb \\
\end{block}
\end{blockarray}
\,,\phantom{aaa}
\label{matn3}
\eeqn
\beqn
M\!\left(\eiksum_{b_4}^-\right)=
\begin{blockarray}{ccccccr}
   \Qa & \bt & \bo & \Qb & \bt & \bo & \\
\begin{block}{(cccccc)r}
 0 &  1 &  0 &  0 &  0 &  0 & \Qa \\
   &  0 &  0 &  0 &  0 &  0 & \bt \\
   &    &  0 &  1 &  0 &  0 & \bo \\
   &    &    &  0 &  0 &  0 & \Qb \\
   &    &    &    &  0 &  1 & \bt \\
   &    &    &    &    &  0 & \bo \\
\end{block}
\end{blockarray}
\;\longrightarrow\;
\begin{blockarray}{ccccr}
   \Qa & \bt & \bo & \Qb  & \\
\begin{block}{(cccc)r}
 0 &  1 &  0 &  0 &  \Qa \\
   &  0 &  1 &  0 &  \bt \\
   &    &  0 &  1 &  \bo \\
   &    &    &  0 &  \Qb \\
\end{block}
\end{blockarray}
\,,\phantom{aaa}
\label{matn4}
\eeqn
so that
\beqn
M\!\left(\eiksum^-\right)&=&
M\!\left(\eiksum_{b_1}^-\right)+
M\!\left(\eiksum_{b_2}^-\right)+
M\!\left(\eiksum_{b_3}^-\right)-
M\!\left(\eiksum_{b_4}^-\right)
\label{Meq4M}
\\*&=&
M\!\left(\eiksum_{b_1}^-\right)+
M\!\left(\eiksum_{b_3}^-\right)\,,
\label{Meq2M}
\eeqn
with the matrices on the r.h.s.~understood to be those obtained
by line-merging.

The present case being derived from that of a $3$-quark-line process, 
we know (see the comment before eq.~(\ref{Mmkdef})) that the equivalent 
dipole graphs must include quark-antiquark eikonals. These are the elements
of the matrices on the l.h.s.~of eqs.~(\ref{matn1})--(\ref{matn3}) 
whose row and column indices have opposite parities. Since these
eikonals cannot stem from $\eiksum^-$, they are subtracted by the
matrix on the l.h.s.~of eq.~(\ref{matn4}), whose elements are all of
this kind. We point out that while this implies that the $c^-$ 
coefficients on the r.h.s.~of eq.~(\ref{eikmvsdip}) do not have
the same sign for the $3$-quark-line process, this is not the case
after line-merging. As we see from the r.h.s.~of 
eqs.~(\ref{matn1})--(\ref{matn4}), such an operation leads to 
degenerate results, and thus to immediate cancellations and to the
final result of eq.~(\ref{Meq2M}), i.e.~to a linear combination of
dipole graphs with coefficients all equal to one.

The rightmost matrices in eqs.~(\ref{matn1}) and~(\ref{matn3}) correspond 
to the dipole graphs $\Gamma_2$ and $\Gamma_1$ of fig.~\ref{fig:qqgg_dipoles}, 
respectively. Using these identifications, those relevant to $\eiksum$ and
$\eiksum^+$ found after eq.~(\ref{eikMCL2ex1mer}) and~(\ref{matp5}), as well 
as eq.~(\ref{MeqMpMm}), we arrive at
\beq
\dipeik\left(\Gamma_6\right)=\dipeik\left(\Gamma_4\right)+
\dipeik\left(\Gamma_3\right)+\dipeik\left(\Gamma_5\right)-
\Big[\dipeik\left(\Gamma_1\right)+\dipeik\left(\Gamma_2\right)\Big]\,,
\eeq
which coincides with eq.~(\ref{qqggeikid}). This is a direct manifestation
of the non-uniqueness of the choices in eq.~(\ref{subdipset}), although
of limited practical interest (since the present example is relevant to 
a single closed flow, rather than to the set of all closed flows that one 
needs to consider in physical applications).

\phantomsection
\addcontentsline{toc}{section}{References}
\bibliographystyle{JHEP}
\bibliography{FKSflows}

\end{document}